%% file: ParametricQuantile3.tex
\documentclass[12pt,indent]{article}
\usepackage{a4,latexsym,amsmath}
\usepackage[latin1]{inputenc}
\usepackage{float}
\usepackage{natbib}
\usepackage{graphics}
\usepackage[english]{babel}
\usepackage{tabularx}
\usepackage{lscape}
\usepackage{multirow}
\usepackage{rotating}
\usepackage{dsfont}
\usepackage{subfig}
\usepackage{bbm}
\usepackage[table]{xcolor}

\usepackage{amsmath}
\usepackage{amsfonts}

\usepackage{calc}
\usepackage{ifthen}

\usepackage{%
  longtable,
  array,
  booktabs,
  dcolumn,
  rotating,
  shortvrb,
  tabularx,
  units,
  multirow
}

\newenvironment{tabularsmall}
{ \footnotesize \sffamily \tabular } {
\endtabular
\normalfont }

\input{def}

\newcommand{\argmin}{\operatorname{argmin}} 
\usepackage{natbib}

\usepackage{setspace}

\begin{document}
\bibliographystyle{chicago}
\sloppy

\makeatletter
\renewcommand{\section}{\@startsection{section}{1}{\z@}%
        {-3.5ex \@plus -1ex \@minus -.2ex}%
        {1.5ex \@plus.2ex}%
        {\reset@font\Large\sffamily}}
\renewcommand{\subsection}{\@startsection{subsection}{1}{\z@}%
        {-3.25ex \@plus -1ex \@minus -.2ex}%
        {1.1ex \@plus.2ex}%
        {\reset@font\large\sffamily\flushleft}}
\renewcommand{\subsubsection}{\@startsection{subsubsection}{1}{\z@}%
        {-3.25ex \@plus -1ex \@minus -.2ex}%
        {1.1ex \@plus.2ex}%
        {\reset@font\normalsize\sffamily\flushleft}}
\makeatother



\newsavebox{\tempbox}
\newlength{\linelength}
\setlength{\linelength}{\linewidth-10mm} \makeatletter
\renewcommand{\@makecaption}[2]
{
  \renewcommand{\baselinestretch}{1.1} \normalsize\small
  \vspace{5mm}
  \sbox{\tempbox}{#1: #2}
  \ifthenelse{\lengthtest{\wd\tempbox>\linelength}}
  {\noindent\hspace*{4mm}\parbox{\linewidth-10mm}{\sc#1: \sl#2\par}}
  {\begin{center}\sc#1: \sl#2\par\end{center}}
}



\def\R{\mathchoice{ \hbox{${\rm I}\!{\rm R}$} }
                   { \hbox{${\rm I}\!{\rm R}$} }
                   { \hbox{$ \scriptstyle  {\rm I}\!{\rm R}$} }
                   { \hbox{$ \scriptscriptstyle  {\rm I}\!{\rm R}$} }  }

\def\N{\mathchoice{ \hbox{${\rm I}\!{\rm N}$} }
                   { \hbox{${\rm I}\!{\rm N}$} }
                   { \hbox{$ \scriptstyle  {\rm I}\!{\rm N}$} }
                   { \hbox{$ \scriptscriptstyle  {\rm I}\!{\rm N}$} }  }

\def\d{\displaystyle}

\title{ Flexible Predictive Distributions from Varying-Thresholds Modelling}
\author{Gerhard Tutz \\{\small Ludwig-Maximilians-Universit\"{a}t M\"{u}nchen}\\
\small Akademiestra{\ss}e 1, 80799 M\"{u}nchen }


\maketitle

\begin{abstract} 
\noindent
A general class of models is proposed that is able to estimate the whole predictive distribution of a dependent variable $Y$ given a vector of explanatory variables $\xb$. The models exploit  that the strength  of explanatory variables to distinguish between  low and high values of the dependent variable may vary across the thresholds that are used to define low and high. Simple linear versions of the models are   generalizations of classical linear regression models but also of widely used ordinal regression models. They allow to visualize the effect of explanatory variables in the form of parameter functions. More general models are based on efficient nonparametric approaches like random forests, which are more flexible and are strong prediction tools. A general estimation method is given that can use all the estimation tools  that have been proposed for binary regression, including  selection methods like the lasso or elastic net. 
For linearly structured models maximum likelihood estimates are derived. The usefulness of the models is illustrated by simulations and several real data set.

\end{abstract}


\section{Introduction}
Thresholds are typically used to distinguish between more and less severe responses.  For example, in survival modeling important thresholds that are often used are three or five years survival rates.
Also when investigating the risk factors associated with low infant birth weight threshold values have been used to define low birth weight \citep{HosLem:89,VenRip:2002}. If thresholds are used one wants to know which explanatory variables determine responses above or below that threshold.
Unfortunately, it is typically ignored that the choice of the threshold can be crucial for the examination of the impact of explanatory variables because for different thresholds different variables may be influential.

The present paper provides tools to investigate the dependence on thresholds. Beyond that a general class of models is proposed that allows for variation of the dependence on covariates over thresholds. The model is very flexible, it can be based on linear predictors, which provide parameter functions that show  the variation over thresholds.  But also nonparametric predictors can be used to obtain a more flexible form of the dependence and better prediction.  All models allow for non-linear effects, if trees are used also interaction effects are taken into account.   

A strength of the model class is that it makes no strict assumptions on the conditional distribution of the dependent variable in a regression setting. The form of the distribution is determined by the variation of parameters or more general predictors. 
The models allow to estimate the whole conditional distribution $Y|\xb$ of a dependent variable $Y$ given a vector of explanatory variables $\xb$ instead of solely the mean $\E(Y|\xb)$, which is typically used in regression.
In the last decades several methods have been proposed that aim at investigating the conditional distribution more closely. Quantile regression 
\citep{koenker2001quantile,fitzenberger2013economic,koenker2017handbook,koenker1994quantile, meinshausen2006quantile} allows to estimate the  quantiles of the dependent variables as a function of explanatory variables. Transformation models estimate the whole predictive distributions under weak restrictions 
 \citep{hothorn2014conditional,hothorn2018most,hothorn2021predictive}.
The  models proposed here also estimate the predictive distribution, but in contrast to transformation models the focus is on the dependence of thresholds on explanatory variables. In basic models the dependence of the variable $Y$ on explanatory variables is visualized by parameter functions that have an easy interpretation.

In Section \ref{sec:vary} the models are motivated and a simple version is derived as an  extension  of the classical linear regression model. It is shown that the models also generalize  widely used models for ordinal responses. A general estimation method is given that uses existing software tools to obtain flexible estimates. For linearly structured models maximum likelihood estimation methods are derived, which yield smoother parameter functions. Then it is demonstrated how variable selection tools and nonparametric approaches can be used to obtain more flexible models. In Section \ref{sec:quant} it is shown how the models may be used as quantile regression methods. After considering bootstrap confidence intervals (Section \ref{sec:boot}) in Section \ref{sec:pred} the ability of the models to obtain bettter predictions is demonstrated.

\section{Varying-Thresholds Models}\label{sec:vary}

\subsection{Motivation and Basic Model}

Let us consider a regression model $Y= \gamma_0 + \xb^T \gammab+ \sigma \varepsilon$, where $Y$ is a metrically scaled response, $\xb$ is a vector of explanatory variables, $\gamma_0, \gammab$ are coefficients, and $\sigma$ is the standard deviation linked to the standardized symmetric noise variable $\varepsilon$, which has distribution function $F(.)$. Main objectives in regression modelling are prediction and investigating the impact of explanatory variables. The latter is usually obtained by examining the size and significance of the entries of the coefficient vector $\gammab$.

If the simple linear regression model holds one obtains a classical binary response model for a fixed threshold. Let, for simplicity $F(.)$ be a symmetric distribution function. Then  one obtains for fixed threshold $\theta$
\begin{equation}\label{eq:basic}
P(Y > \theta| \xb) = F(\frac{\gamma_0-\theta}{\sigma} + \xb^T \frac{\gammab}{\sigma}) = F({\beta_0} + \xb^T {\betab}),
\end{equation}
where $\beta_0=(\gamma_0-\theta)/{\sigma}, \betab={\gammab}/{\sigma}$. Thus,  for the binary variable $Y_{\theta} = 1$  if $Y > \theta$, and  $Y_{\theta} = 0$  otherwise,  the binary regression model
\[
P(Y_{\theta} = 1| \xb) =  F({\beta_0} + \xb^T {\betab})
\]
holds. Fitting of the binary model yields estimates of $\betab$, which is a scaled version of $\gammab$. Also inference on the relevance of  covariates 
can  be obtained from estimates of  $\betab$ although one uses less information than by fitting the original regression model. More interestingly, one can fit binary regression models for several thresholds and compare the inferences drawn from them to investigate the dependence structure more closely.

The objective here is not to derive alternative inference tools for the parameters of the continuous regression model but to consider a much more general model, which provides diagnostic tools but is more generally able to draw inferences about  the conditional distribution of $Y|\xb$ without assuming a fixed distribution. 

The simplest parametric version is the  \textit{varying-thresholds model} 
\begin{equation}\label{eq:parthr}
P(Y> \theta| \xb) =  F({\beta_0}{(\theta)}+ \xb^T {\betab}{(\theta)}),
\end{equation}
where the parameters $\beta_0{(\theta)}, {\betab}{(\theta)}$ vary across thresholds. It is a direct generalization of model (\ref{eq:basic}) but with thresholds depending on parameters.  Though a fixed distribution function $F(.)$ is used to fit models, the distribution of $Y|\xb$ is much more flexible due to the dependence of the parameters on $\theta$. For simplicity we refer to the model as a varying-thresholds model \textit{with a linear predictor} but it is definitely not a linear model, not because of the transformation function $F(.)$ but because of the semi-parametric structure of the predictor, which makes the impact on the dependent variable non-linear.  


Advantages of the varying-thresholds  model are:
\begin{itemize}
\item The linear term in the model ${\beta_0}^{(\theta)}+ \xb^T {\betab}^{(\theta)}$ allows to investigate the importance of explanatory variables at specific thresholds. Typically it will vary over thresholds.
\item The flexibility of the model allows to estimated the whole distribution function of $Y|\xb$ and therefore investigate the quantiles as a function of explanatory variables in a very flexible way.
\item The method works without assuming a distribution for the dependent variable given covariates and provides an alternative to quantile regression.
\item By using selection methods as the lasso or elastic net one gets easy to obtain variable selection in quantile regression
\item Prediction can substantially be improved by using a model that does not assume that the distribution of the dependent variable is known.
\end{itemize}

Model (\ref{eq:parthr}) is the simplest version of the varying coefficient model that can be used as a diagnostic tool for investigating if a linear regression model really captures the essential impact structure. The class of models is much wider and more flexible. A general version of the varying-thresholds model  
has the form 
\begin{equation}\label{eq:genthr}
 P(Y> \theta| \xb) =  F(\eta{(\theta,\xb)}),
\end{equation}
where $\eta{(\theta,\xb)}$ is a predictor function depending on $\theta$ and $\xb$. It can be a linear predictor as in the parametric model considered above but can be much more flexible. For example, it can be a smooth predictor as in generalized additive models,
$
 \eta^{(\theta,\xb)}= {\beta_0}{(\theta)} + \sum_j s(x_j,\theta),
$
where $s(x_j,\theta)$ is a two-dimensional smooth function of $x_j$ and $\theta$. However, $\eta{(\theta,\xb)}$ can also be estimated by using nonparametric approaches as  random forests, which typically yields good prediction results. 
Then one gains flexibility but the simple interpretation of the linear term gets lost.

It should be noted that the varying-thresholds model is a varying-coefficients model but quite different from classical concepts of varying-coefficients models. 
Classical varying-coefficients models, which have been considered by \citet{HasTib:93,cai2000efficient,fan1999statistical,park2015varying}, assume a semi-parametric predictor $\eta =\beta_0(z_0)+x_1\beta_1(z_1)+\dots+x_p\beta_p(z_p)$ that allows that the effects of predictors are modified by other variables, the so-called effect modifiers. More concise, the effect of $x_j$ is modified by $z_j$, and it is typically assumed that the function $\beta_j(z_j)$ is smooth. 

In model (\ref{eq:parthr}) the varying coefficients have a quite different meaning. The coefficients $\beta_1(\theta),\dots,\beta_p(\theta)$ on predictors $x_1,\dots,x_p$ indicate how well a predictor variable is able to distinguish between low and high values of the dependent variable, where low and high is defined by $Y \le \theta$ and $Y > \theta$ for fixed value $\theta$. The ability to distinguish can distinctly  vary over thresholds. There is no external effect modifier as in classical varying-coefficients models. This has several advantages. In particular one has not to find and identify the variables that modify effects, which is no easy task. In addition the variation of effects is always one-dimensional. In  classical varying-coefficients models a parameter $\beta_j(\zb_j)$ can depend on a vector of covariates $\zb_j$, which calls for higher dimensional smoothers, which are hard to handle in dimensions larger than two.

There are strong links between the general model (\ref{eq:genthr}) and the so-called transformation models proposed by \citet{hothorn2014conditional}, \citet{hothorn2018most}, which have been extended to transformation forests more recently by \citet{hothorn2021predictive}. Transformation models assume $P(Y\le y)=F(h(y))$, where $h(.)$ is an unknown increasing transformation function. The transformation function can be additive \citep{hothorn2014conditional} or nonparametric \citep{hothorn2021predictive}. Although the models look quite similar there are some crucial differences. In transformation models with a parameterized transformation function $h(y)= \ab(y)^T\deltab$ the function $\ab(.)$ is expanded in basis functions to obtain a flexible predictor. With the focus on  thresholds,  in linear varying-thresholds models  basis functions are used to obtain flexible parameter functions $\betab(\theta)$ instead of using fixed parameters on transformed $y$-values.
While transformation models have been fitted by variants of componentwise boosting the varying-thresholds models are fitted either by using available software tools that fit binary models or, in the linear case, by maximum likelihood with expanded parameter functions.

\subsection{Linearly Structured Predictors}

A particularly interesting version of the varying-coefficient model is  model (\ref{eq:parthr}), which uses a linear predictor $\eta{(\xb)}= \beta_0{(\theta)} + \xb^T \betab{(\theta)}$. It is a generalization of  models that are used in metric but also in ordinal regression. 

\subsubsection*{Classical Linear Regression}
As shown in the Appendix  the linear regression model is equivalent to a varying-coefficient model with a linear intercept function $\beta_0{(\theta)}$.  If the regression model $Y= \gamma_0 + \xb^T \gammab+ \sigma \varepsilon$ with symmetric distribution function $F(.)$ holds one obtains  the varying-coefficient model $P(Y > \theta|\xb)=F(\beta_0(\theta)+\xb^T\betab)$ with
$\beta_0(\theta)$  given by $\beta_0(\theta)=\gamma_0/\sigma - \theta\sigma$. Conversely, if a varying coefficients model with a \textit{linear} intercept function,
\[
P(Y > \theta|\xb)=F(\alpha_0 + \alpha\theta+\xb^T\betab) 
\]
holds, one obtains the linear regression model $Y= \gamma_0 + \xb^T \gammab+ \sigma \varepsilon$ with distribution function $F(.)$ for $\varepsilon$ with the variance given by $\sigma=\var_F/\alpha^2$, where $\var_F$ is the variance of $F(.)$. In summary, the varying-thresholds model with  a linear intercept function and a linear term $\xb^T \betab$ is equivalent to a classical linear regression model. 

If one  fits a varying-thresholds model and constrains the estimates such that the intercept function is linear one fits a classical linear model. If one lets the parameter functions vary freely one fits a much more flexible model, which can be used to diagnose deviations from the linear model.
If the predictor has varying-coefficients, $\eta{(\xb)}= \beta_0{(\theta)} + \xb^T \betab{(\theta)}$  with $\betab(\theta)$ varying over values $\theta$ the model is definitely not a linear regression model. The dependence of the mean on the covariates can take  quite different, in particular non-linear forms.  
Plots of the varying coefficients indicate how far the underlying dependence structure is away from the classical linear model.  


\subsubsection*{Ordinal Regression} 

Since McCullagh's seminal article \citet{McCullagh:80} ordinal regression methods have become widely used tools that take the scale level of the dependent variable seriously and avoid problems that occur from using classical regression models that falsely assume that the  dependent variable is continuous, see, for example, \citet{Agresti:2009}. The most prominent model is the proportional odds model and the versions that derive from it. 

Let the dependent $Y \in \{1,\dots,k\}$ denote the ordinal response. Then the basic \textit{cumulative model}  has the form 
\begin{equation}\label{eq:cum}
P(Y > r|\xb) = F (\beta_{0r}+\xb^T\betab),                                                                             
\end{equation} 
where $F(.)$ is a distribution function. If $F(.)$ is chosen as the  logistic distribution function one obtains the \textit{proportional odds model}, which has easy to interpret effects, see \citet{Agresti:2009, TutzBook2011}. It is easily seen that the cumulative model is a varying-thresholds model. Since the dependent variable is discrete, only a finite number of thresholds can be distinguished. In $P(Y> \theta| \xb) =  F({\beta_0}{(\theta)}+ \xb^T {\betab}{(\theta)})$ all thresholds   $\theta \in [r,r+1)$ yield the same probabilities. Therefore, only thresholds $\theta_1,\dots,\theta_{k-1}$ given by $\theta_r=r$  are needed.

If one assumes that the parameter function linked to the explanatory variables do not vary over the categories, that is, $\betab(\theta)=\betab$, one obtains the ordinal cumulative model (\ref{eq:cum}). Thus, the cumulative model is a varying-coefficients model, for which ${\beta_0}{(\theta)}$ varies across categories, but ${\beta_j}{(\theta)}, j=1,\dots,p$ do not depend on $\theta$.

The link between varying-coefficients models and ordinal regression models is even stronger. The cumulative model (\ref{eq:cum}) has been extended to allow for predictors $\eta_r=\beta_{0r}+\xb\betab_r$, in which some or all of the effects of explanatory variables depend on the categories. These extension are often referred to as partial proportional odds models, see, for example, \citet{Brant:90}, \citet{PetHar:90}, \citet{Cox:95},   \citet{liu2009graphical},  \citet{williams2016understanding}. The varying-coefficients model for ordered categorical responses can be seen as the general cumulative model with possibly category-specific effects for all explanatory variables.

It is somewhat surprising that a model can be used for metric responses as well as for categorical responses without violating the demands of the scale level.
The reason is that the varying-coefficients model is essentially a split model that distinguishes between a response below or above a threshold. That approach can be used for all kinds of  dependent variables that contain order information, that means continuous metrically scaled variables but also ordered categories.
The main difference is that for continuous variables the parameter functions are also continuous while they are discrete in the case of ordered categories.

\subsection{Fitting Varying-Coefficients models by Using Binary Models}\label{sec:algo}

We first consider a general concept to estimate the conditional  distributions $Y|\xb$ for varying-thresholds models. The method applies for all sorts of predictors. More elaborate estimation concepts that apply in linear varying-thresholds models are considered  in Section  \ref{sec:ML}.

Let a grid of thresholds from the range of $Y$, $\theta_0 < \dots < \theta_k$ with $\theta_0= \min\{Y_1,\dots, Y_n\}$, $\theta_k =\max\{Y_1,\dots, Y_n\}$, be chosen, yielding $k$ intervals.

For threshold $\theta_r$ define binary variables 
\[
Y^{(r)} = 1 \text{ if } Y > \theta_r,  Y^{(r)} = 0 \text{ otherwise. }
\]         
A general method to obtain estimates of $Y|\xb$ is the following:

\begin{itemize}
\item[-] 
Fit binary models $P(Y^{(r)}=1|\xb) = F(\eta_r{(\xb)})$ to obtain predictors $\eta_r{(\xb)}$.
Estimate the conditional distribution function of $Y|\xb$ at points $\theta_1, \dots , \theta_{k-1}$ by 
\[
\hat F_{Y|\xb}(\theta_r|\xb) = \hat P(Y \le \theta_r|\xb)= 1- F(\hat\eta_r{(\xb)}).
\] 

\item[-]
The estimate $\hat F_{Y|\xb}(.)$ is obtained as a linear interpolation  between thresholds with $\hat F(\theta_0)=0$, $\hat F(\theta_k)=1$.

\item[-]
Before interpolation one has to account for the restriction $\hat F_{Y|\xb}(\theta_1|\xb) \le \dots \le F_{Y|\xb}(\theta_{k-1}|\xb)$, which holds if
$\hat\eta_1{(\xb)} \ge \dots \ge \hat\eta_{k-1}{(\xb)}$. For $\xb$ values  for which the restriction  does not hold monotone regression is used.
\end{itemize}

When fitting a linear model by using $\eta_r{(\xb)}= \beta_0^{(r)} + \xb^T \betab^{(r)}$, the fitting procedure yields conditional distribution functions $\hat F_{Y|\xb}(.|\xb)$, as well as parameter functions $\hat\beta_0^{(r)}, \hat\betab^{(r)}$. The distribution functions can be used to compute relevant indicators, for example the mean, the median, the standard deviation of $Y|\xb$. In particular the latter can vary across the range of the dependent variable, which is  excluded in classical regression models. 

When fitting the varying-thresholds model an essential restriction is that $\hat F(\theta_{j+1}|\xb) \ge \hat F(\theta_{j}|\xb)$ has to hold for all $\xb$.
In the general algorithm we use monotone regression to obtain admissible estimates $\eta_r{(\xb)}$ whatever the functional form of $\eta_r{(\xb)}$ is by using monotone regression provided by the \textit{isoreg} function in R. By regressing $\eta_r{(\xb)}$  on $r$ one obtains predictors for which $\hat\eta_1{(\xb)} \ge \dots \ge \hat\eta_{k-1}{(\xb)}$ holds.  


The obtained values of the conditional distribution function $\hat F_{Y|\xb}(\theta_r|\xb)$ are used to obtain the estimated distribution function 
\[
\hat F_{Y|\xb}(\theta|\xb)= \hat F_{Y|\xb}(\theta_j|\xb) + \frac{\hat F_{Y|\xb}(\theta_{j+1}|\xb)-\hat F_{Y|\xb}(\theta_j|\xb)}{\theta_{j+1}-\theta_{j}}(\theta-\theta_{j}) \quad \theta \in [\theta_{j},\theta_{j+1}]
\]  
and $\hat F_{Y|\xb}(\theta|\xb)=0$ for $\theta \le \theta_j$, $\hat F_{Y|\xb}(\theta|\xb)=1$ for $\theta \ge \theta_k$.


\subsubsection*{Illustrative Simulation: Classical Model Holds}

Let us consider the simple case of a classical regression model  $Y= \gamma_0 + \xb^T \gammab+ \sigma \varepsilon$ with $\varepsilon$ following a standard normal distribution, and parameters $\gamma_0=1, \gamma_0=0.5, \gamma_0=1$, $\sigma=1$. The number of observations was $n=100$.

\begin{figure}[H]
\centering
\includegraphics[width=0.45\textwidth]{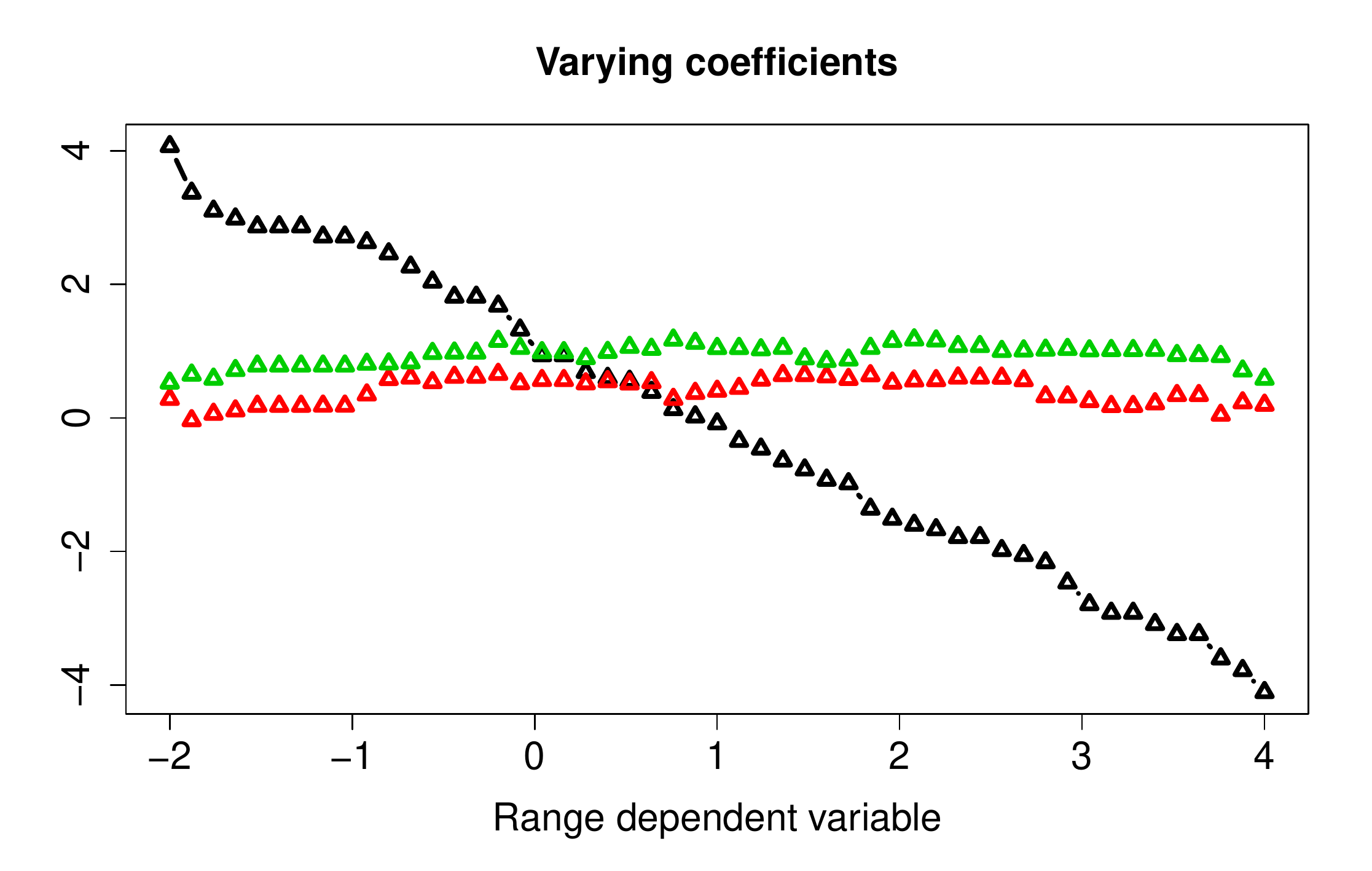}
\includegraphics[width=0.45\textwidth]{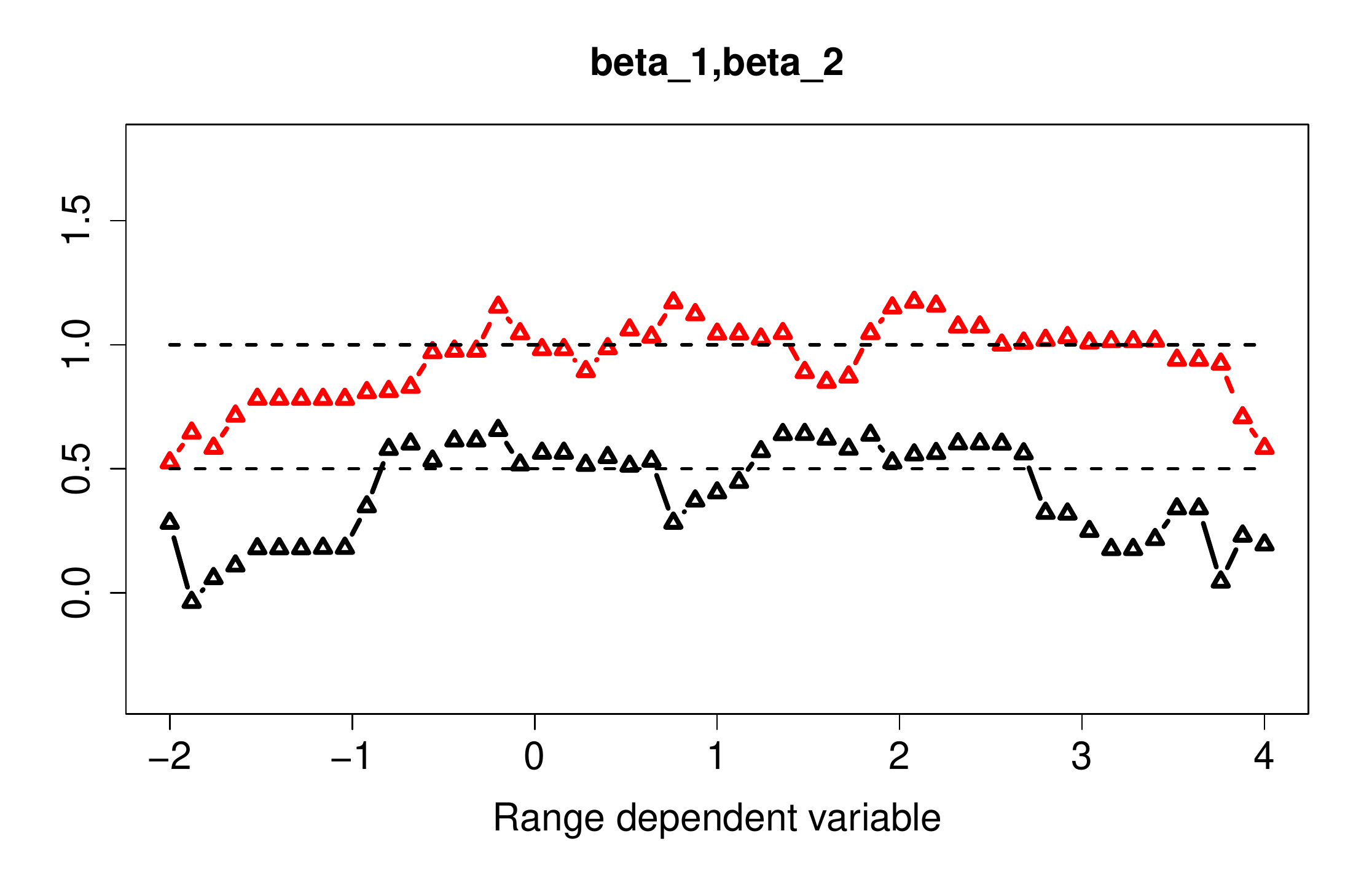}
\includegraphics[width=0.45\textwidth]{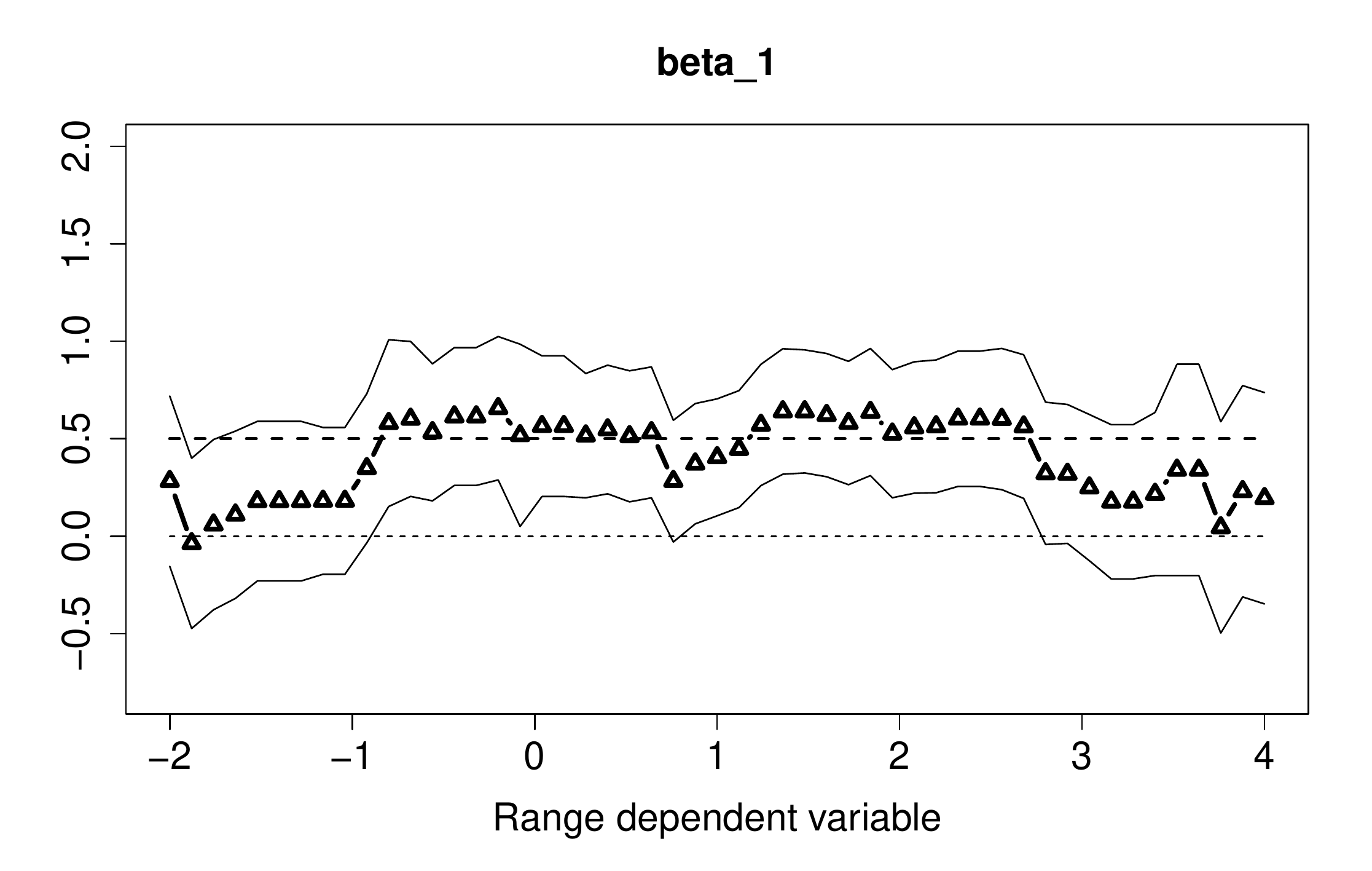}
\includegraphics[width=0.45\textwidth]{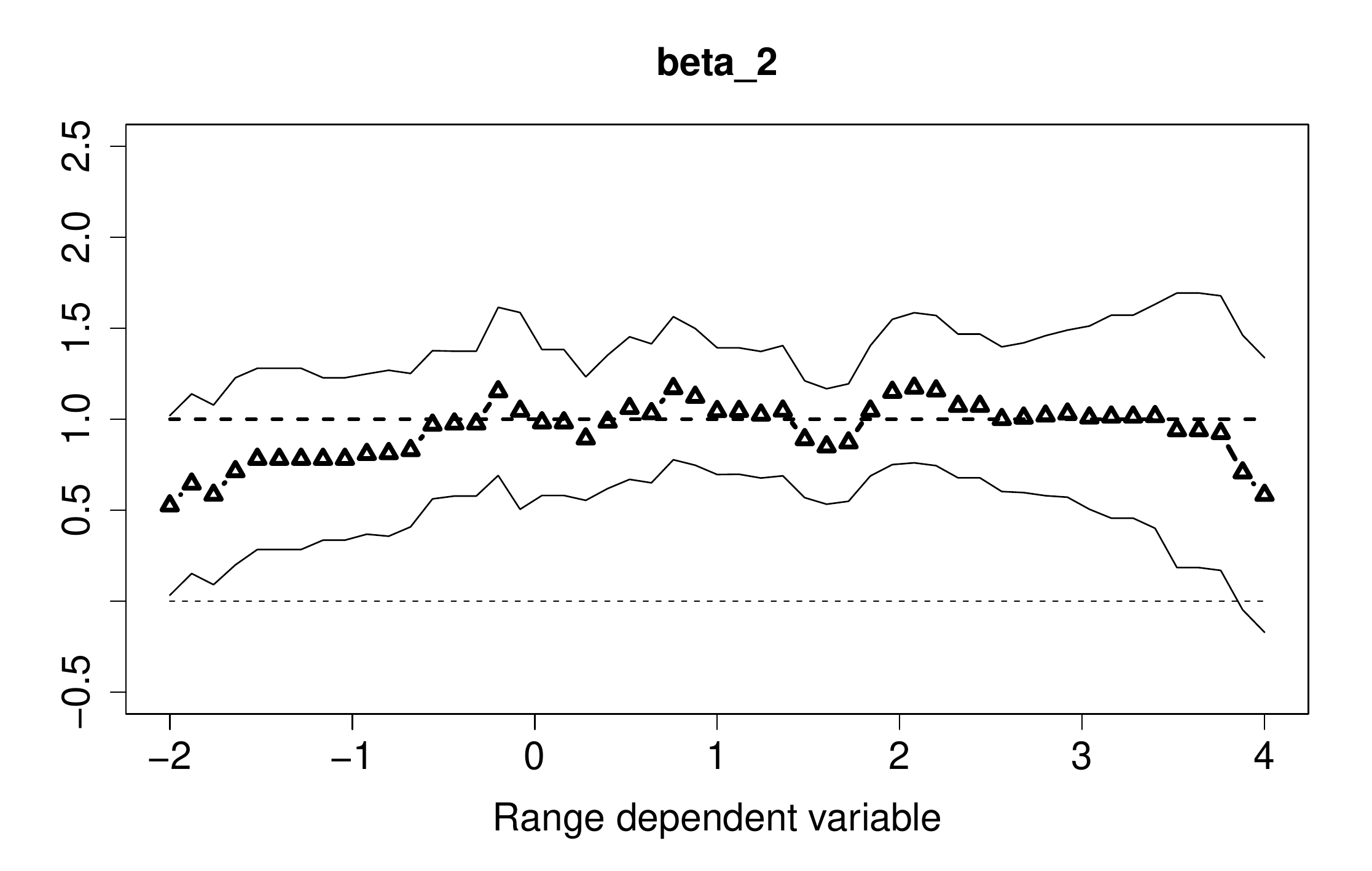}
\caption{Varying coefficients for simulation. Left upper panel: all coefficients, right upper panel: only $\beta_1, \beta_2$, left lower panel:  $\beta_1$ with error bounds +- 1.96 stderr, right lower panel:  $\beta_2$ with error bounds +- 1.96 stderr, smooth versions}
\label{fig:sim1}
\end{figure}

The upper panels of Figure \ref{fig:sim1} show the  estimated parameter function across the range of $Y$ if  $F(.)$ is chosen as the standard normal distribution. In the left panel the intercept is included,  the right panel is without intercept to show the variation of coefficients linked to explanatory variables. As expected the intercept is decreasing  since the number of observations with response '1' decrease over the range of $Y$. In addition, the decrease is approximately linear, which is to be expected from the derivations given in the preceding section.
The lower panels shows the two coefficients together with 0.95 pointwise confidence intervals computed at threshold values.  It is seen that over a wide range the estimated parameters are  stable and close to the true values (0.5 for variable 1 and 1.0 for variable 2).

\begin{figure}[H]
\centering
\includegraphics[width=0.45\textwidth]{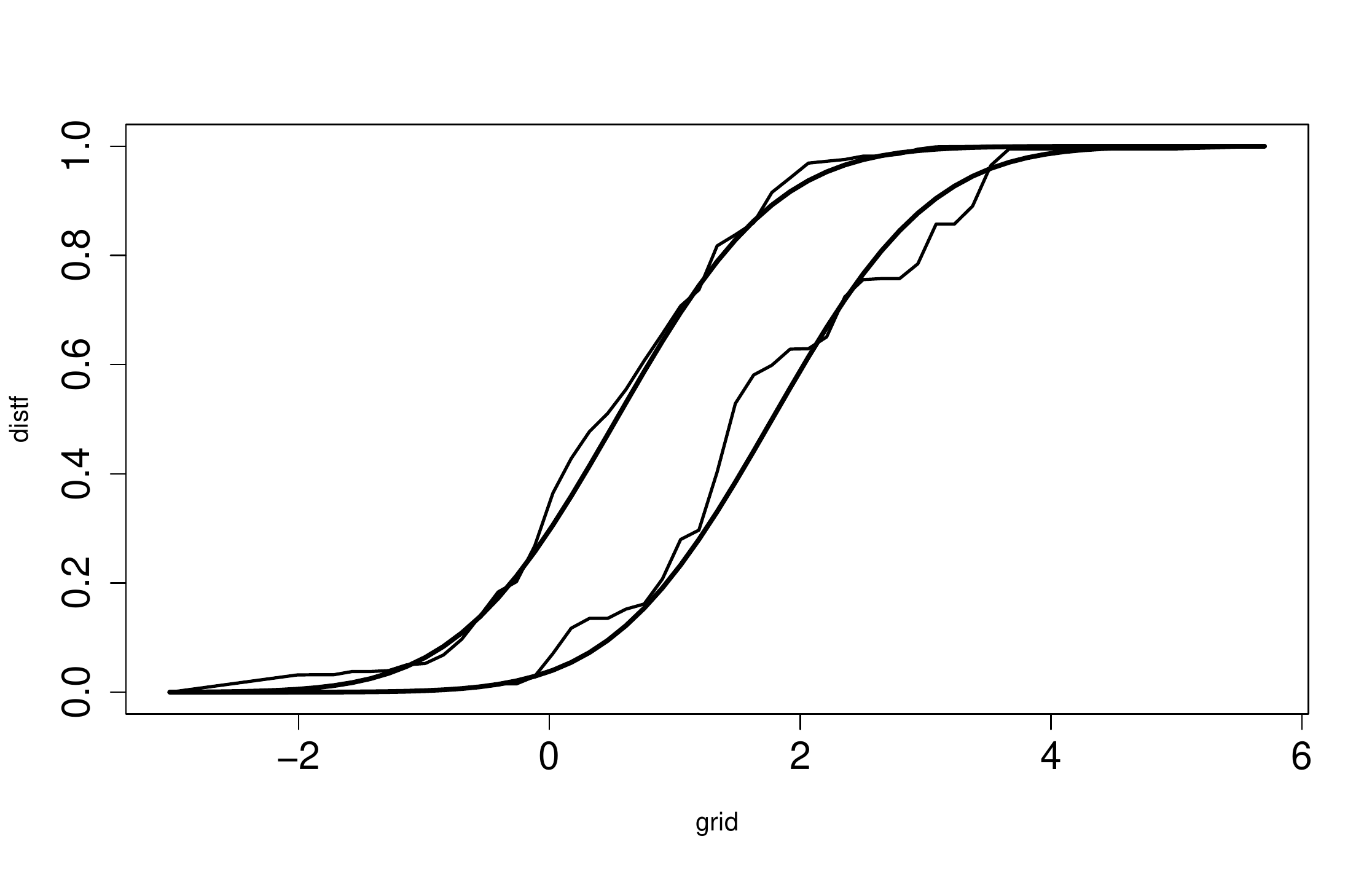}
\caption{Fitted and true distribution functions for two observations with mean values 0.53 (left) and 1.77 (right)}
\label{fig:simobs}
\end{figure}

Figure \ref{fig:simobs} shows the fitted  distribution functions together with the true function for two observations with mean values 0.53 (left) and 1.77 (right). It is seen that the estimated distribution functions are quite close to the true functions, in particular since the semi-parametric model that is used   does not assume that the distribution function is normal. 

\subsubsection*{Illustrative Application: Birth Weight Data}

In the \textit{lobwt} data set  contained in the R package \textit{rpartOrdinal} the dependent variable is birth weight (bwt). Explanatory variables are 
age (age of mother in years), lwt  (weight of mother at last menstrual period in Pounds),
smoke (Smoking status during pregnancy, 1: No, 2: Yes), ui (presence of uterine irritability),
ht (history of hypertension, 0: No, 1: Yes),
ftv (number of physician visits during the first trimester,1: None, 2: One, 3: Two, etc), race of the mother (white, black, or other), which, following \citep{hosmer1997comparison} has been coded into two design variables using white race as the referent group (Race1, Race2)).
The data set has been used before in binary and categorical regression after using thresholds, for example by \citet{hosmer1997comparison,galimberti2012classification}. However, the binary or categorical response was defined by fixed thresholds, which are considered as given. In contrast, here the thresholds are part of the modeling and the response is considered continuous, which yields differing results.

\begin{figure}[H]
\centering
\includegraphics[width=0.45\textwidth]{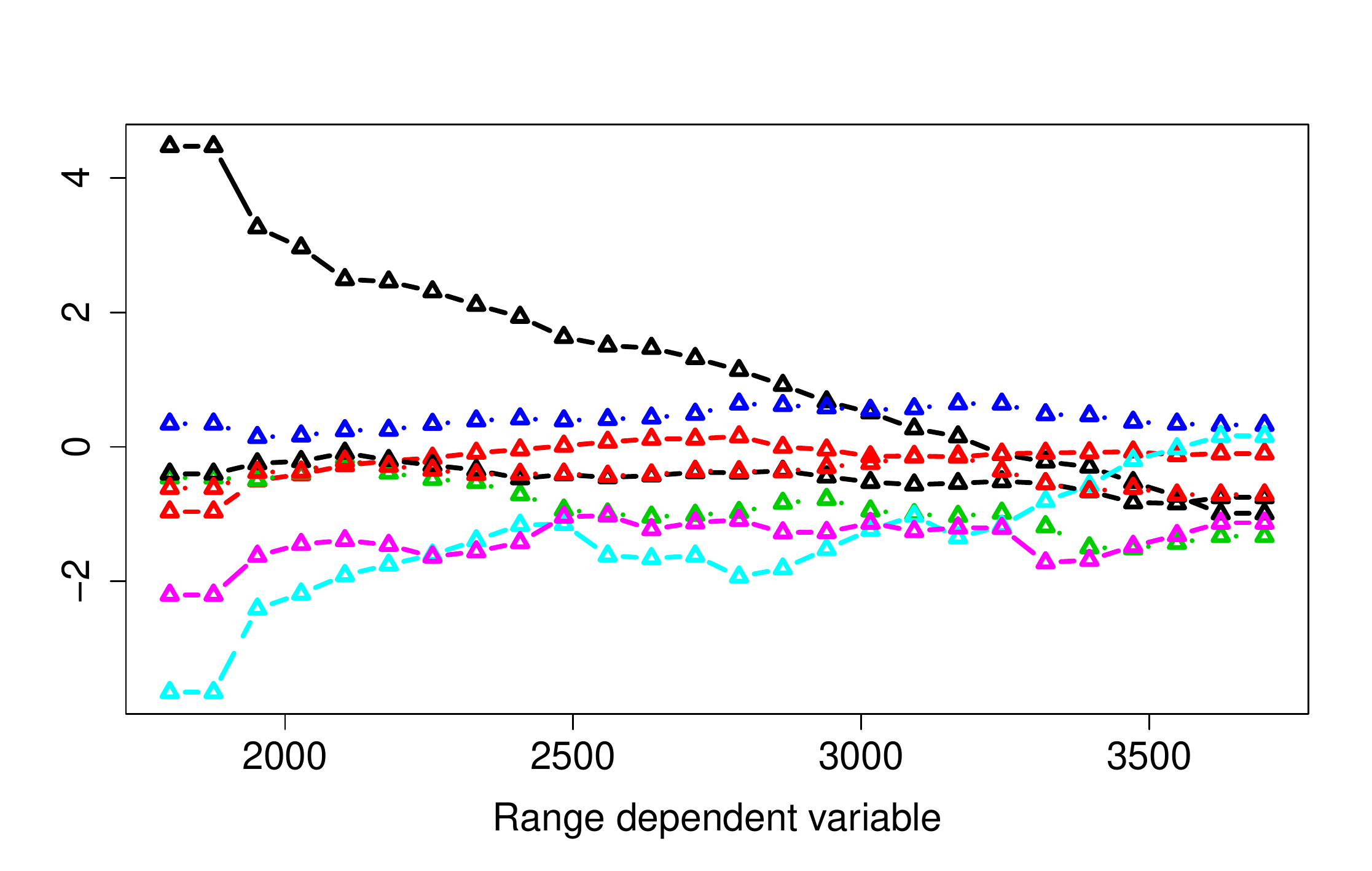}
\includegraphics[width=0.45\textwidth]{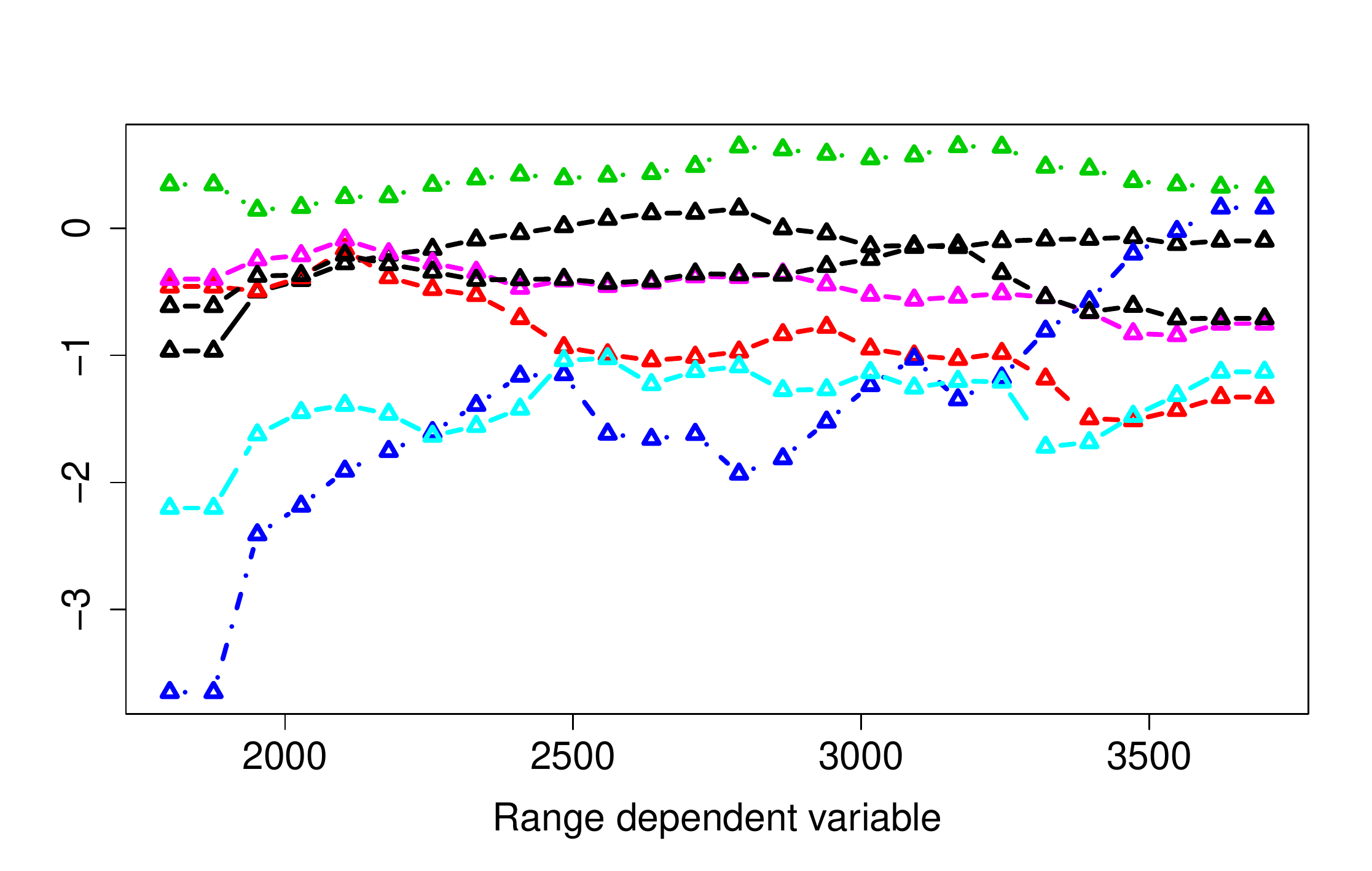}
\caption{Parameter functions for birth weight data with standardized predictors, in the left picture the intercept function is included, the right picture shows only coefficients linked to explanatory variables.}
\label{fig:birth1}
\end{figure}

\begin{figure}[H]
\centering
\includegraphics[width=0.45\textwidth]{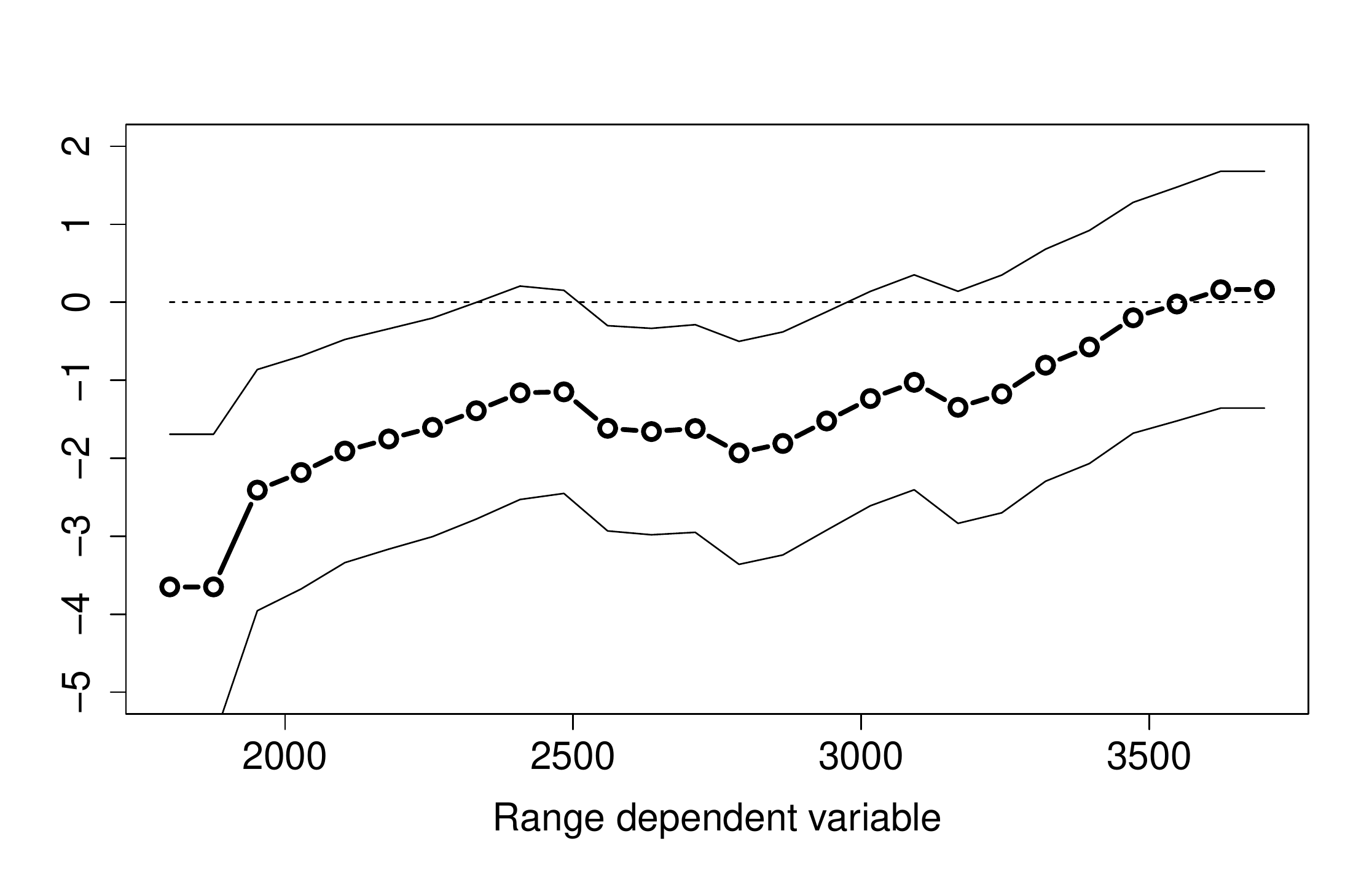}
\includegraphics[width=0.45\textwidth]{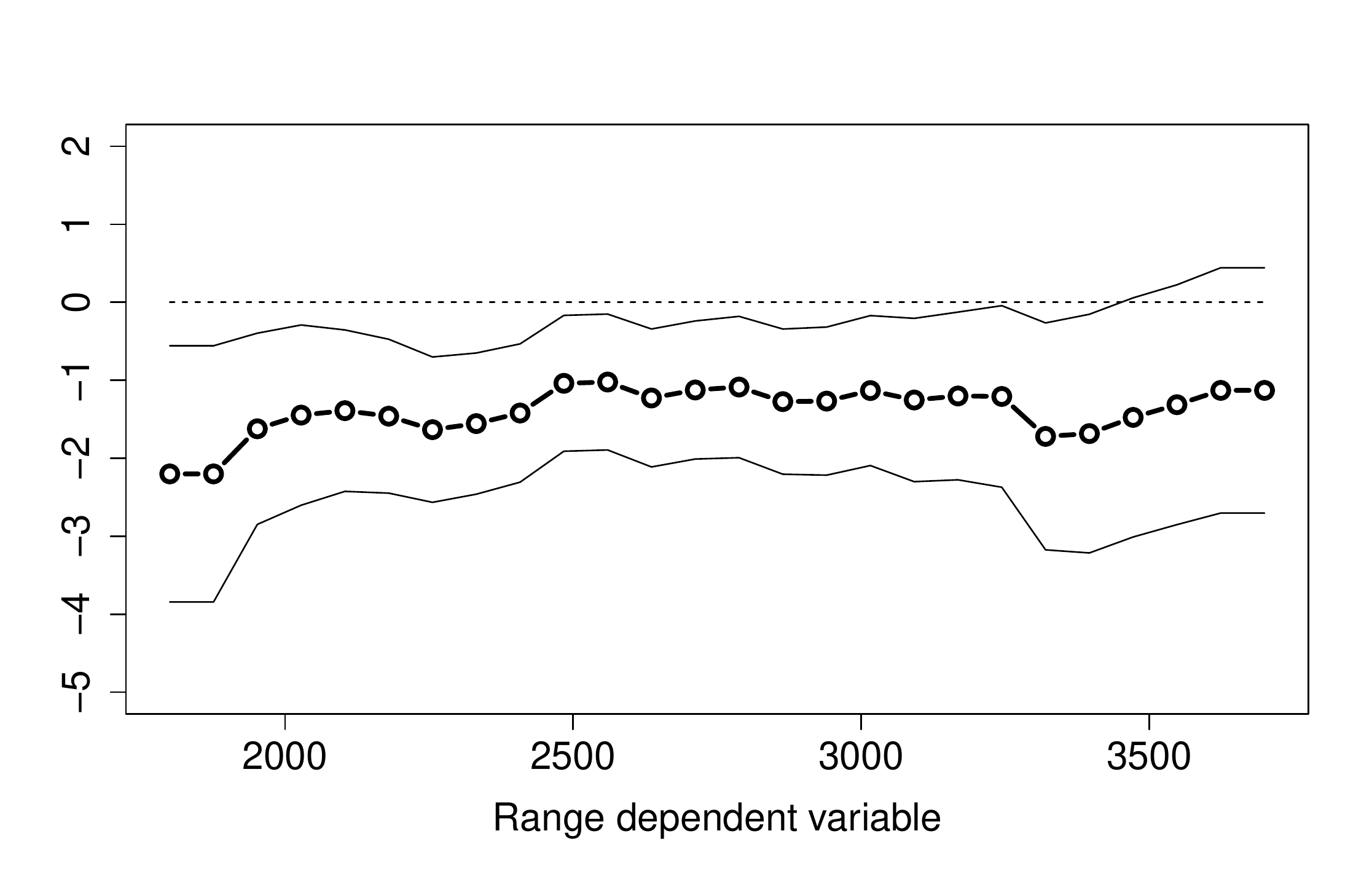}
\includegraphics[width=0.45\textwidth]{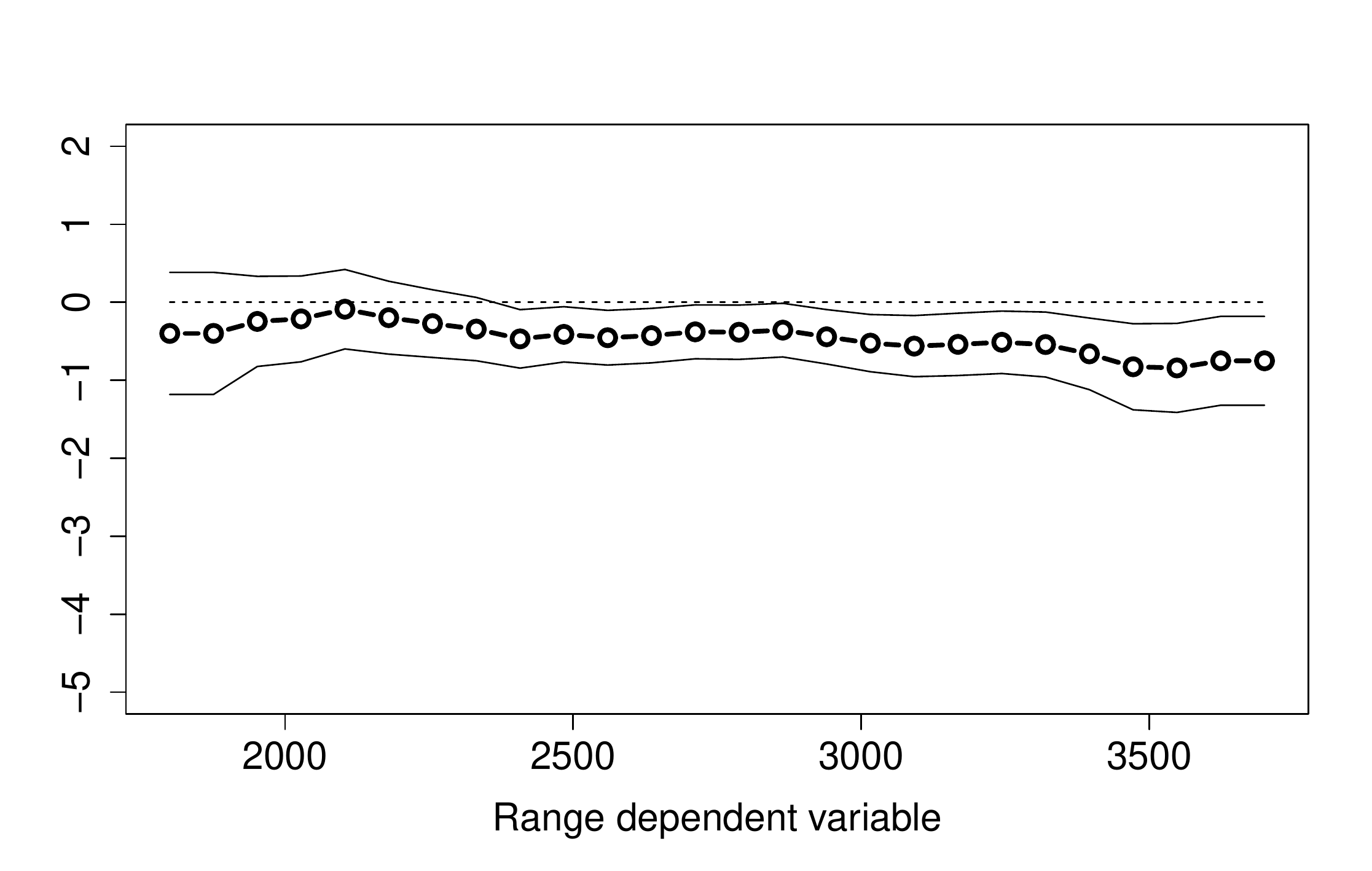}
\includegraphics[width=0.45\textwidth]{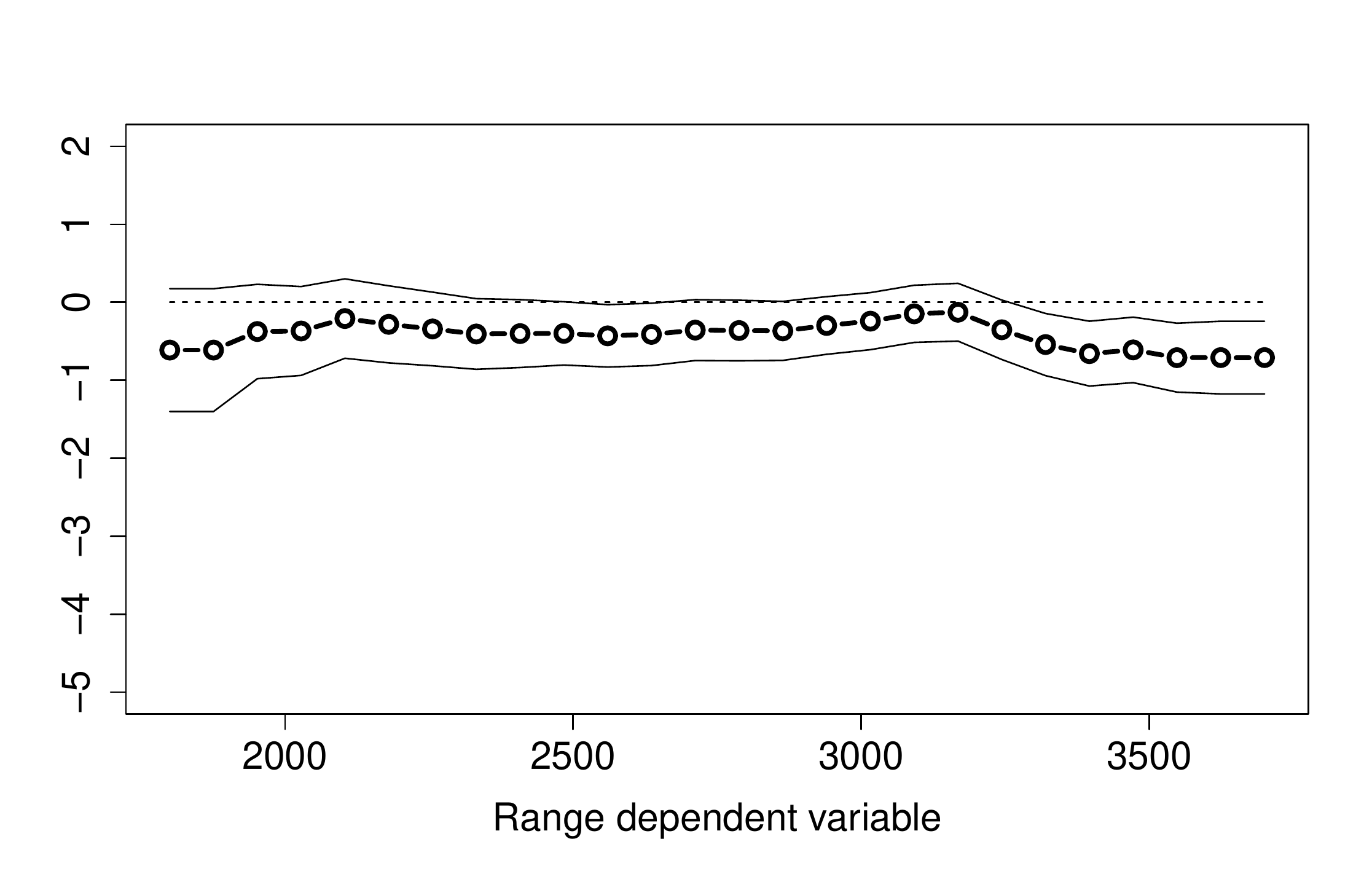}
\caption{Coefficients for variables hypertension,  uterine irritability, race1 and race2   for birth weight data.}
\label{fig:birth2}
\end{figure}

Figure \ref{fig:birth1} shows the variation of parameter estimates over the dependent variable for standardized variables when $F(.)$ is the logistic distribution function. In the left panel the intercept is included (decreasing function), in the right panel only variable coefficients are shown. It is seen that some parameter estimates vary strongly over the range of $Y$. Figure \ref{fig:birth2}
shows the  parameter estimates for hypertension, uterine irritability,and race1,race2 with confidence intervals. With the exception of hypertension all coefficients are very stable. The hypertension coefficient is increasing. the effect is strong  for low values but comes close to zero for high values.  That means, hypertension as a risk factor can distinguish between high and low levels in the lower range of the dependent variable.  The effects of race are stable and small but are not to be neglected. 

\subsection{Maximum Likelihood Estimation in Models with Linear Predictor}\label{sec:ML}

The general algorithm given in Section \ref{sec:algo}  can be used for all kinds of predictors. In the model with a linear predictor, $P(Y> y| \xb) =  F({\beta_0}{(y)}+ \xb^T {\betab}{(y)})$, an attractive alternative estimation method is (constrained) maximum likelihood estimation. From the distribution function 
\[
F_{Y|\xb}(y) =P(Y \le y| \xb) =  1-F({\beta_0}{(y)}+ \xb^T {\betab}{(y)}) 
\]
one obtains the density
\[
f_{Y|\xb}(y)=\frac{\partial F_{Y|\xb}(y)}{\partial y} = - f({\beta_0}{(y)}+ \xb^T {\betab}{(y)})(\beta_0{(y)}'+ \xb^T {\betab}{(y)}'),
\]
where $f(.)$ is the density corresponding to $F(.)$, and $\beta_0{(y)}', {\betab}{(y)}'$ are derivatives of the parameter functions.
Let the parameter functions be  specified by  $M$ basis functions. These can be radial basis functions or B-splines as propagated by \citep{EilMar:96,eilers2021practical}. With basis functions $\Phi_{j1},\dots, \Phi_{jM}$  one has
\[
\beta_j(y) = \sum_{l=1}^B \alpha_{jl}\Phi_{jl}(y)= \Phib_{j}(y)^T\alphab_{j},
\]
where $\Phib_{j}(y)^T=(\Phi_{j1}(y),\dots,\Phi_{jM}(y))$,  $\alphab_{j}^T=(\alpha_{j1},\dots,\alpha_{jM})$. Then one obtains for observation $(y_i,\xb_i)$
\[
f_{Y|\xb}(y_i) = - f(\sum_{j=0}^p x_{ij} \Phib_{j}(y_i)^T\alphab_{j})( \sum_{j=0}^p x_{ij} \Phib_{j}'(y_i)^T\alphab_{j})
\]
where $x_{i0}=1$, and $\Phib_{j}'(y_i)^T=(\Phi_{j1}'(y),\dots,\Phi_{jM}'(y))$ contains the derivatives of the basis functions.

Maximization of the corresponding likelihood is constrained by the restriction that ${\beta_0}{(y_i)}+ \xb^T {\betab}{(y_i)}$ is non-increasing for all observations $(y_i,\xb_i), i=1,\dots,n$, that is,
\[
\sum_{j=0}^p x_{ij} \Phib_{j}'(y_i)^T\alphab_{j} \le 0.
\]
Following \citet{EilMar:96} we use penalized splines, which means we use cubic B-splines as basis functions and maximize the penalized likelihood 
\[
l_p = l -  \lambda \sum_{j=1}^p \sum_{l=1}^{M-1} (\alpha_{j,l+1}-\alpha_{jl})^2,
\]
where $l$ is the usual likelihood, and $\lambda$ is a smoothing parameter. The difference penalty $\sum_{l} (\alpha_{j,l+1}-\alpha_{jl})^2$
ensures that the estimated functions are smooth. It should be noted that only the parameter functions $\beta_1{(.)},\dots,\beta_p{(.)}$ are penalized, not the
parameter function of the intercept $\beta_0{(.)}$. For small $\lambda$ the parameter functions are not restricted while for large $\lambda$ one fits the model
where $\beta_j{(.)}, j=1,\dots,p$ are constants, that is, the classical  regression model.

\begin{figure}[H]
\centering
\includegraphics[width=0.45\textwidth]{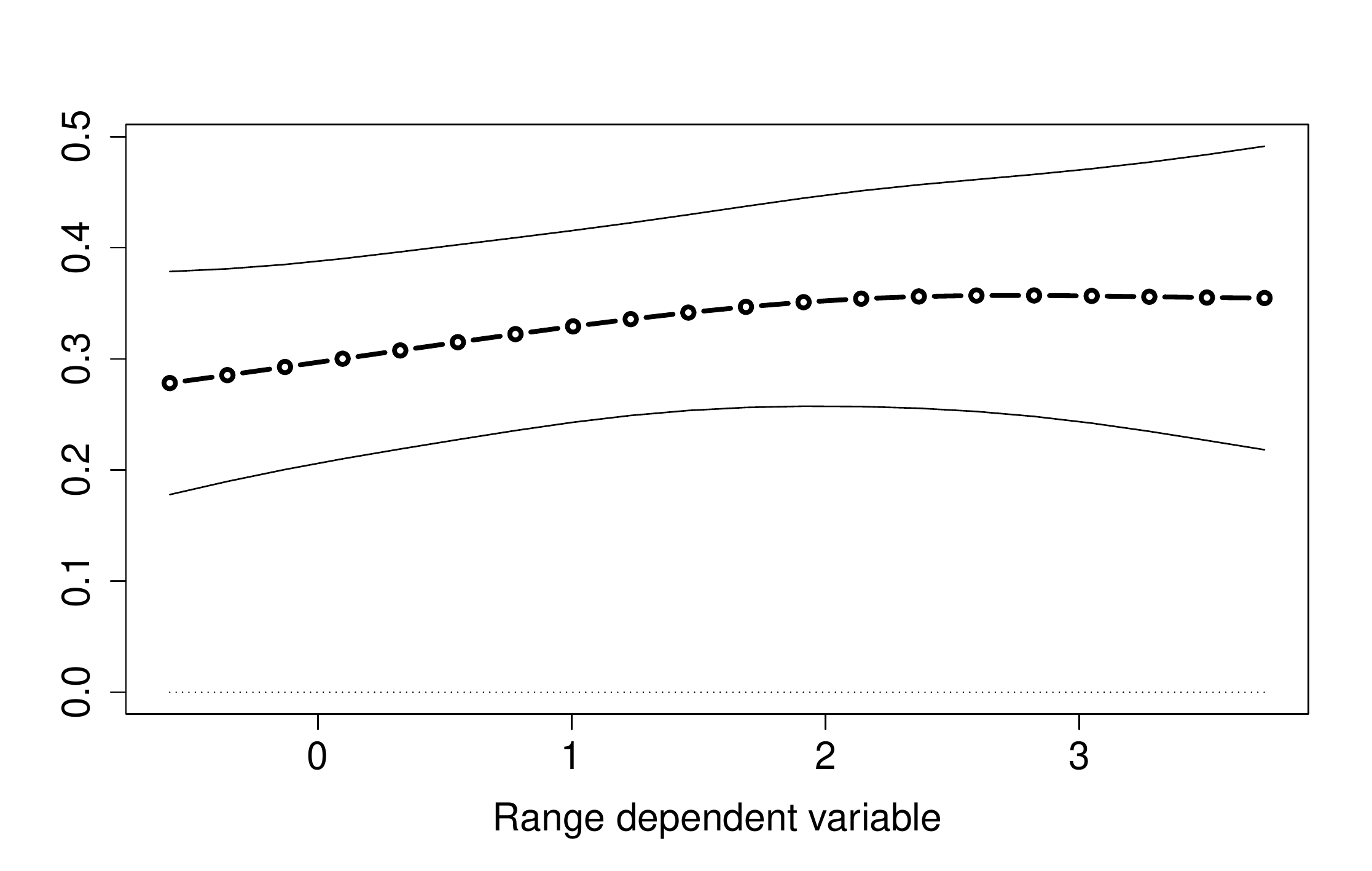}
\includegraphics[width=0.45\textwidth]{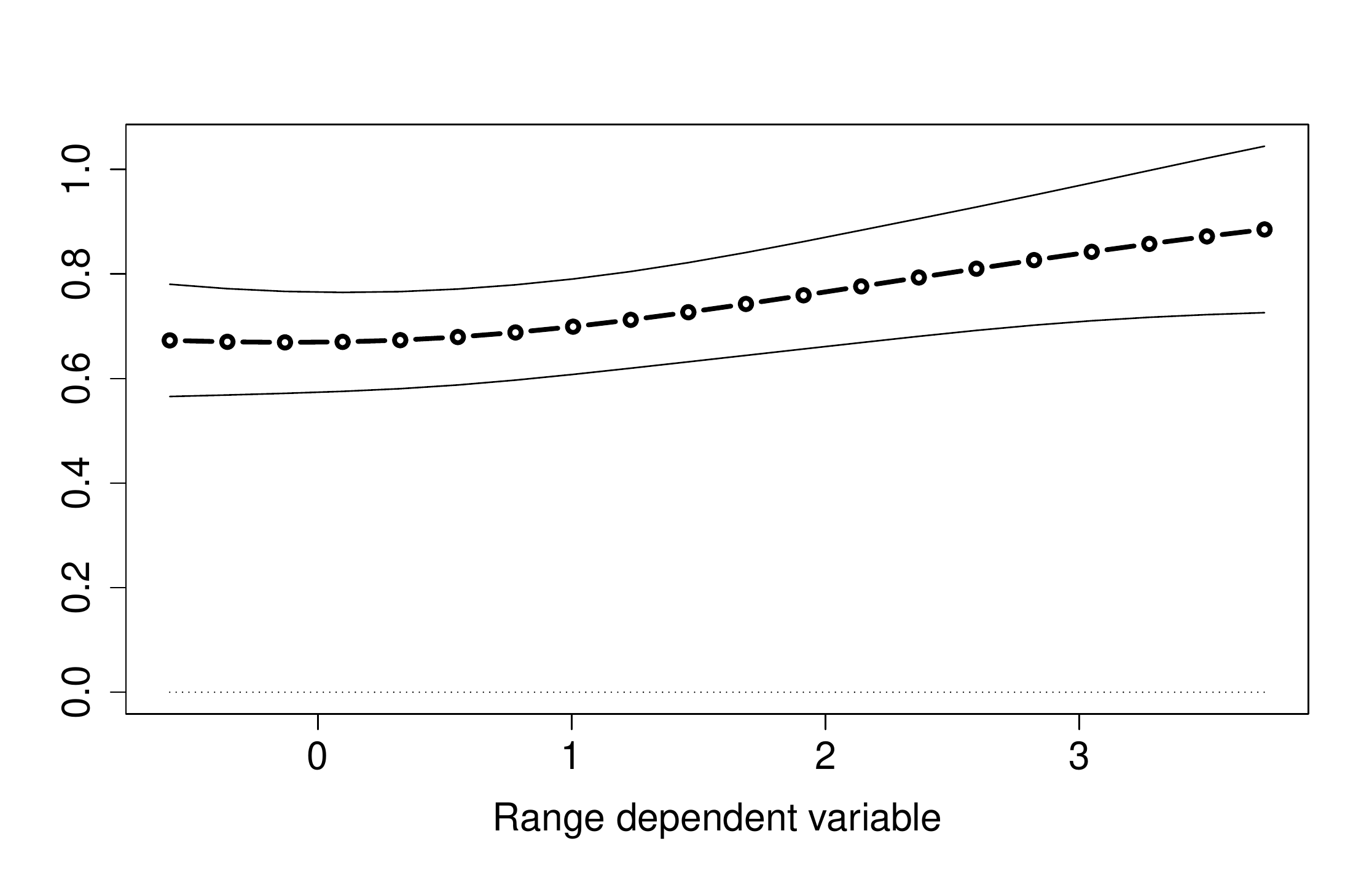}
\caption{Maximum likelihood estimates of parameter curves  for simulation data with smoothing parameter chosen by cross validation, left: variable 1, right: variable 2. }
\label{fig:simpen}
\end{figure}

Figure \ref{fig:simpen} shows the fitted functions and confidence intervals for variables 1 and 2 for the simulation setting with $\lambda$ chosen by 5-fold cross validation ($\lambda=14.05$) and ten cubic B-splines. It is seen that the functions are almost constant, which they should be since the generating model is a linear regression model. If $\lambda$ is chosen very large the the resulting  functions are very similar but strictly constant (not shown).

\begin{figure}[H]
\centering
\includegraphics[width=0.45\textwidth]{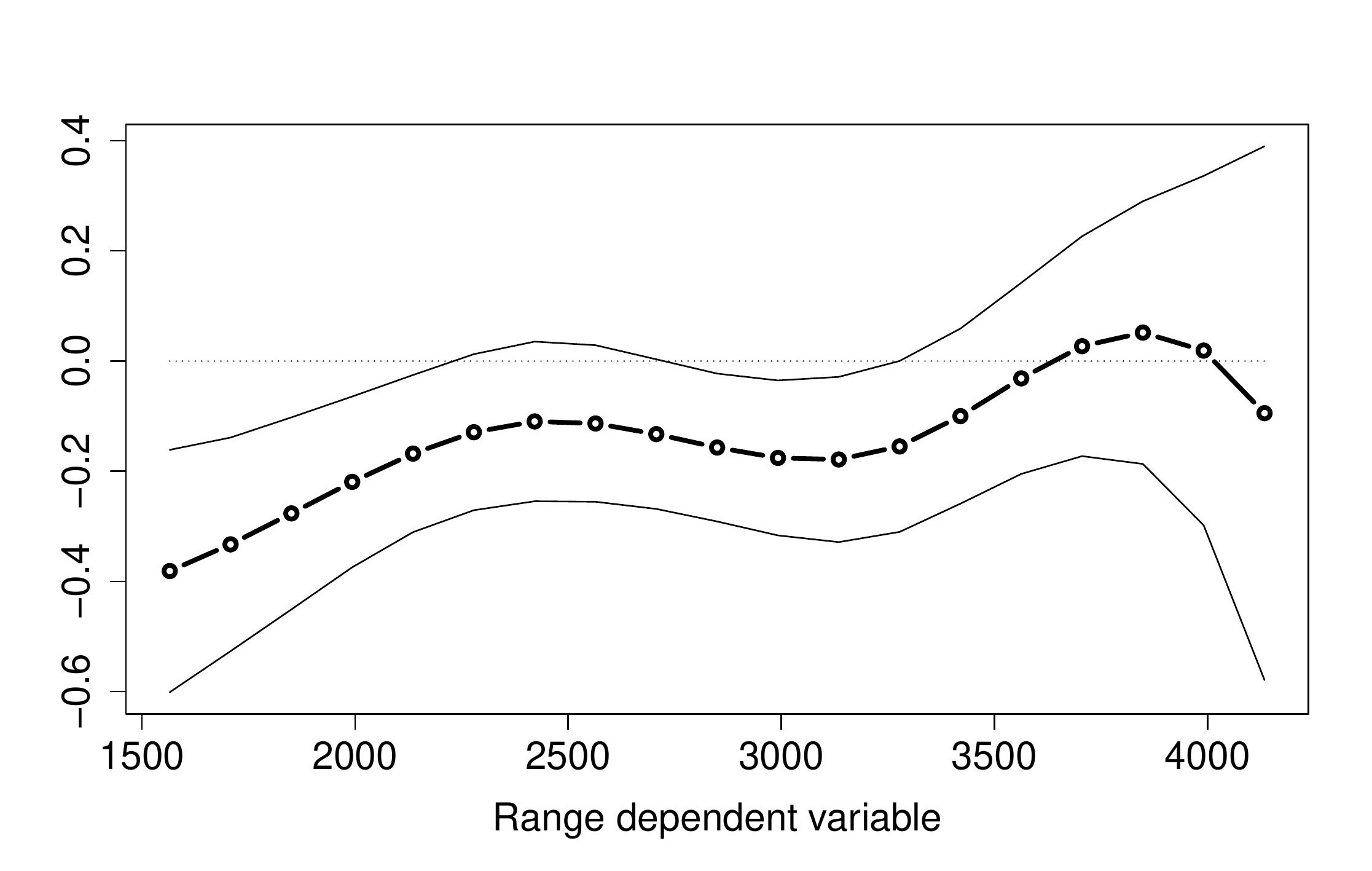}
\includegraphics[width=0.45\textwidth]{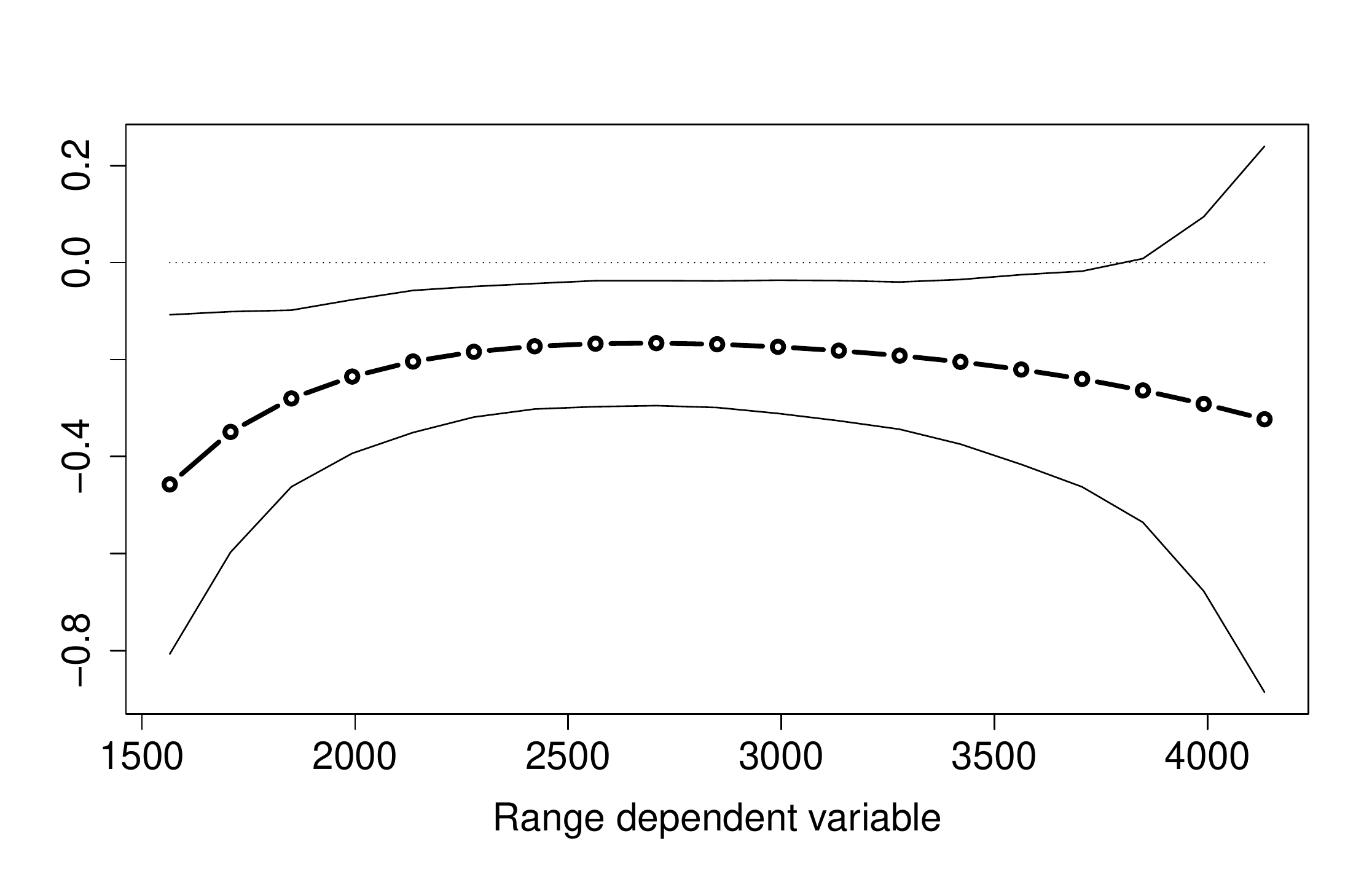}
\includegraphics[width=0.45\textwidth]{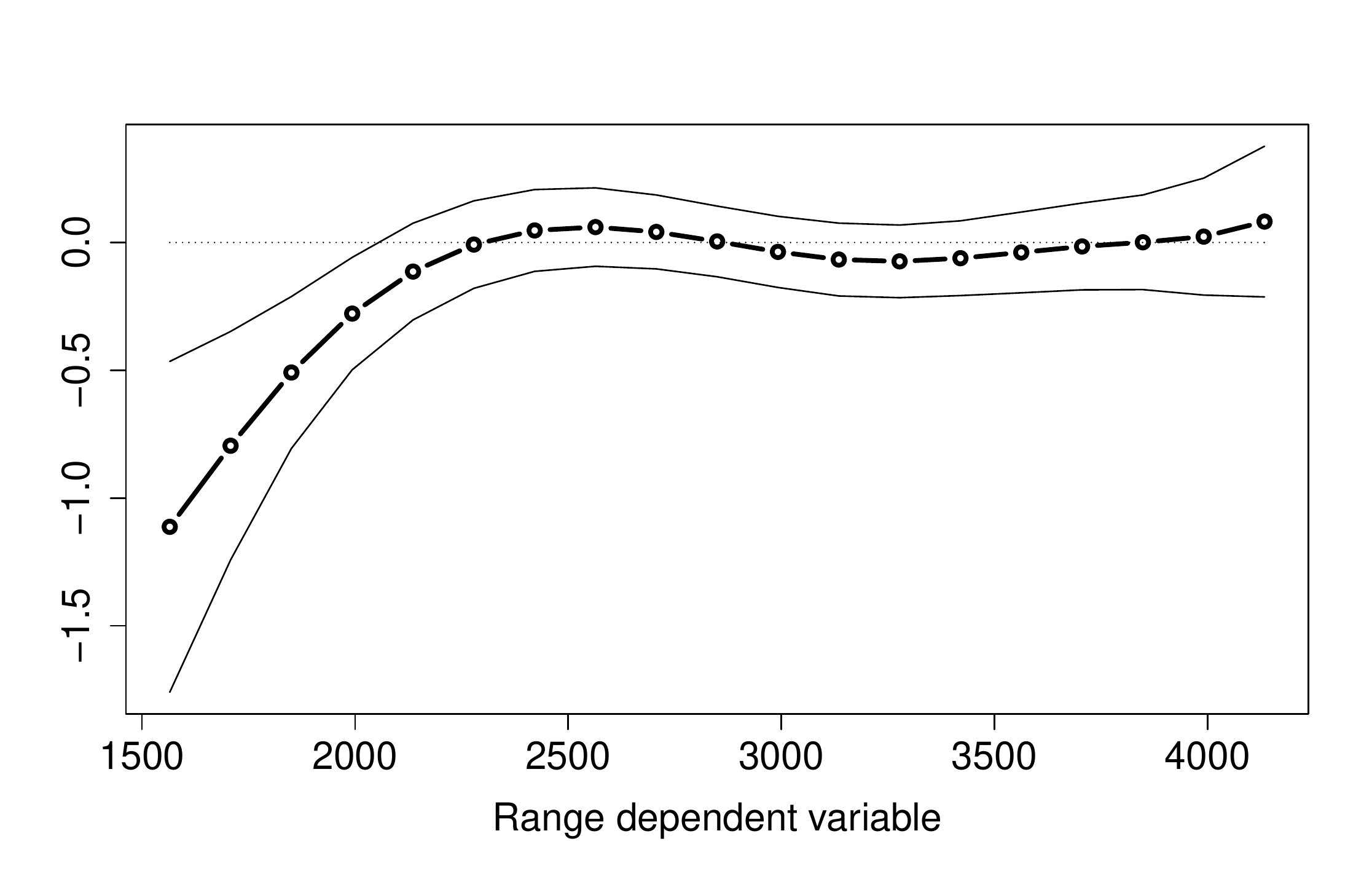}
\includegraphics[width=0.45\textwidth]{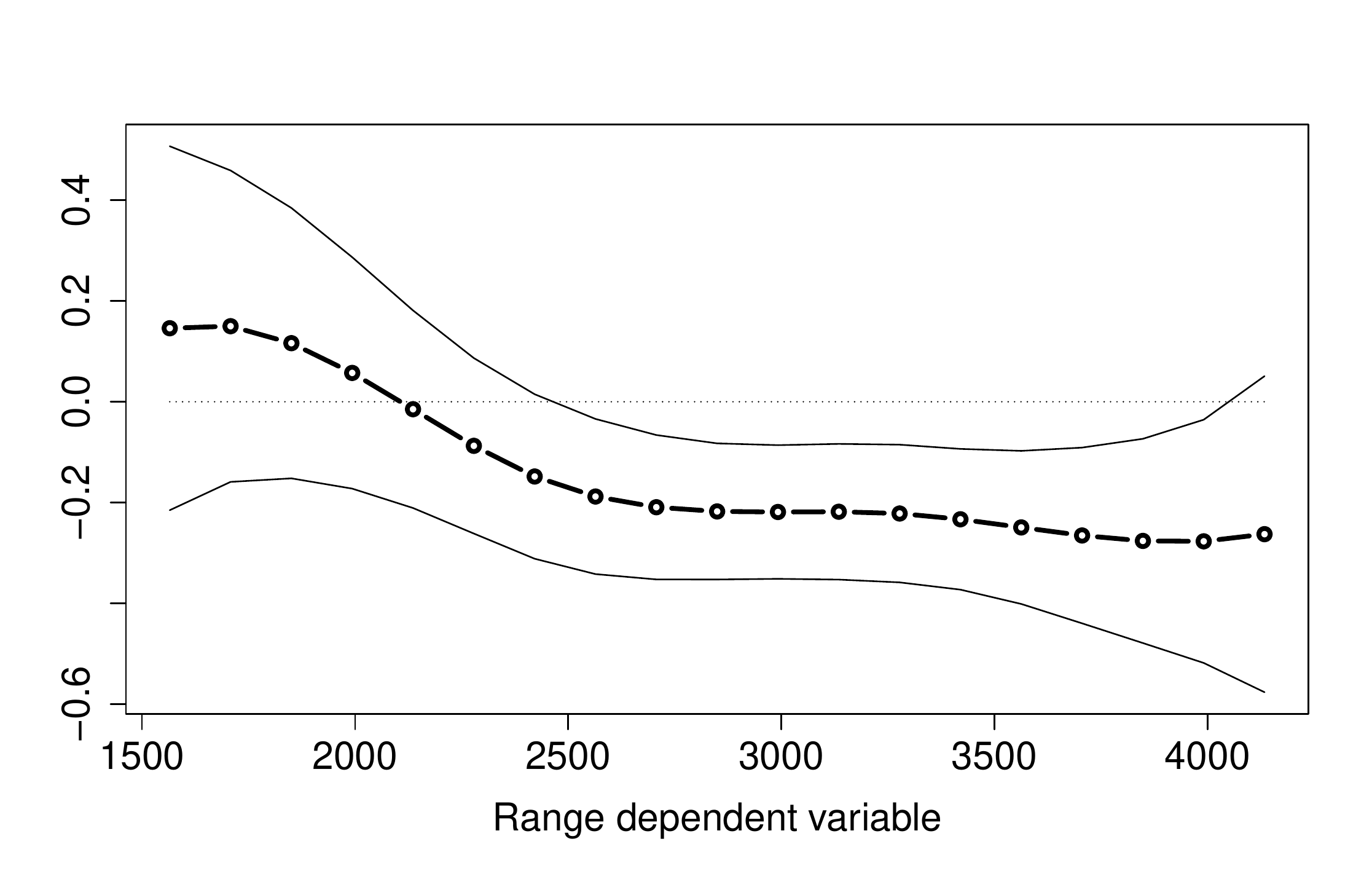}
\caption{Maximum likelihood estimates of parameter curves  for birth weight  data with smoothing parameter chosen by cross validation for hypertension, uterine irritability, age, smoke.  }
\label{fig:birthMLpaths}
\end{figure}

Figure \ref{fig:birthMLpaths} shows selected parameter functions for the birth weight data. In the upper panels  the same variables as in Figure \ref{fig:birth2}
are shown (hypertension and uterine irritability). It is seen that the parameter  functions are very similar to that in Figure \ref{fig:birth2}, but due to the use of basis functions are much smoother. Since one does not fit at specific thresholds but uses  a closed estimation method  the information in the data is used more efficiently. In addition the variables age and smoke given in the lower panels turn out to be relevant. While age seems to distinguish between high and low values in the low range, smoke seems influential in the higher range. 

\subsection{More Flexible Predictors: Variable Selection and Nonparametric Approaches}

A strength of the general estimation method from Section \ref{sec:algo} is that various state of the art modeling tools can be used to estimate the conditional distribution. This includes in particular methods that allow for efficient variable selection and nonparametric methods like random forests.

\subsubsection*{Including Selection Tools: Air Quality Data}
We will use daily air quality measurements in New York available from the R package airquality. 
The dependent variable is 
    Ozone (mean ozone in parts per billion from 1300 to 1500 hours at Roosevelt Island).
    Explanatory variables are 
    Solar.R (solar radiation in Langleys in the frequency band 4000--7700 Angstroms from 0800 to 1200 hours at Central Park),
    Wind: (average wind speed in miles per hour at 0700 and 1000 hours at LaGuardia Airport)
    Temp (maximum daily temperature in degrees Fahrenheit at La Guardia Airport).

\begin{figure}[H]
\centering
\includegraphics[width=0.45\textwidth]{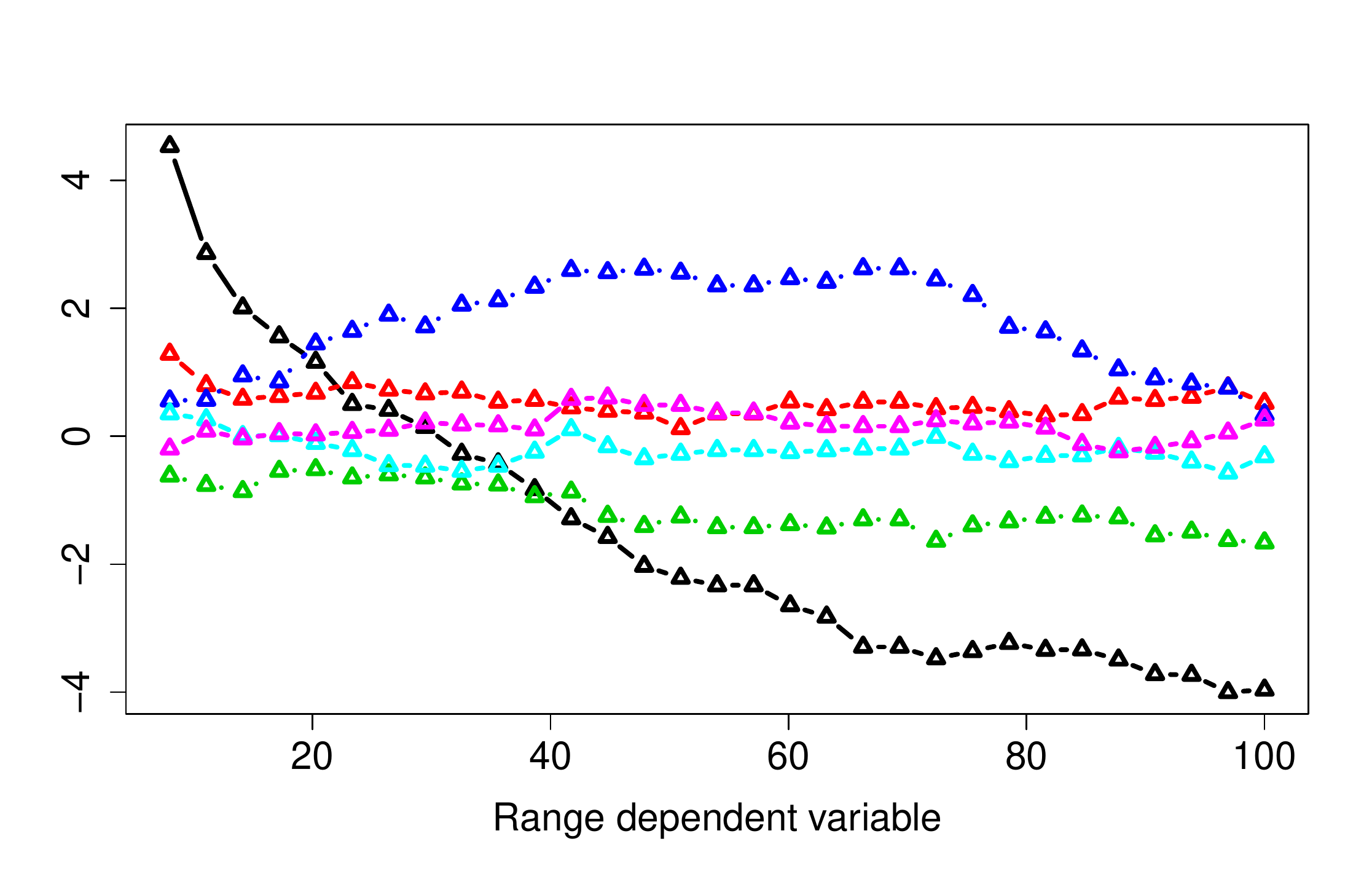}
\includegraphics[width=0.45\textwidth]{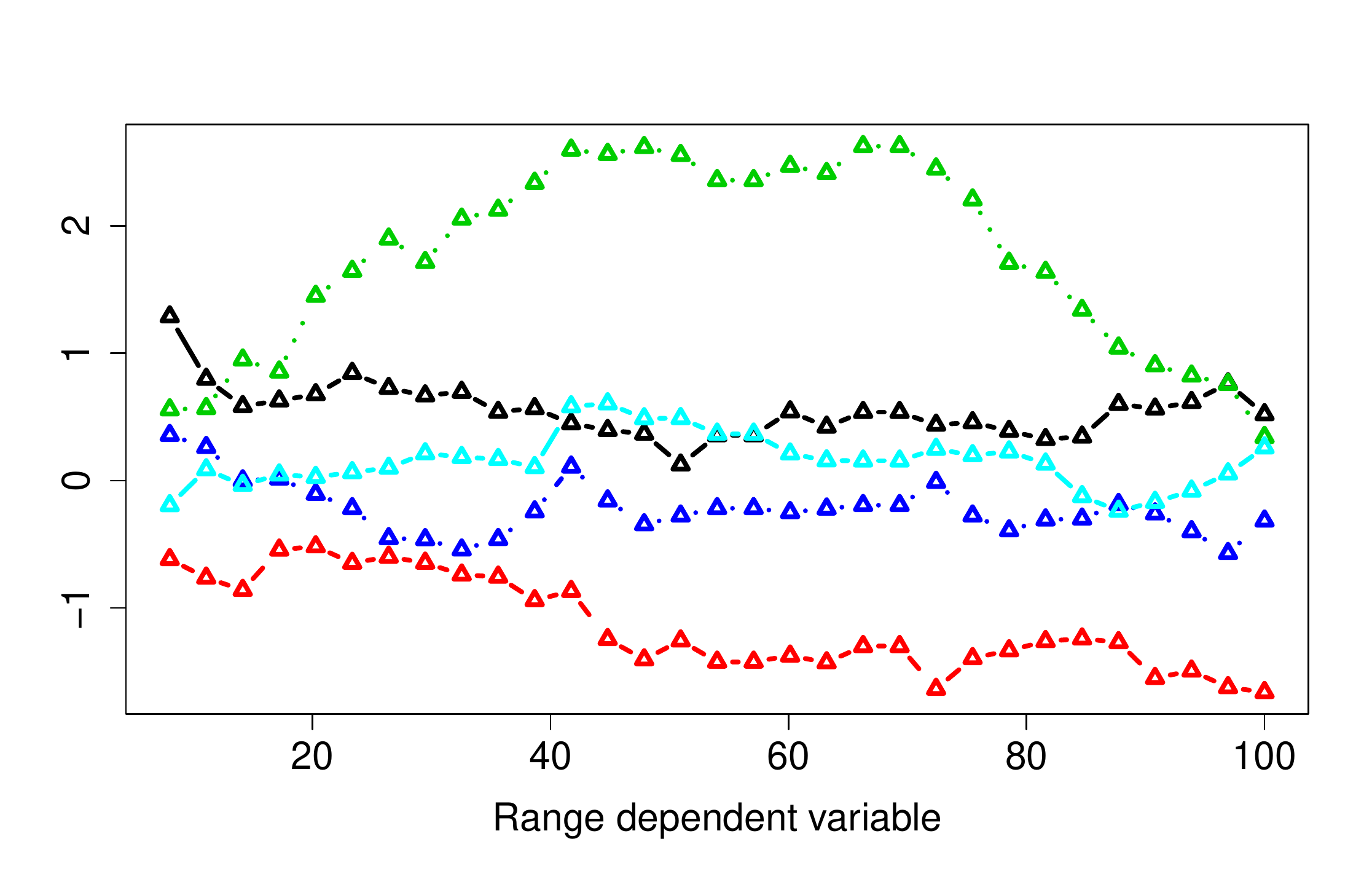}
\caption{Parameter functions of air quality data with standardized predictors; left: all coefficients including the intercept, right: only coefficients of explanatory variables. }
\label{fig:airn1}
\end{figure}

\begin{figure}[H]
\centering
\includegraphics[width=0.45\textwidth]{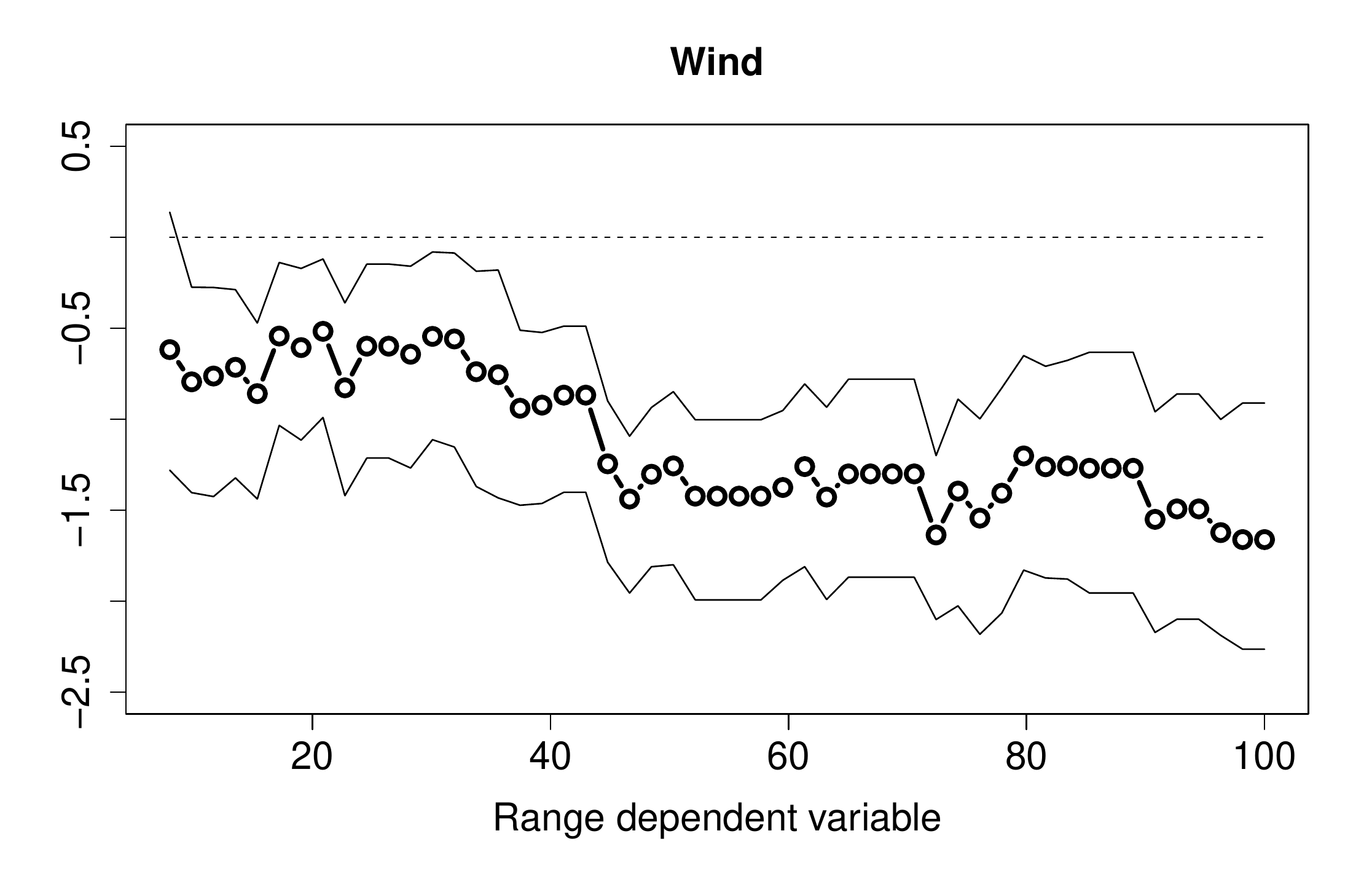}
\includegraphics[width=0.45\textwidth]{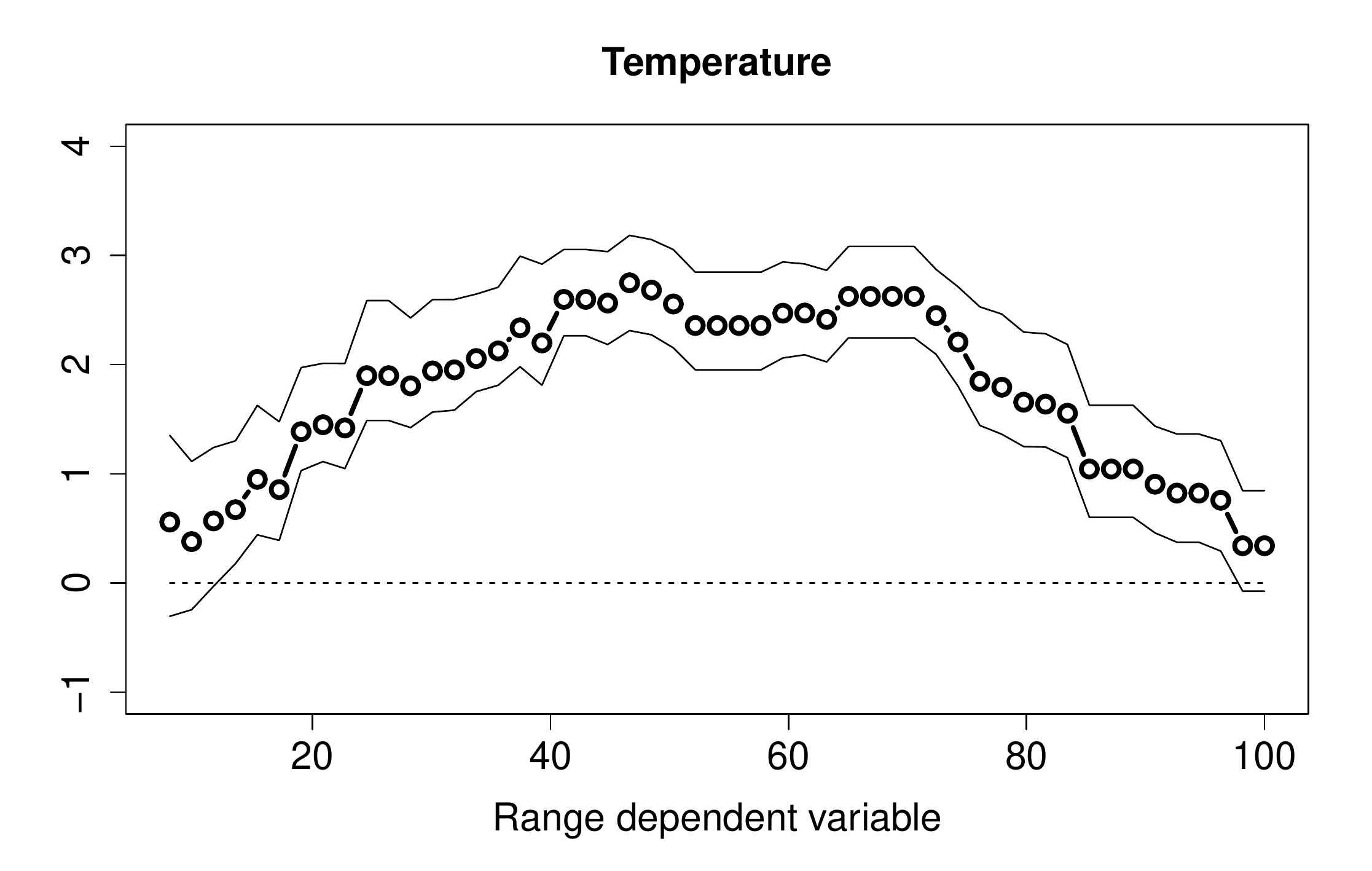}
\caption{Parameter functions for wind and temperature (air quality data with standardized predictors).}
\label{fig:airconf}
\end{figure}

Figure \ref{fig:airn1} shows the variation of parameter estimates over the dependent variable for standardized variables. In the left panel the intercept is included (decreasing function), in the right panel only variable coefficients are shown.  Figure \ref{fig:airconf} shows the   parameter estimates of the two most influential variables wind and temperature. It is seen that wind speed decreases the ozone level, but there seem to be two levels. The parameter function is 
close to -0.5 up to ozone levels of 40, but then is on a lower level around -1.5. Temperature increases ozone levels in particular in the middle range but the effect is much weaker at the boundaries. It suggests that covariate effects are definitely nonlinear, which is also supported by the analysis of the regression structure in Section \ref{sec:quant} and the performance in terms of accuracy of prediction (Section \ref{sec:pred}).

In most applications some of the available variables have no effect on the dependent variables  and should be excluded.
A variety of efficient variable selection tools have been developed in the last decades, including penalized likelihood approaches as the lasso or elastic net, SCAD and the Dantzig selector, for overviews see \citet{HasTibFri:2001,buehlmann2011statistics}. It is straightforward to use these selection tools to obtain variable selection within the varying-thresholds framework. In the following this is demonstrated for the air quality data when using lasso methods.

\begin{figure}[H]
\centering
\includegraphics[width=0.45\textwidth]{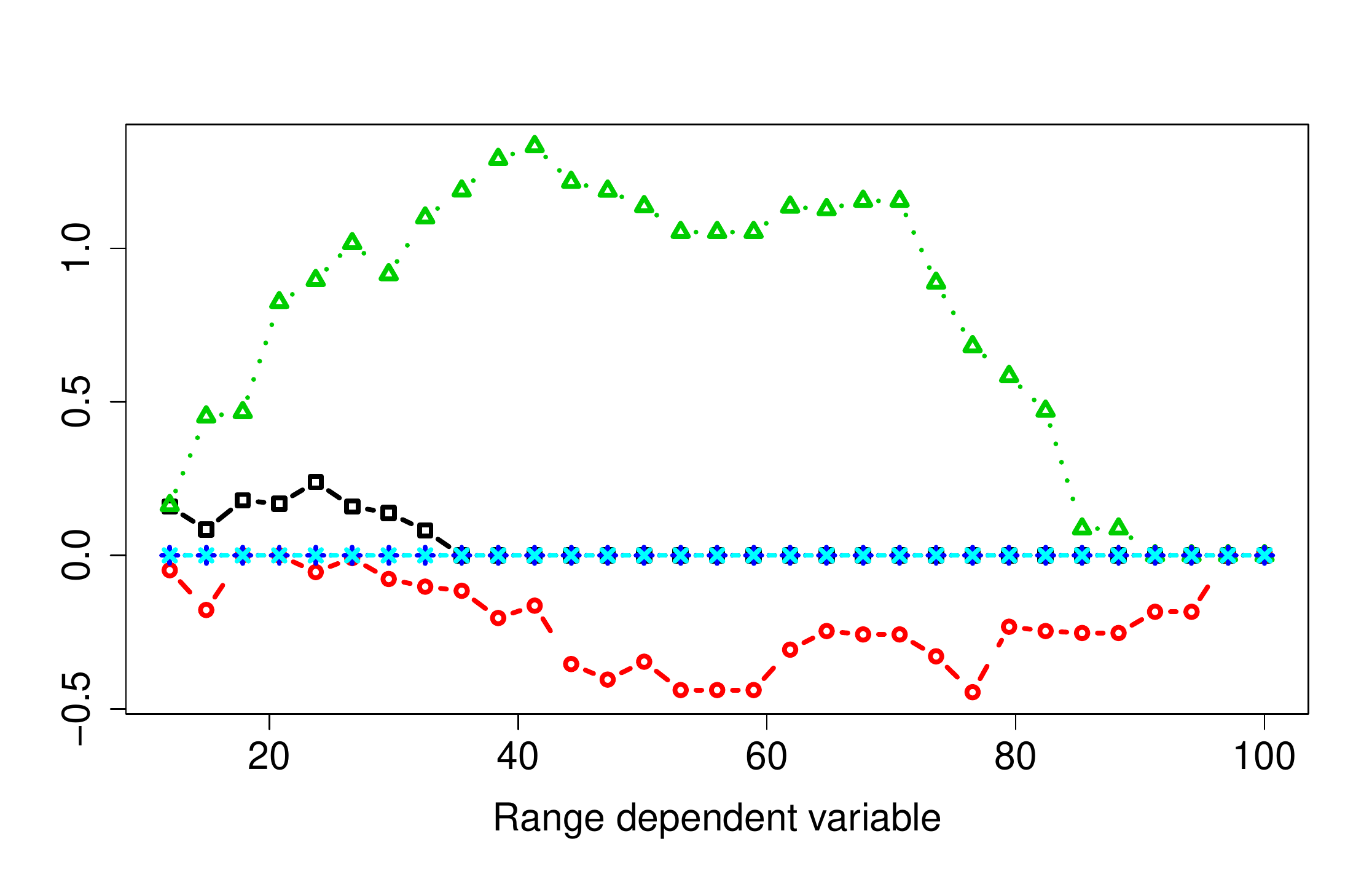}
\includegraphics[width=0.45\textwidth]{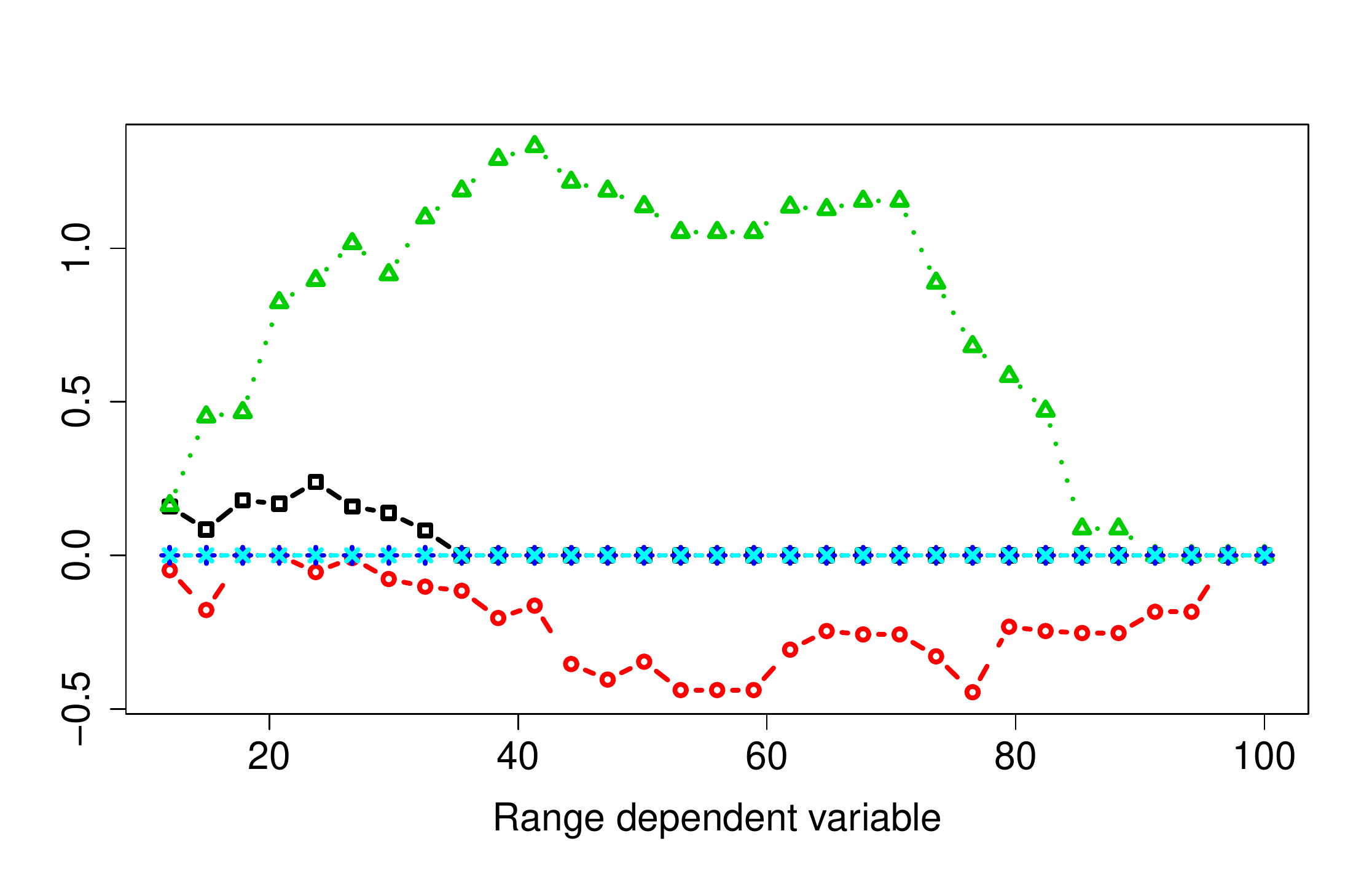}
\caption{Coefficients of air quality data obtained from using lasso with very weak penalty parameter ($\lambda=0.05$) and stronger penalty ($\lambda=0.1$), coefficients unequal zero are temperature (triangles), wind (dots), and solar radiation (squares).  }
\label{fig:airlas}
\end{figure}

Figure \ref{fig:airlas} shows the coefficients if lasso is used to select the relevant variables. Lasso uses the penalty term $\lambda\sum_j |\beta_j^{(r)}|$.
If $\lambda=0$ one obtains the maximum likelihood estimate while large values of $\lambda$ result in strong variable selection. It is seen  that only three variables have coefficients unequal zero. The strongest is temperature (triangles), which over a wide range increases the probability of high ozone levels. The variable wind (dots) decreases the probability of high ozone levels over a wide range, while solar radiation can only discrimnate between high and low levels of ozone if levels are very low.

\subsubsection*{Nonparametric Approaches}

Recursive partitioning methods, often simply referred to as trees, are nonparametric tools that are quite efficient in modeling interaction effects, see 
\citet{BreiFrieOls:84} for more classic approaches, \citet{HotLau:03,StrBouAug:2007} for the maximally selected statistics approach. An extensive overview was given by \citet{Strobetal:2009}. Single trees can be very unstable and affected by few observations. Much more stable methods are random forests  \citep{Breiman:98, Breiman:2001a}. They  are  examples of a wider group of methods, called {ensemble
methods}.  When using ensembles one fits several models, in the case of random forests several trees, and combines them to infer on the distribution of the dependent variable. Random forests have been shown to be among the best predictors, see, for example, \citet{DiaAnd:2006}. Moreover, random forests have a built-in selection procedure that selects the most important predictors, and therefore   can be also applied when the predictor space is high dimensional.

By using random forests for the binary splits flexible estimates of the varying-thresholds model are obtained. 
The only disadvantage of random forests is that the contribution of single
variables in the classification rule gets lost. Importance measures
try to measure this contribution for random forests. A permutation
accuracy importance measure, proposed by \citet{Breiman:2001a} can be used to evaluate the difference in prediction accuracy
between the trees based on the original observations and trees that
are built on data where one predictor variable is randomly permuted,
thereby breaking the original association with the response
variable. 
Various importance measures for random forests have been proposed, see, for example, \citet{strobl2007bias,strobl2008conditional,hapfelmeier2014new,gregorutti2017correlation,hothorn2015partykit}. An example for such a measure  is the "Gini importance" available in the randomForest package, which is used in the following.
It describes the improvement in the "Gini gain" splitting criterion.

\begin{figure}[H]
\centering
\includegraphics[width=0.45\textwidth]{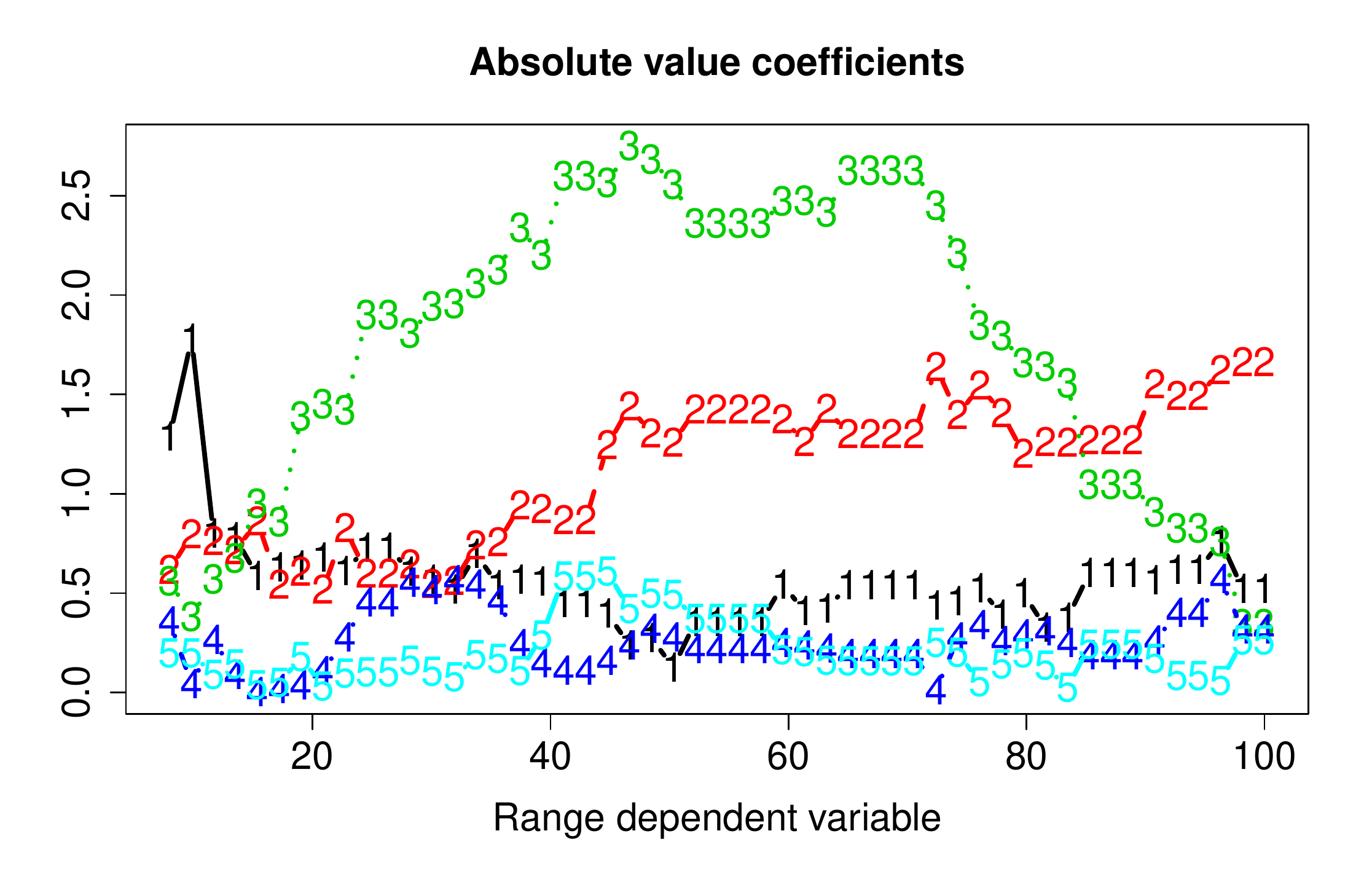}
\includegraphics[width=0.45\textwidth]{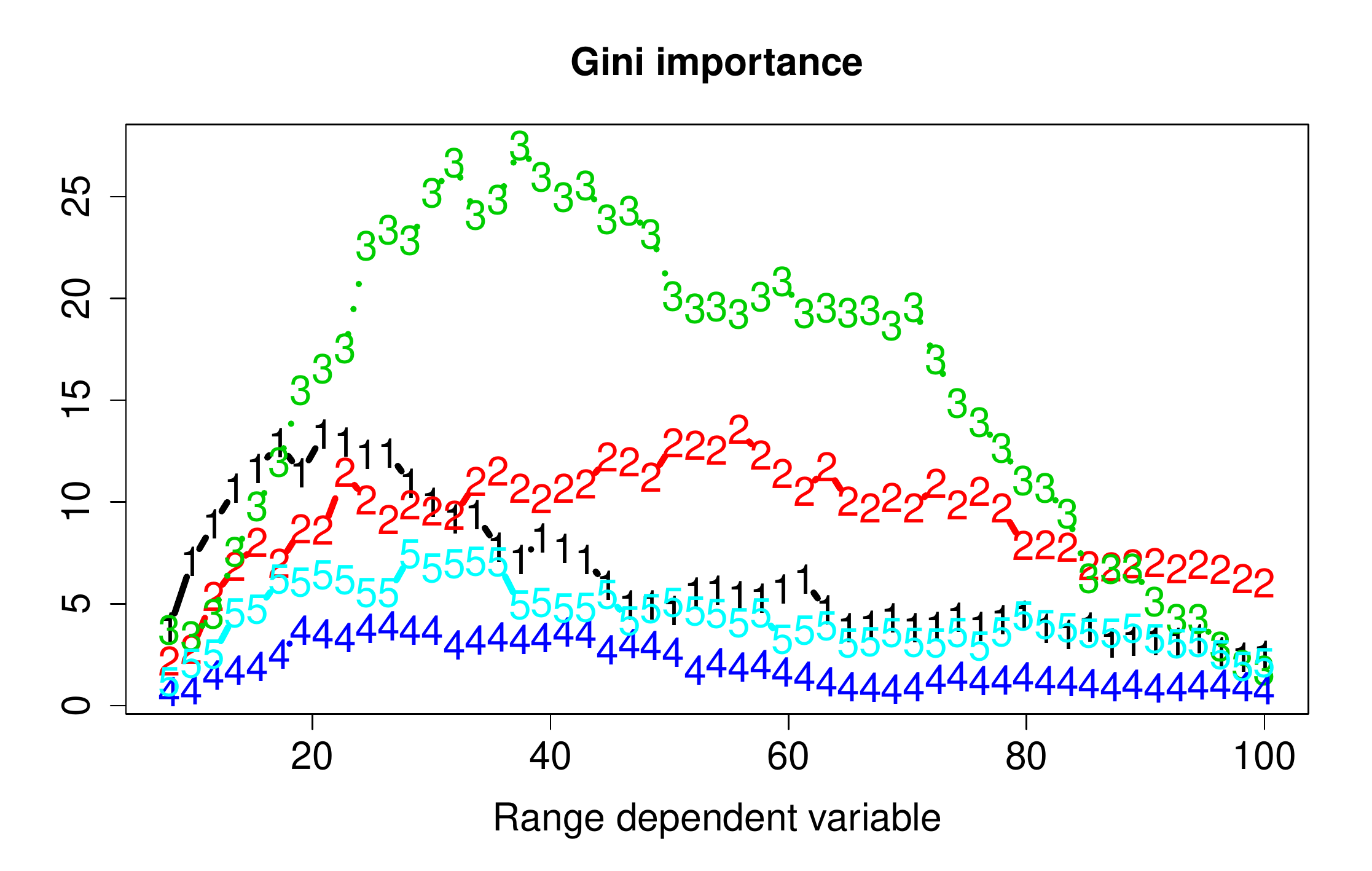}
\caption{Absolute values of coefficients of air quality data with standardized predictors (left), Gini importance measure (right); variables are  Solar.R (1), Wind (2), Temp (3), Month (4), Day (5). }
\label{fig:air1}
\end{figure}

Figure \ref{fig:air1} shows results for the air quality data. the left panel shows the coefficients obtained from fitting a model with linear predictor. Instead of showing the coefficient functions themselves the absolute values are shown since they indicate the strength of the impact of explanatory variable (with standardized predictors). The variables are represented by numbers, Solar.R (1), Wind (2), Temp (3), Month (4), Day (5). The right panel shows the Gini importance resulting from using random forests. It is seen that both approaches indicate that temperature and wind are the most influential predictors. Although  the use of the more flexible random forest yields similar results regarding the impact of variables one should keep in mind that random forests also account for possible interaction effects. The benefits of this is not seen from the importance measures but when investigating the accuracy of prediction (Section \ref{sec:pred}).

Let us consider a further  example, which contains more explanatory variable, since then the built-in variable selection property of random forests becomes more obvious. We use the housing data for 506 census tracts of Boston in the version BostonHousing2, which contains the corrected version of the original data by Harrison and Rubinfeld (1979) and included additional spatial information. The dependent variable is cmedv
(corrected median value of owner-occupied homes in USD 1000's). 
Explanatory variables are crim
(per capita crime rate by town, var 1), 
lstat
(percentage of lower status of the population, var 2)
zn (proportion of residential land zoned for lots over 25,000 sq.ft, var 3),
nox
(nitric oxides concentration inparts per 10 million, var 4),
rm
(average number of rooms per dwelling, var 5),
dis
(weighted distances to five Boston employment centres, var 6),
rad
(index of accessibility to radial highways, var 7),
tax
(full-value property-tax rate per USD 10,000, var 8),
ptratio
(pupil-teacher ratio by town, var 9),
b
(proportion of blacks by town, var 10),
indus
(proportion of non-retail business acres per town, var 11),
age
(proportion of owner-occupied units built prior to 1940, var 12).  

\begin{figure}[H]
\centering
\includegraphics[width=0.45\textwidth]{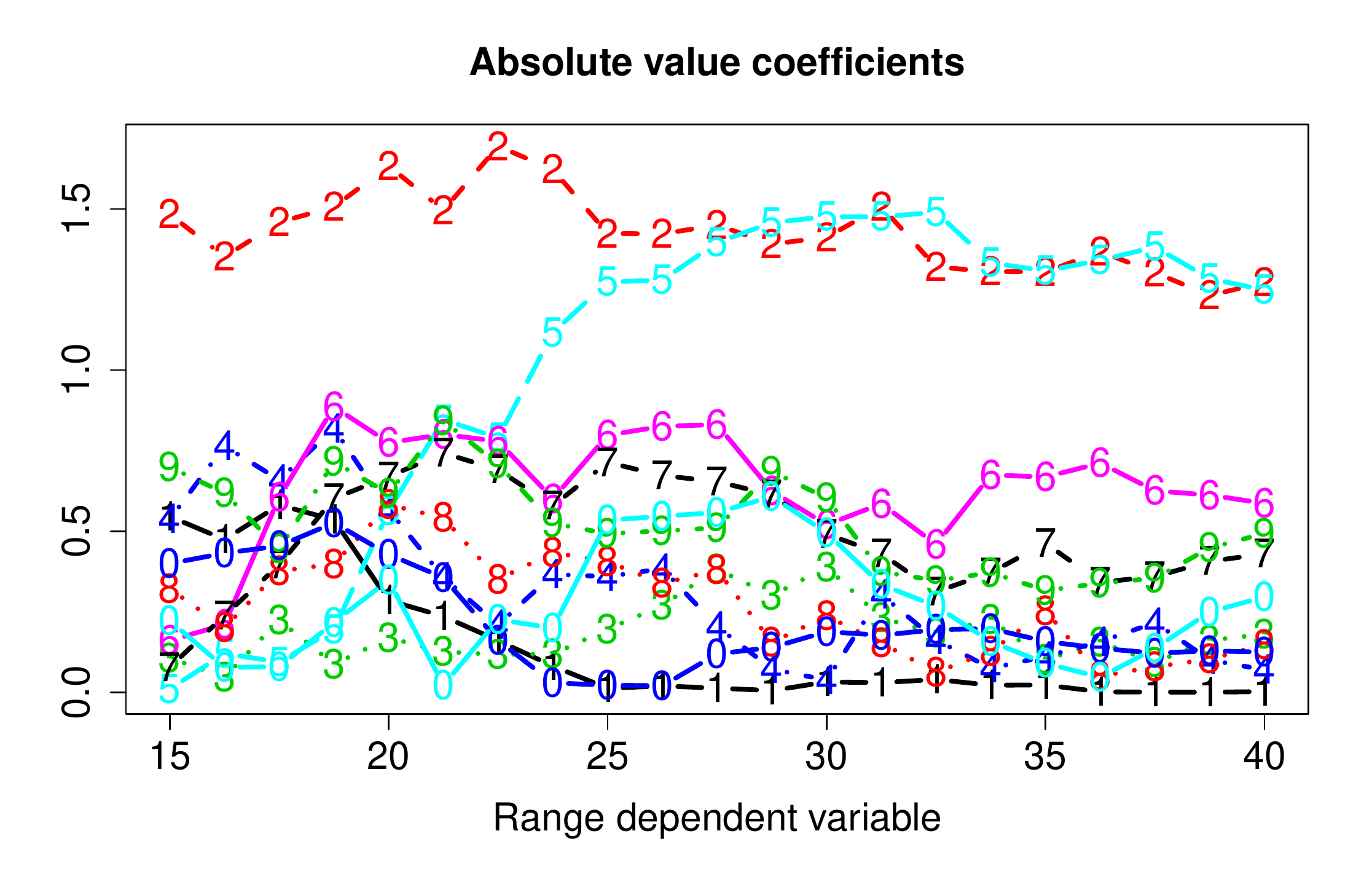}
\includegraphics[width=0.45\textwidth]{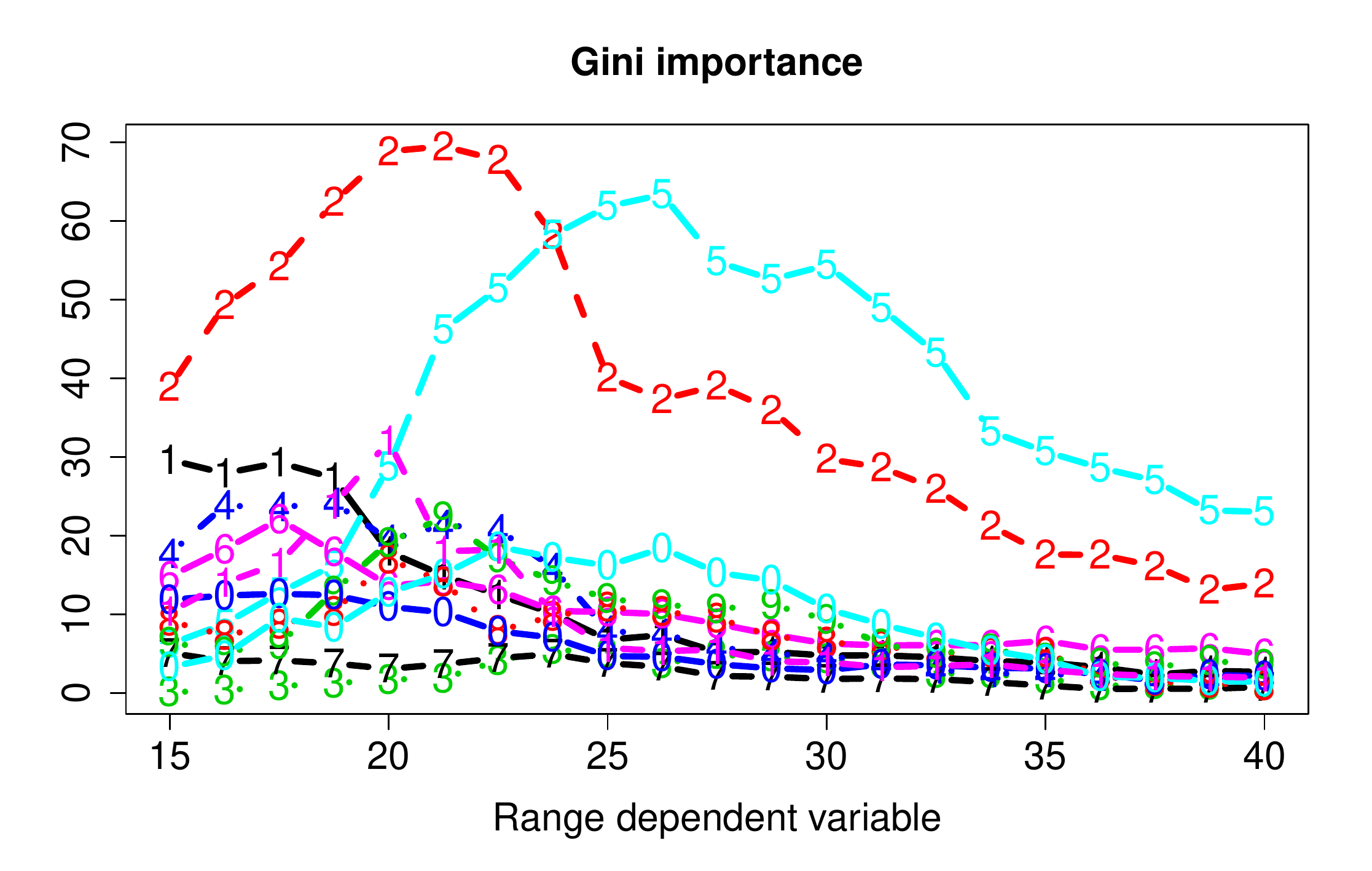}
\caption{Absolute values of coefficients of housing data with standardized predictors (left), Gini importance measure (right), variables are crim (1), lstat (2), zn (3), nox (4), rm (5), dis (6) rad (7), tax (8), ptratio (9), indus (0), age (0)}
\label{fig:house1}
\end{figure}

\begin{figure}[H]
\centering
\includegraphics[width=0.45\textwidth]{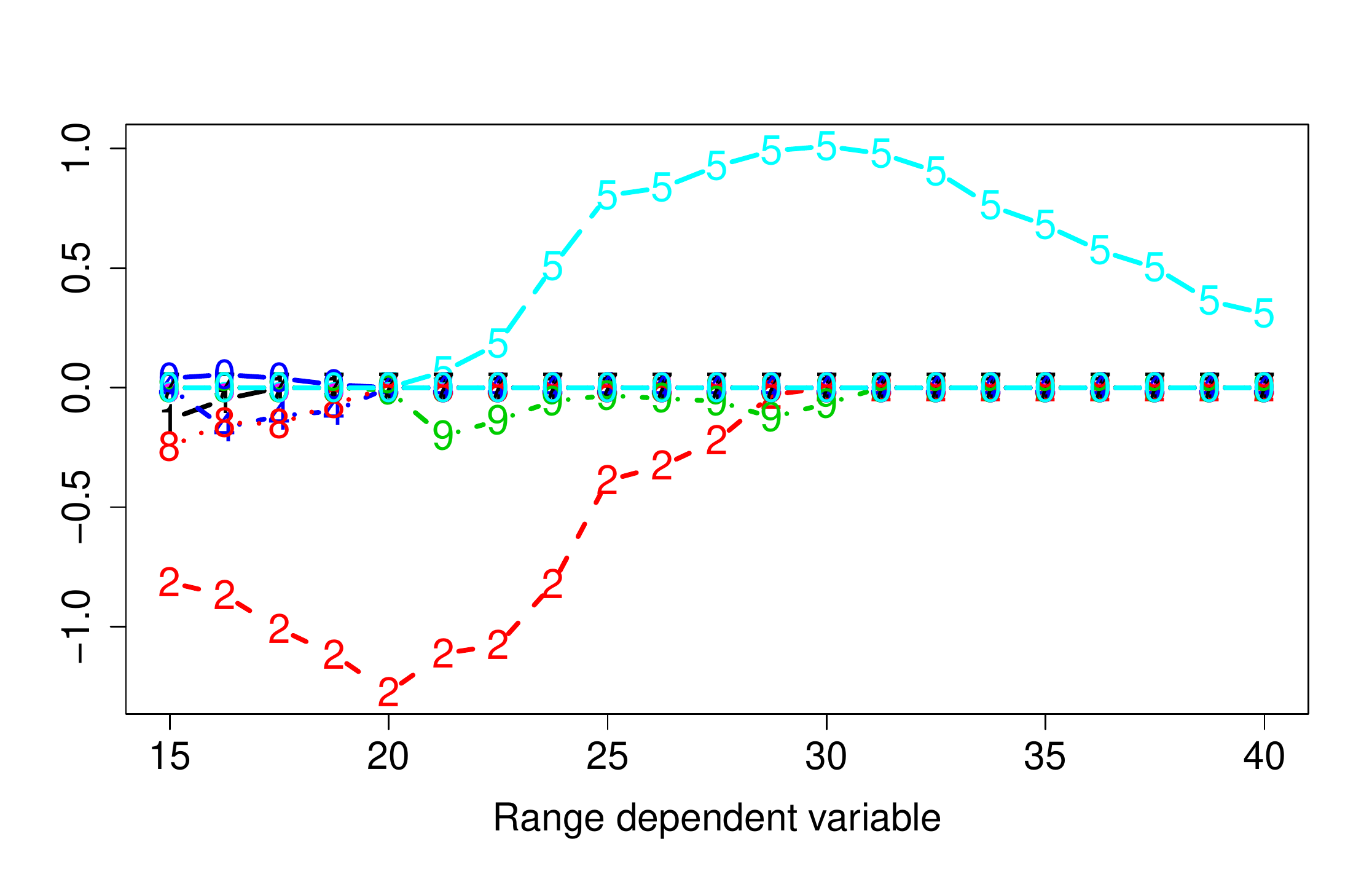}
\includegraphics[width=0.45\textwidth]{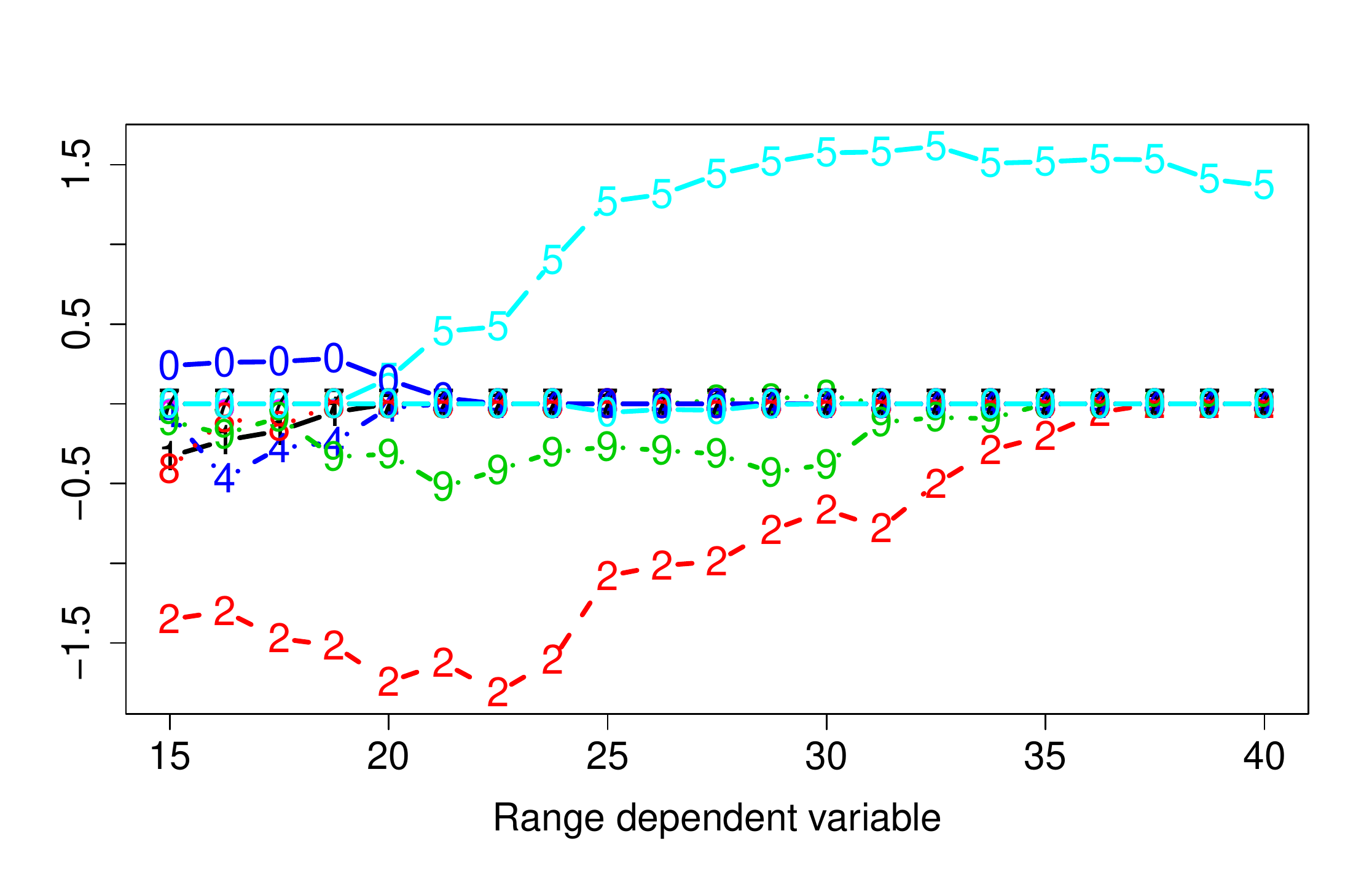}
\caption{Parameter functions for housing data (lasso with linear predictor) with variables  crim (1), lstat (2), zn (3), nox (4), rm (5), dis (6) rad (7), tax (8), ptratio (9), indus (0), age (0), left: $\lambda=1$, right: $\lambda=0.3$}
\label{fig:house2}
\end{figure}

Figure \ref{fig:house1} shows the absolute values of the coefficient curves for the linear model (left) and the Gini importance measure obtained from the use of random forests. In both models variables 2 and 5 turn out to be strongly influential. However, most of the variables have almost no impact in particular for higher values, which becomes obvious in the right picture, in which the variable selection mechanism in random forests is at work. For comparison, in Figure  
\ref{fig:house2} the coefficient paths for lasso are shown. They also indicate that variables 2 and 5 are the most influential.

\section{Using the Predictive Distribution: Quantile Regression}\label{sec:quant}

Quantile regression has been an area of intensive research throughout the last decades, see, for example, \citet{koenker2001quantile,yu1998local,fitzenberger2013economic,koenker2017handbook,koenker1994quantile,waldmann2013bayesian}. 
Quantile regression can be framed as an optimization problem. For $1 < \alpha <1 p$ let the loss function be defined by 
\[
L_{\alpha}(y,q)=\begin{cases}
\alpha |y-q| & \mbox{if} \; y > q\\
(1-\alpha) |y-q| & \mbox{if} \; y \le q
\end{cases}
\]
While the mean is obtained by minimizing the squared loss $\alpha$-quantiles are obtained by minimizing the expected loss function $\E(L_{\alpha}(y,q))$. In quantile regression one typically assumes a parametric function $\mu(\xb, \betab )$ and considers the conditional quantiles 
$
q_{\alpha}(\xb) = \argmin_{\betab} \E(L_{\alpha}(Y,\mu(\xb, \betab ))).
$
Estimates are obtained by using a random sample $(Y_i, \xb_i), i=1,\dots,n$ and minimizing $\sum_{i}(L_{\alpha}(Y_i,\mu(\xb_i, \betab )))$.
Typically the parametric function $\mu(\xb, \betab )$ is chosen as a linear function. Then, the minimization problem can be solved  efficiently by linear programming methods. More elaborate estimation methods are needed in smooth additive versions  \citep{koenker1994quantile,waldmann2013bayesian}.
Variable selection tools have been proposed by  \citet{wu2009variable}. 

Within the varying-thresholds model framework  estimates of conditional quantiles are obtained by computing  the quantiles of the estimated distribution function  
$\hat F(\theta|\xb)$. The advantage is that the effect of covariates on the dependent variable can be modeled very flexibly. In contrast to classical quantile regression it is not necessary to assume that the predictor is linear. By assuming varying-thresholds a non-linear effect on the quantiles is obtained. Selection of variables can be obtained by regularization methods and extremely flexible modeling of interactions by using random forests.

\subsubsection*{Simulation Data}
\begin{figure}[H]
\centering
\includegraphics[width=0.45\textwidth]{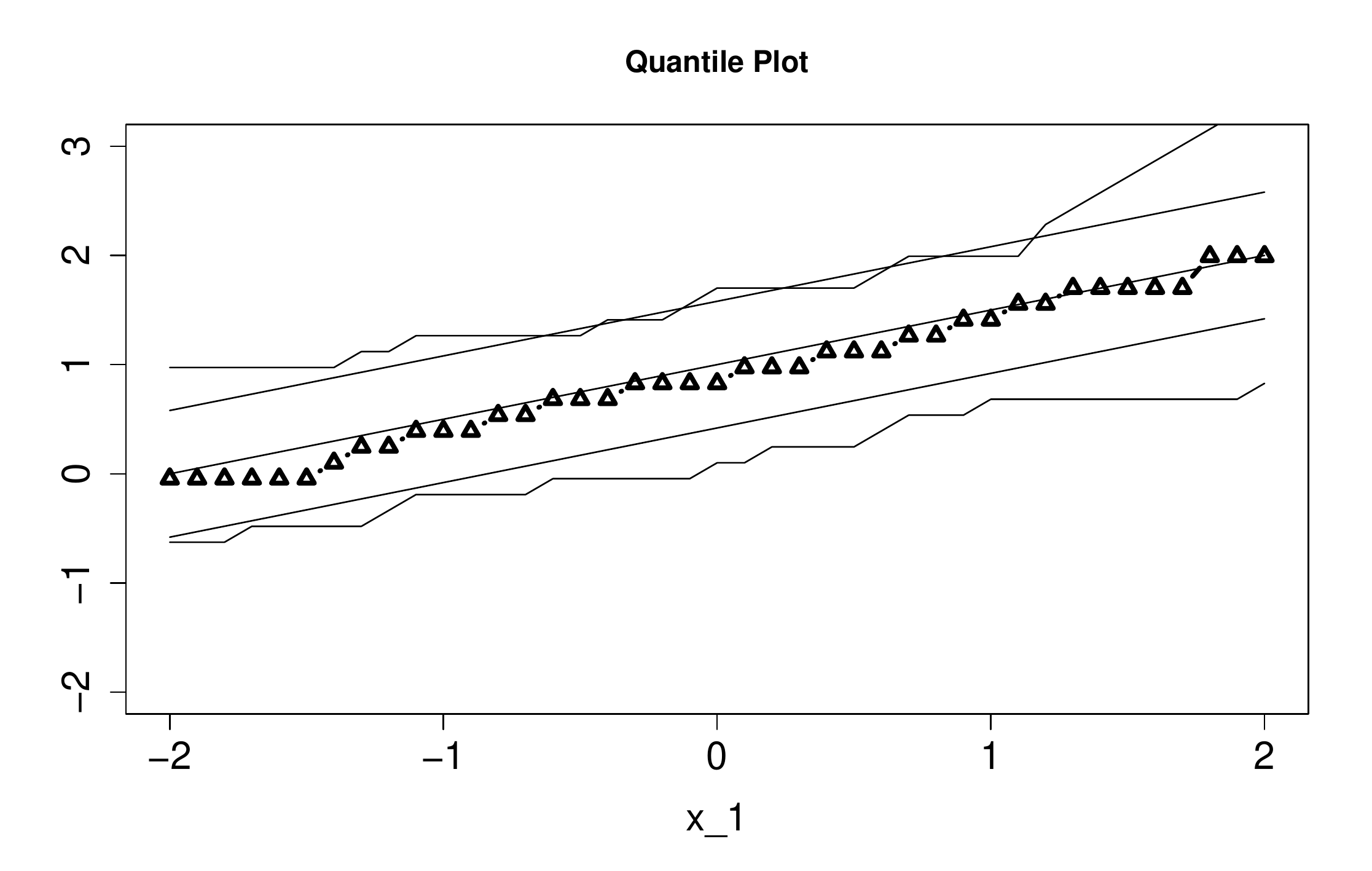}
\includegraphics[width=0.45\textwidth]{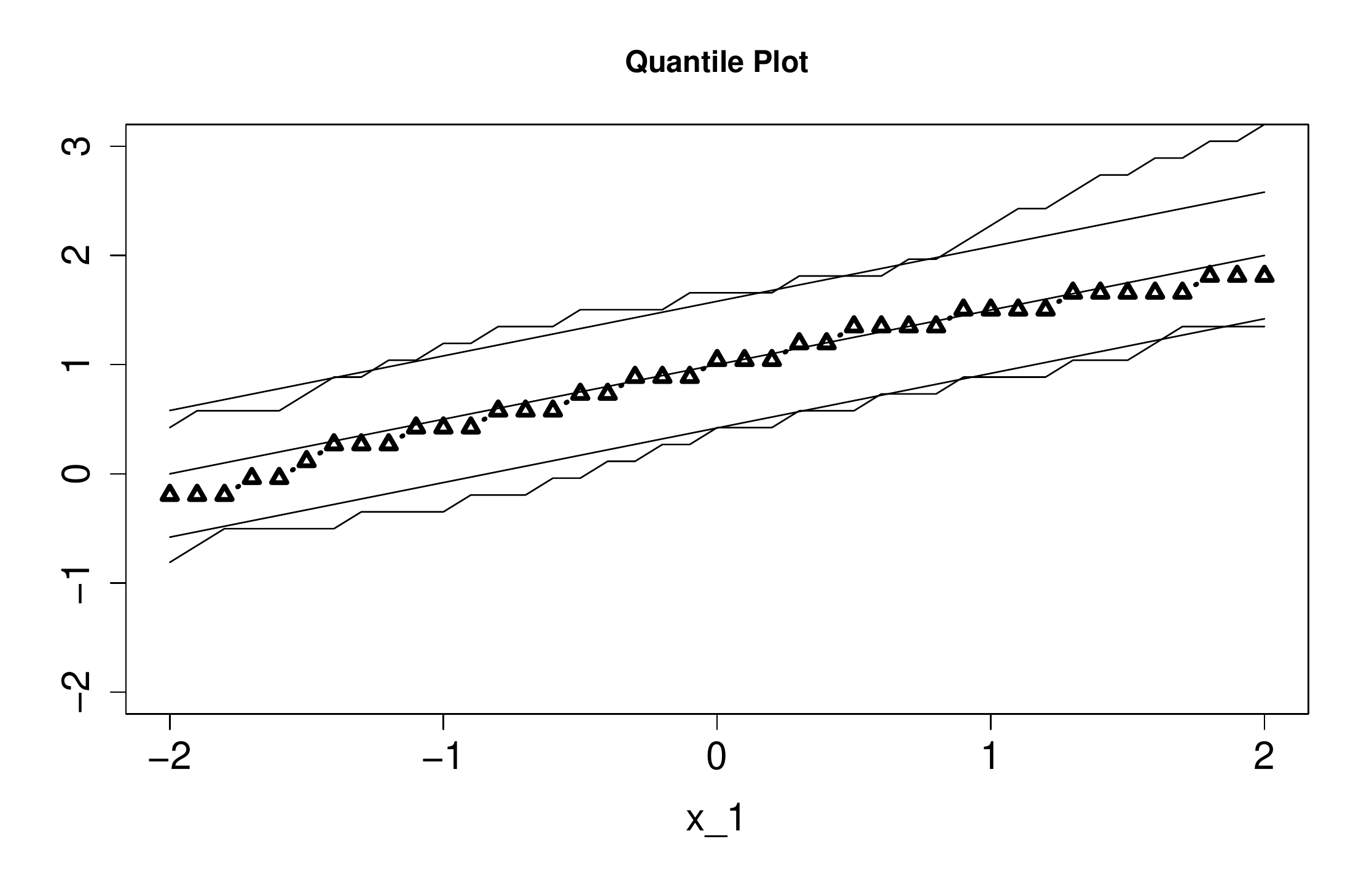}
\caption{Median, .25 and .75 quantiles for simulation data, straight lines represent true values, left: n=100, right: n=300. }
\label{fig:sim3}
\end{figure}

For illustration we use again the simulation data with a normal response. 
Figure \ref{fig:sim3} shows the fitted 0.5, 0.25 and 0.75 quantiles for variable 1, for sample size $n=100$ (left) and sample size $n=300$ (right).
In addition to the estimated quantile functions the true mean and the true quantiles are included (straight lines in the panels). It is seen that over a wide range the quantiles are approximated quite well by the threshold approach although no distribution function is assumed.

\subsubsection*{Quantile Regression for Data Sets}
\begin{figure}[H]
\centering
\includegraphics[width=0.45\textwidth]{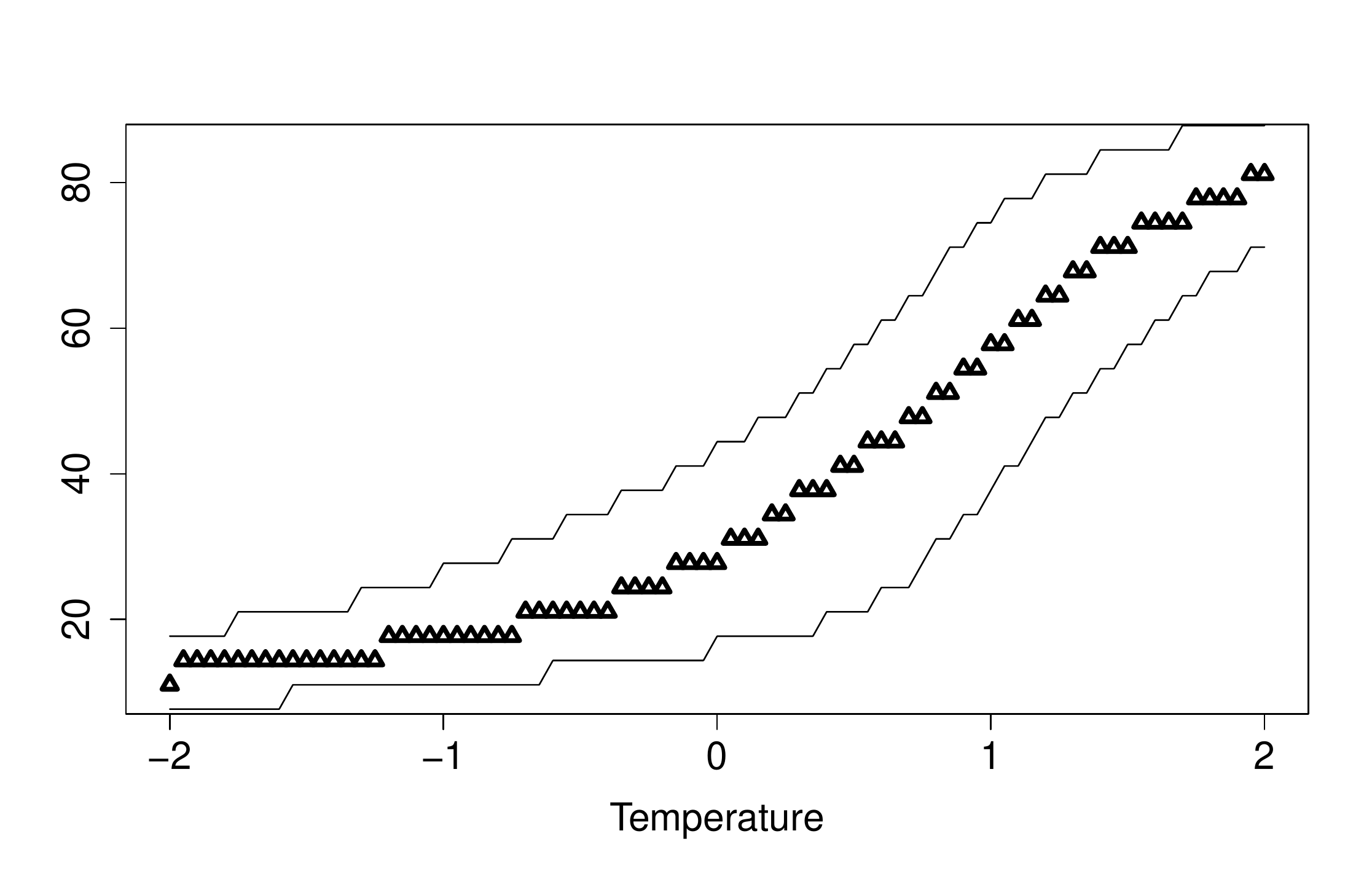}
\includegraphics[width=0.45\textwidth]{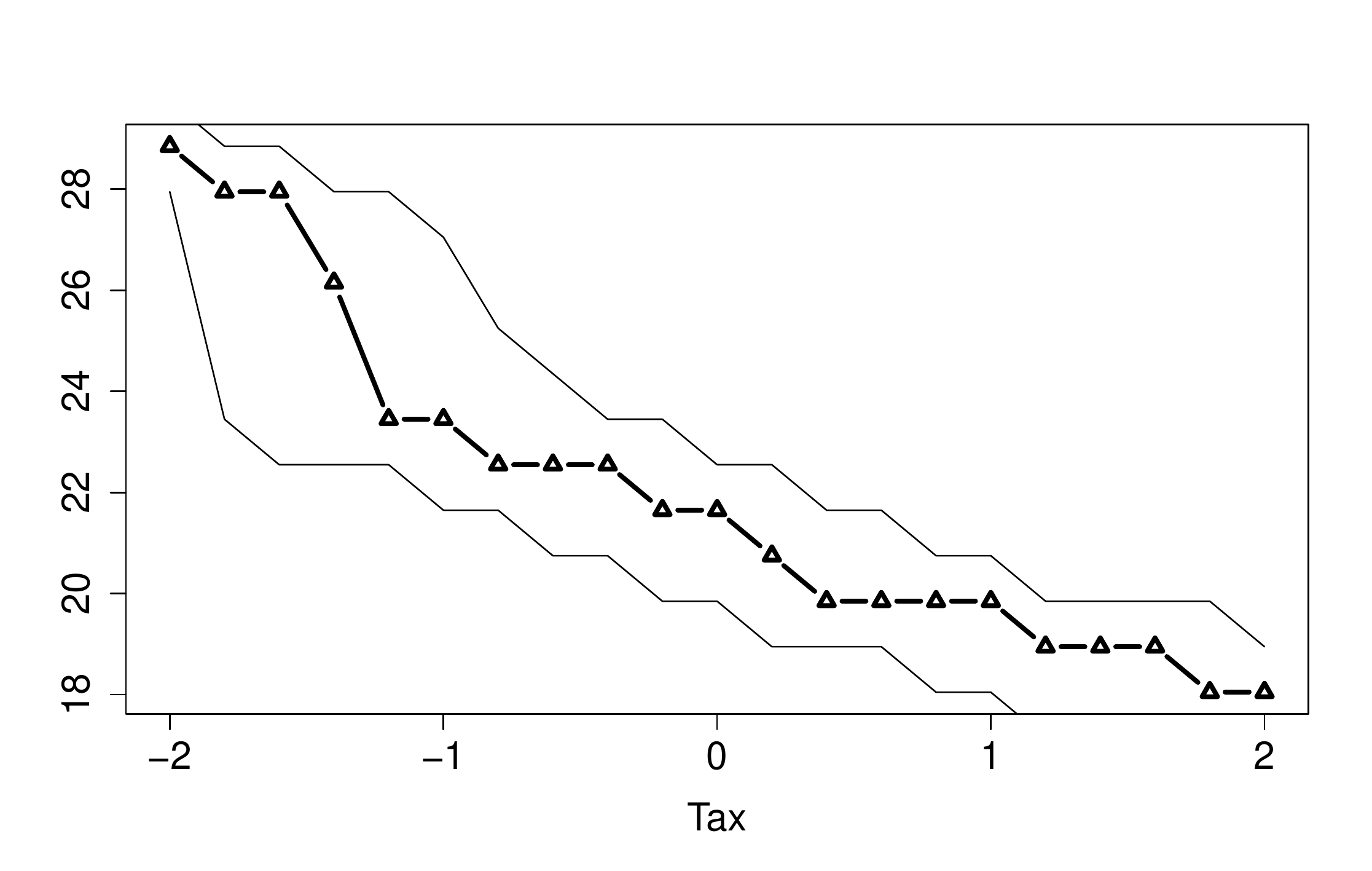}
\caption{Left:  median, .25 and .75 quantiles for air data with dependent variable ozone and explanatory variable temperature; right: median, .25 and .75 quantiles for housing data data with dependent variable median value of homes and explanatory variable tax rate. }
\label{fig:datasetsquant}
\end{figure}

The left panel of Figure \ref{fig:datasetsquant} shows the .25, .5, .75 quantiles of ozone as a function of temperature when all other variables are fixed at their mean value. It is seen that the dependence of  quantiles on temperature is non-linear and S-shaped. The median as well as the quartiles increase only in the middle range but are rather constant for high and low values of temperature. The right panel shows the corresponding quantiles for the variable tax from the housing data. It is seen that the median value of homes decreases non-linearly with increasing tax rate.

\section{Inference Issues: Bootstrap}\label{sec:boot}

\begin{figure}[H]
\centering
\includegraphics[width=0.45\textwidth]{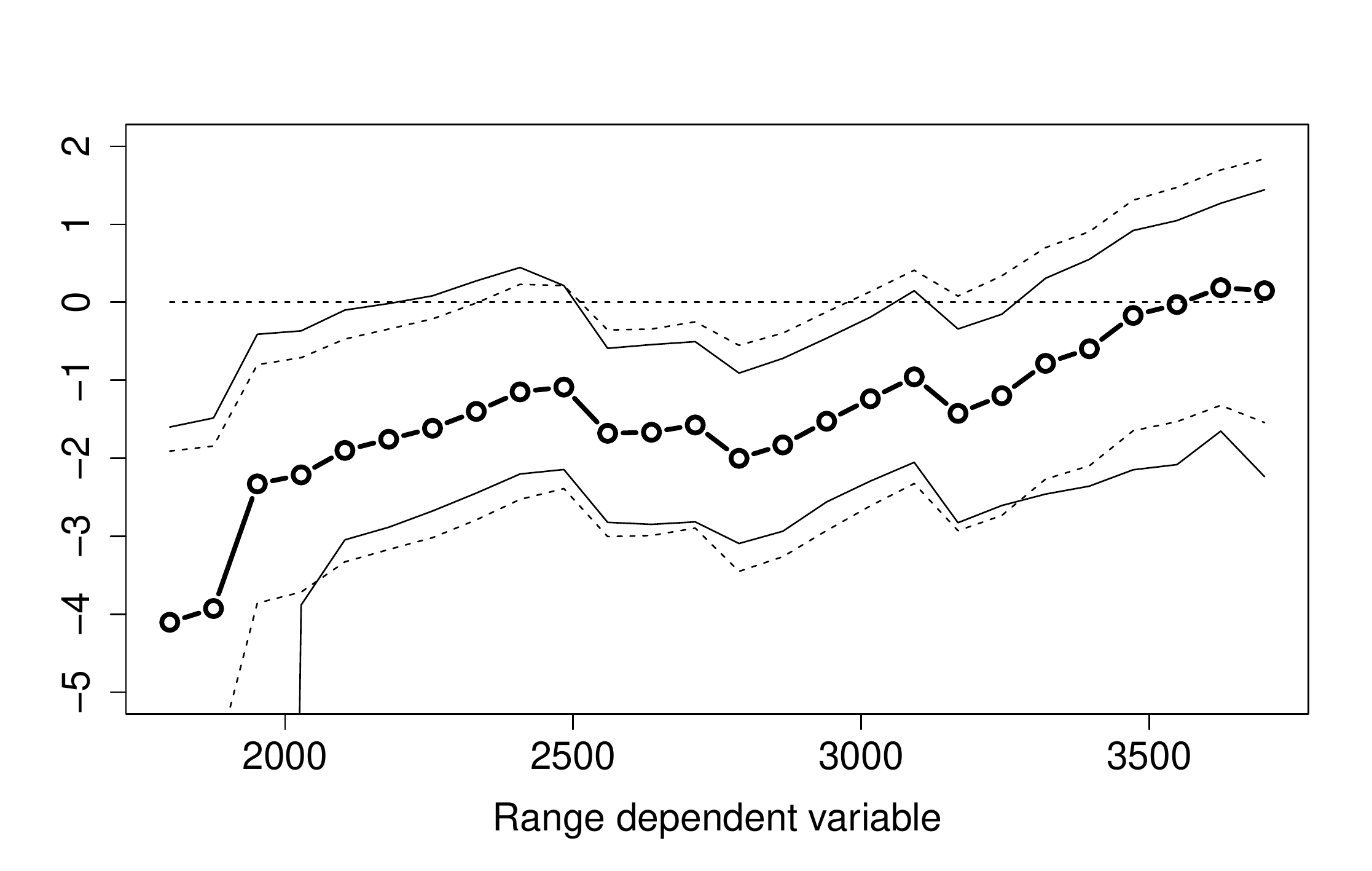}
\includegraphics[width=0.45\textwidth]{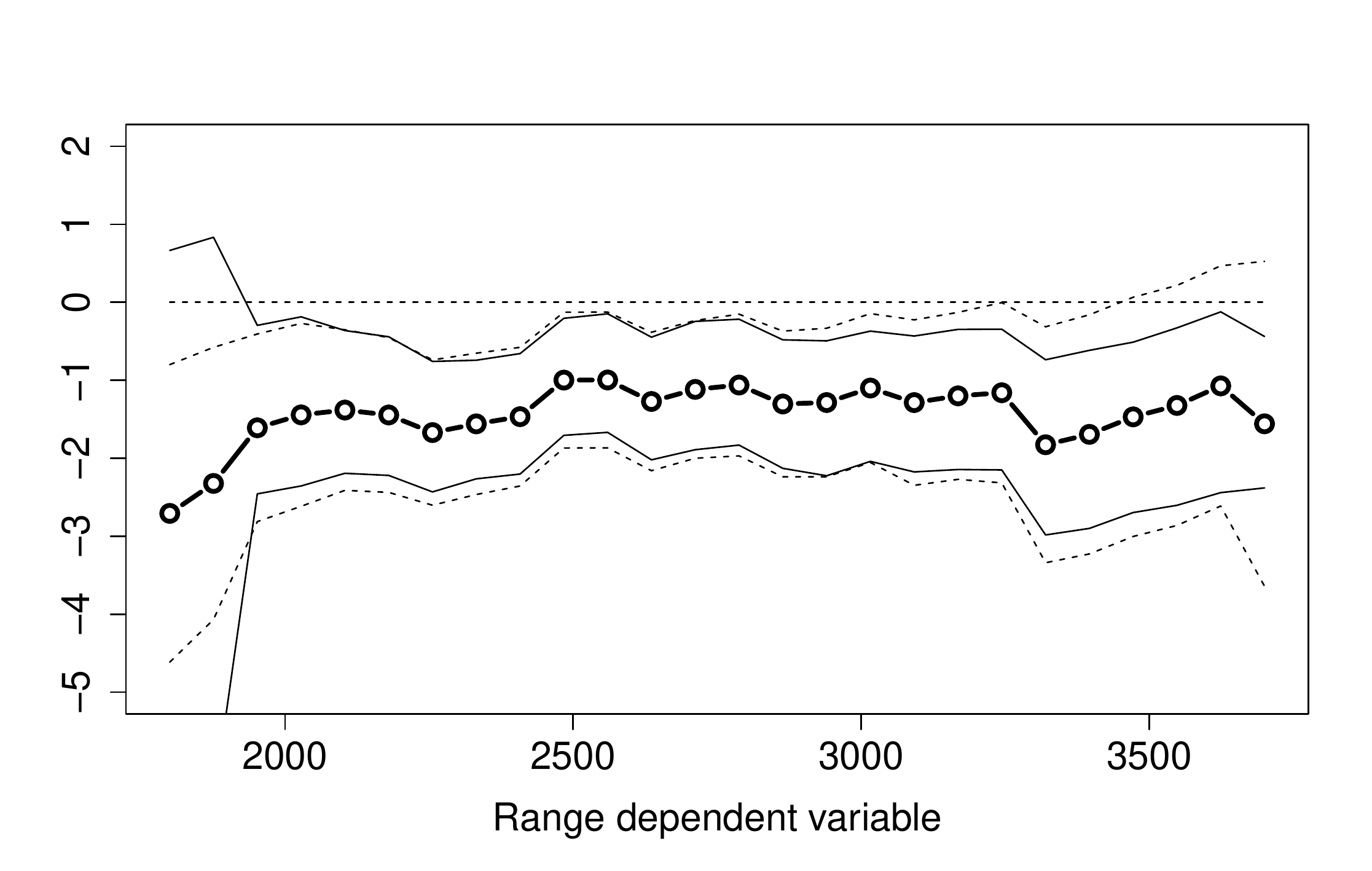}
\caption{Pointwise confidence intervals for variables hypertension (left) and uterine irritability (right) for birth weight data, solid lines are bootstrap intervals, dashed lines are +-1.96 standard errors from original estimates.}
\label{fig:birthboot}
\end{figure}

In the case of a linear predictor with varying coefficients the fitting of  binary models that is used in the general algorithm yields standard errors for the parameters, which provide variability bands as in Figure \ref{fig:birth2}. 
However,  a monotonicity correction is used that can  slightly modify the original estimates. Alternatively one can use maximum likelihood estimates as considered in Section \ref{sec:ML}. A further alternative that also applies in more advanced model structures are bootstrap based confidence intervals \citep{
diciccio1996bootstrap}, which are straightforward to compute. Figure \ref{fig:birthboot}  shows the bootstrap generated pointwise .95 confidence intervals for the variables hypertension  and uterine irritability obtained from the general algorithm. The solid lines are the bootstrap based intervals (1000 replications), the doted lines are the intervals obtained from the original estimates. It is seen that the intervals are very similar over a wide range, only at the boundaries bootstrap intervals are wider indicating stronger uncertainty of estimates.

\begin{figure}[H]
\centering
\includegraphics[width=0.45\textwidth]{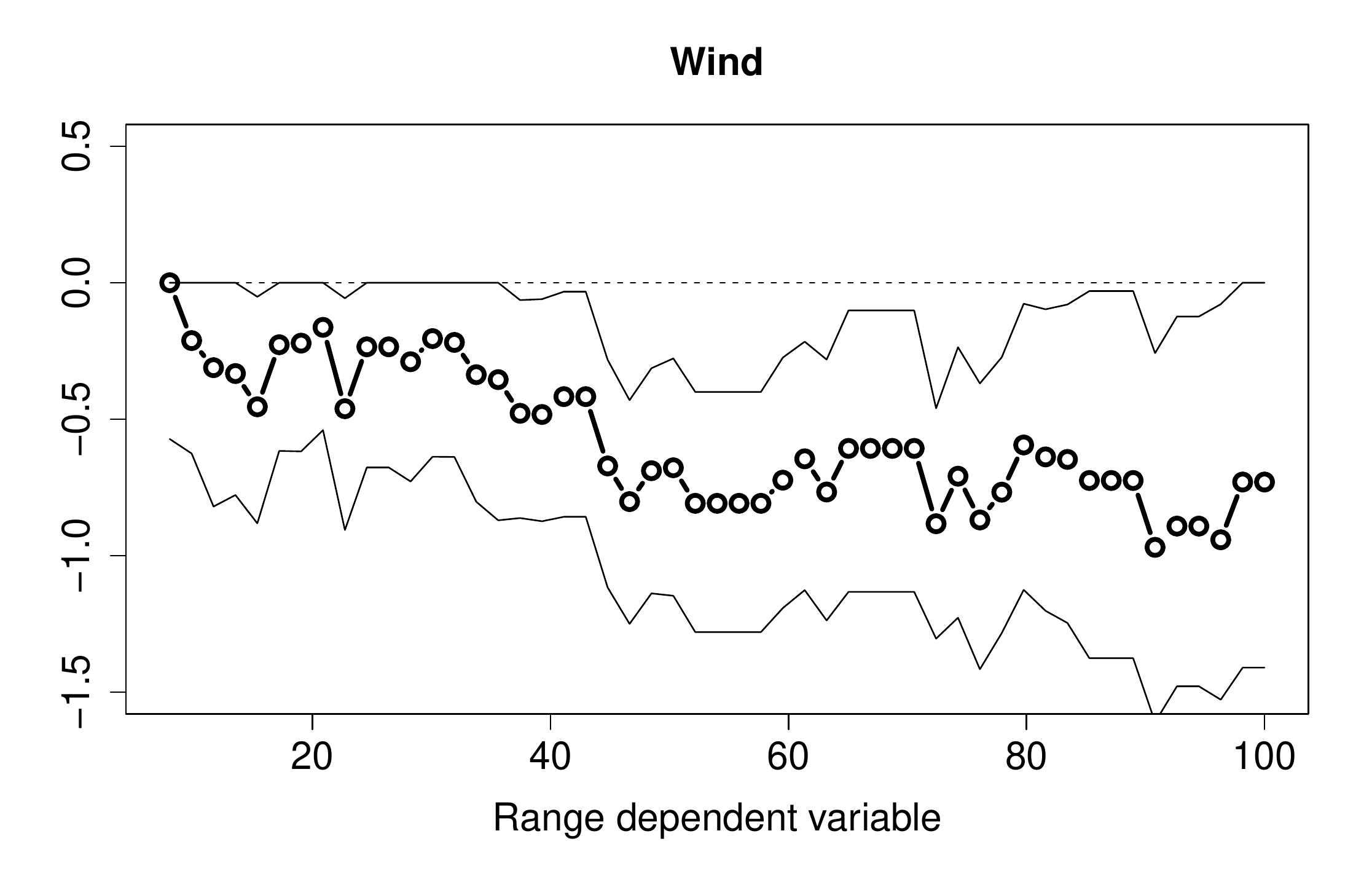}
\includegraphics[width=0.45\textwidth]{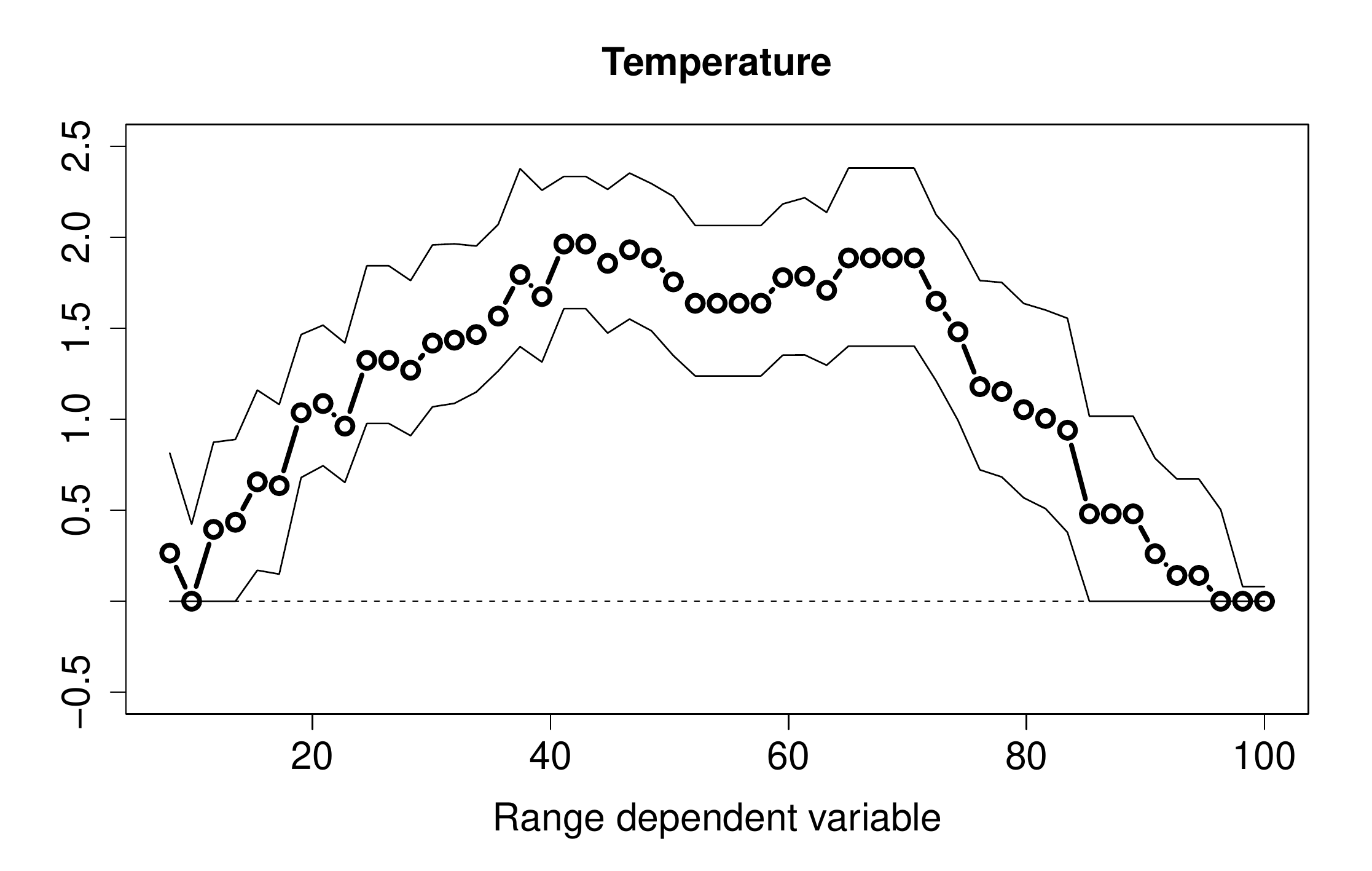}
\includegraphics[width=0.45\textwidth]{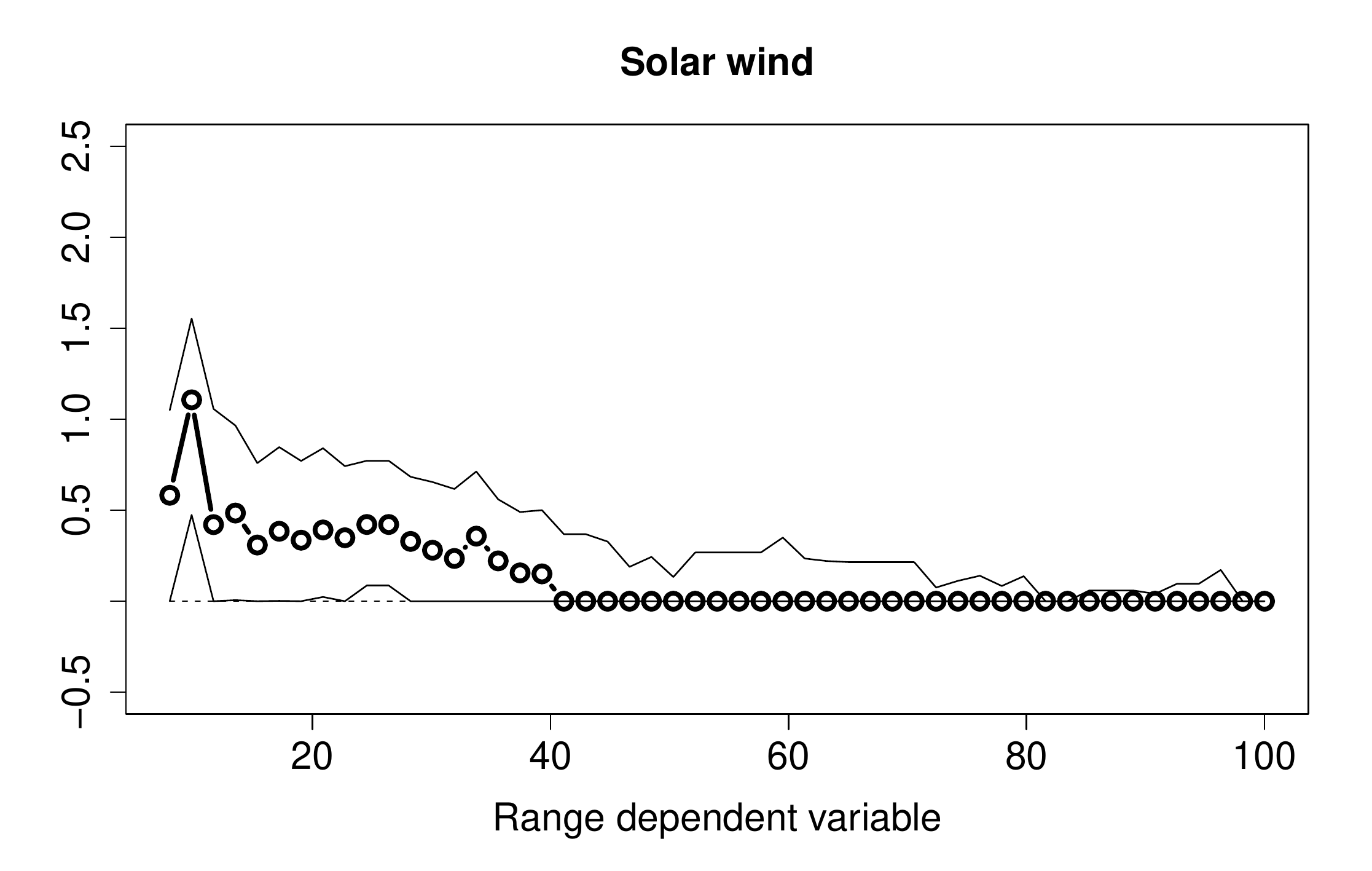}
\includegraphics[width=0.45\textwidth]{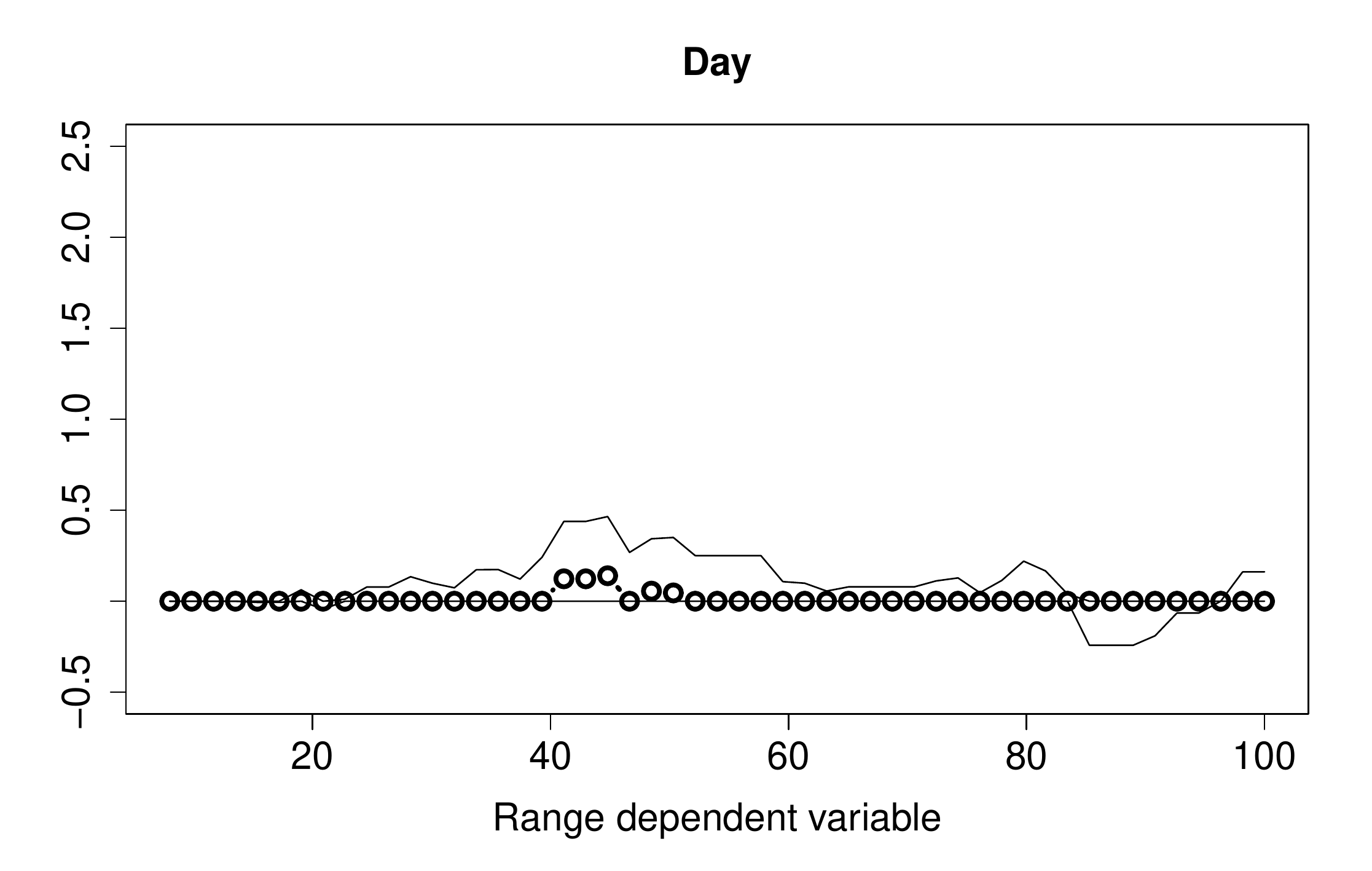}
\caption{Pointwise confidence intervals for variables ... for air quality data when using the lasso to obtain a reduced set of predictors.}
\label{fig:airboot}
\end{figure}

If selection tools like the lasso are used to select variables confidence intervals are much harder to obtain since the fitting of binary models with regularization does not provide standard errors. Then, bootstrap intervals suggest themselves.
Figure \ref{fig:airboot} shows the obtained coefficient curves together with bootstrap based intervals for the air quality data if lasso is used to select relevant variables. It is seen that the variables wind and temperature are distinctly different from zero over a wide range. We also included two other curves, which show that solar wind is included by lasso only for small values while  the variable day is to be considered as irrelevant over the whole range.
It is straightforward to obtain  more pleasing smoother parameter functions and confidence intervals by using, for example, 
local modeling methods \citep{CleDev:88a, CleLoa:96}. We did not use smoothing methods because they conceal the variation of the original estimates.

\begin{figure}[h!]
\centering
\includegraphics[width=0.45\textwidth]{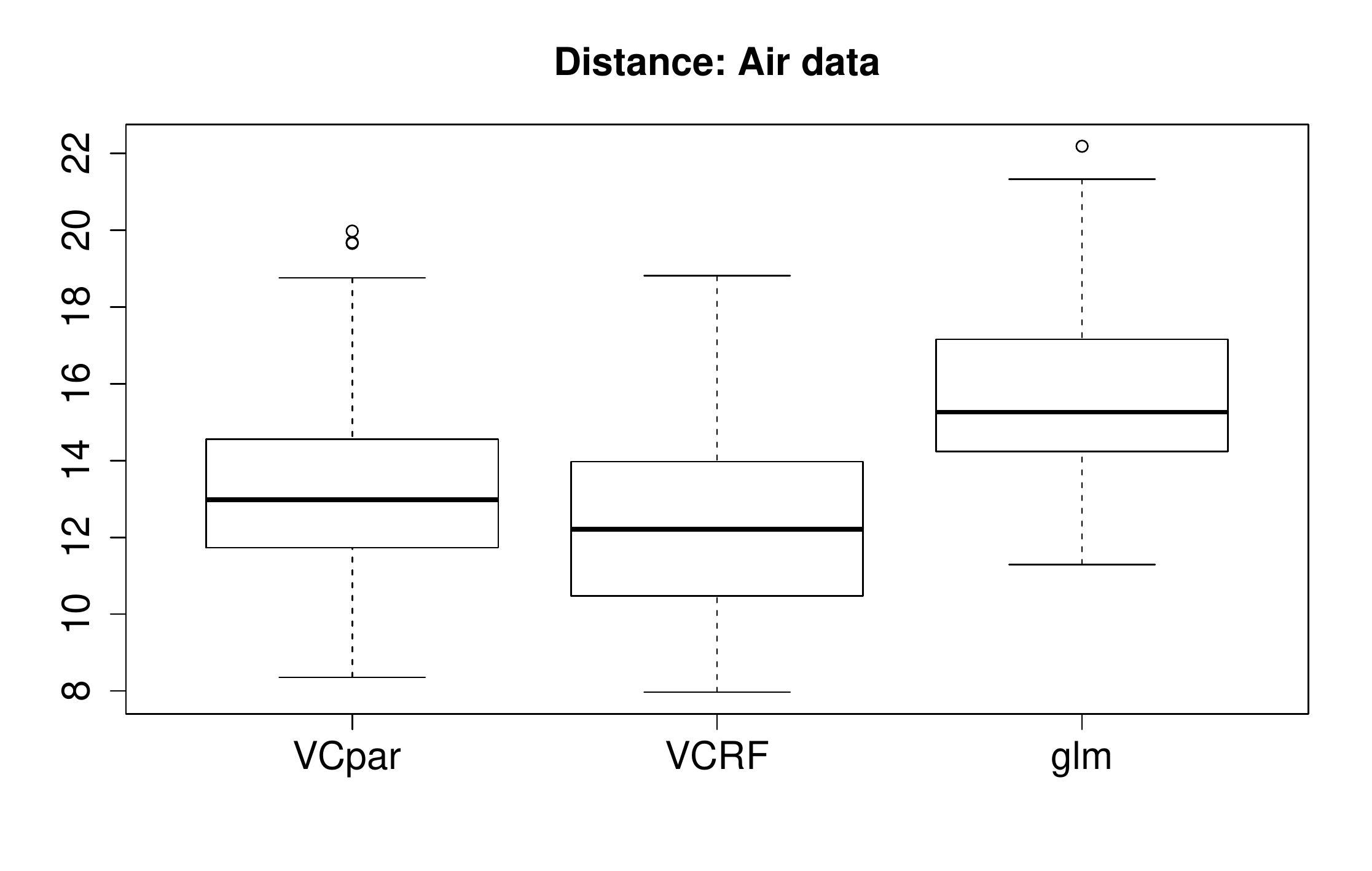}
\includegraphics[width=0.45\textwidth]{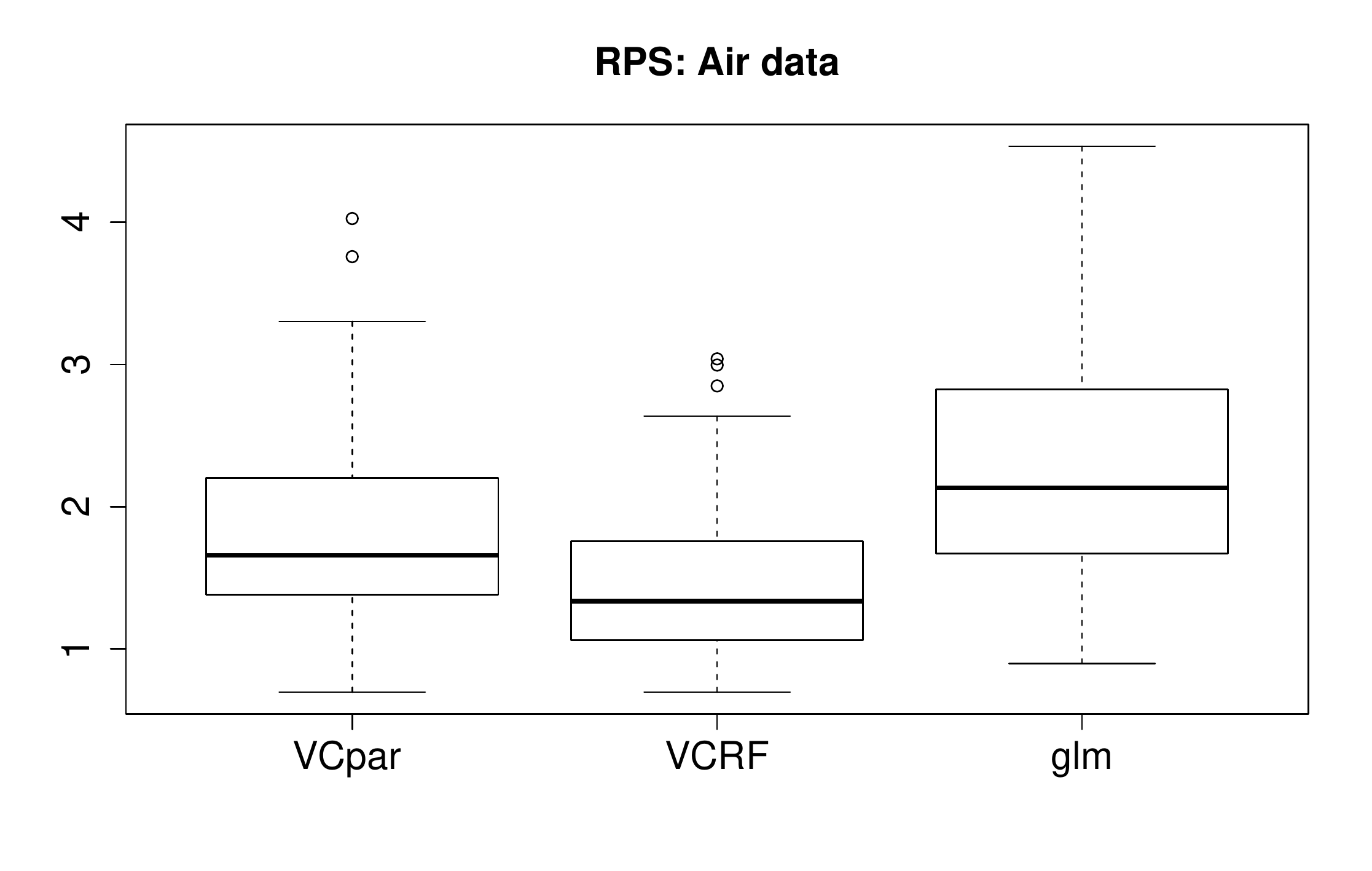}
\includegraphics[width=0.45\textwidth]{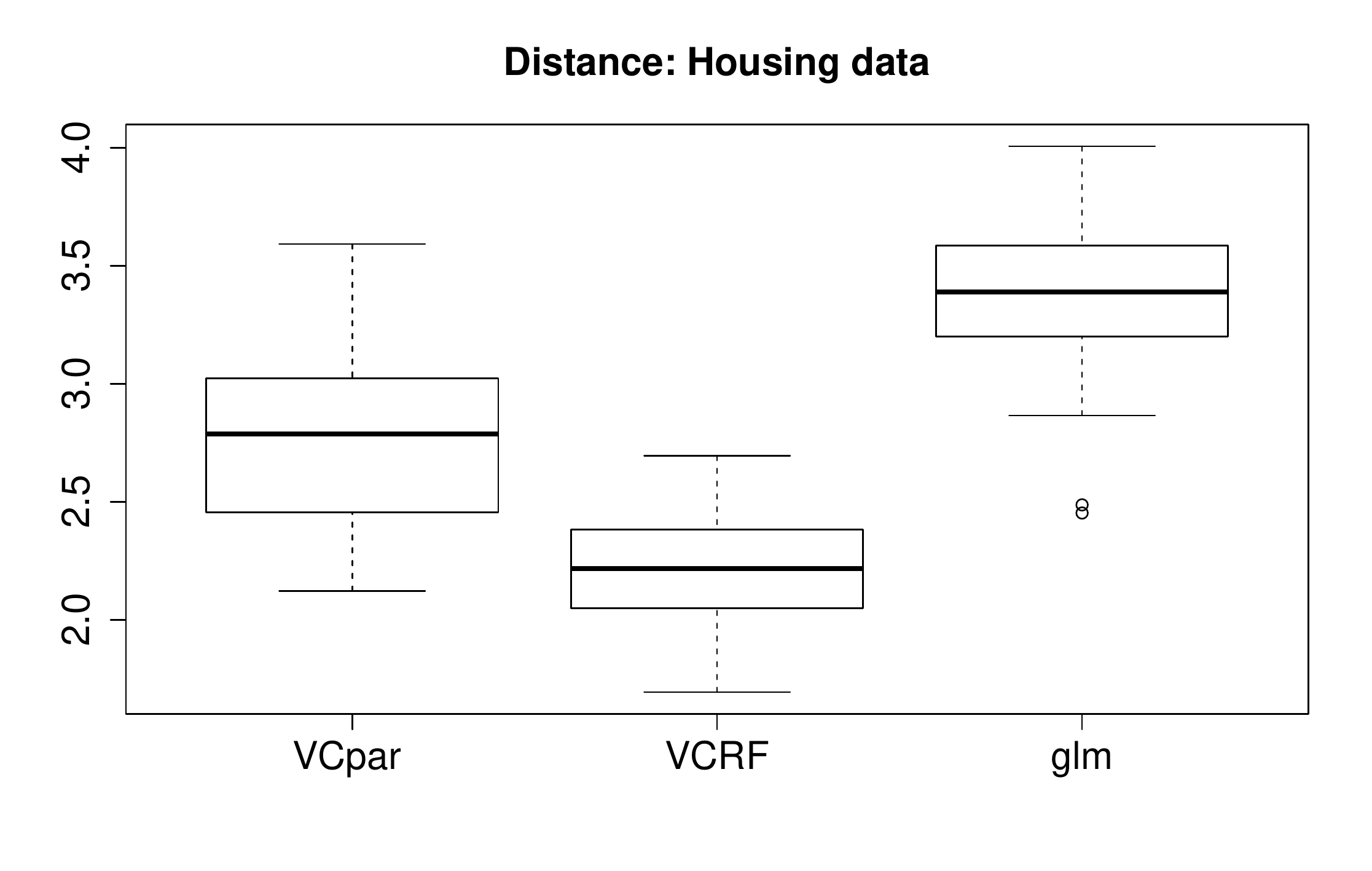}
\includegraphics[width=0.45\textwidth]{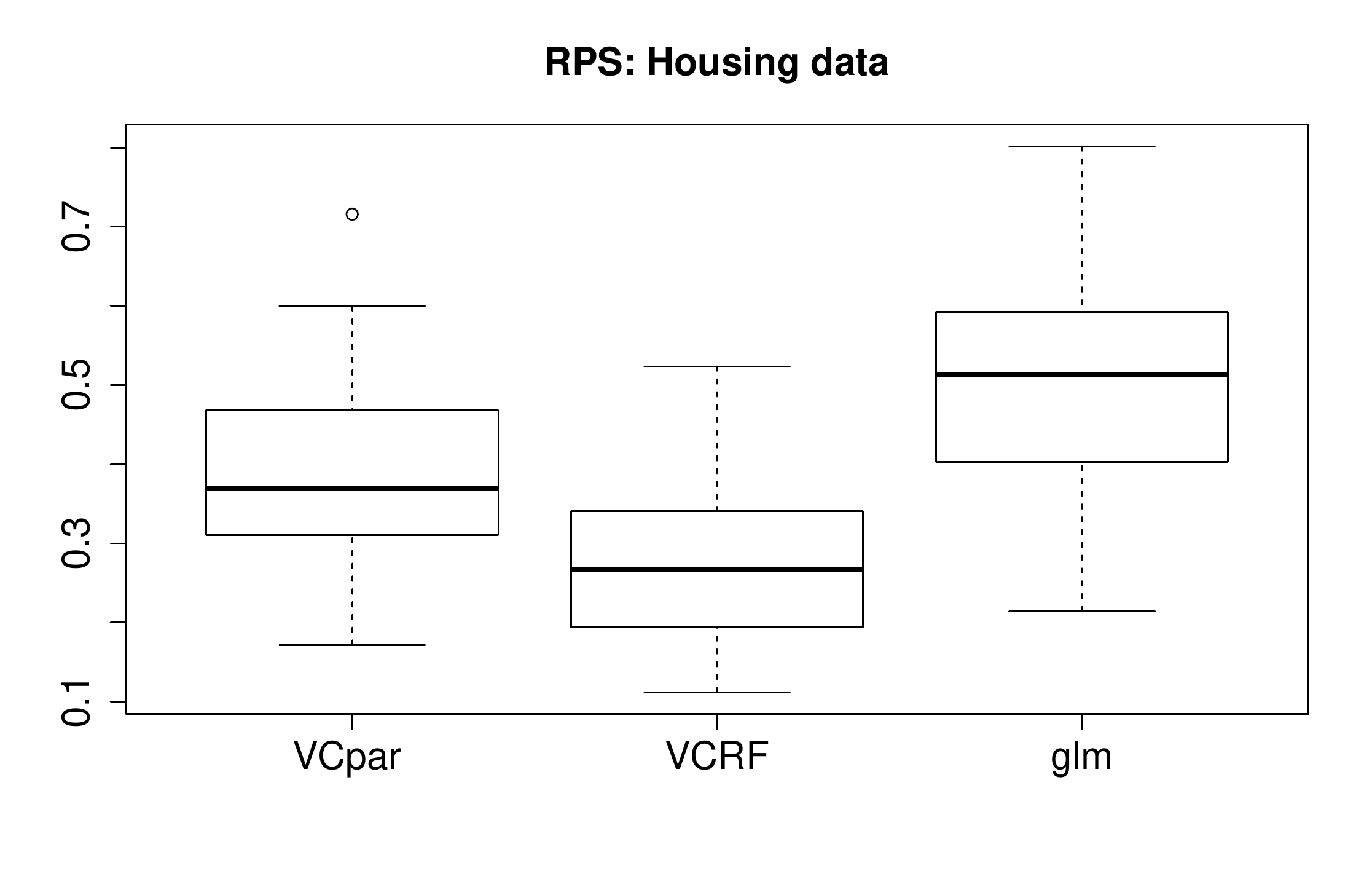}
\includegraphics[width=0.45\textwidth]{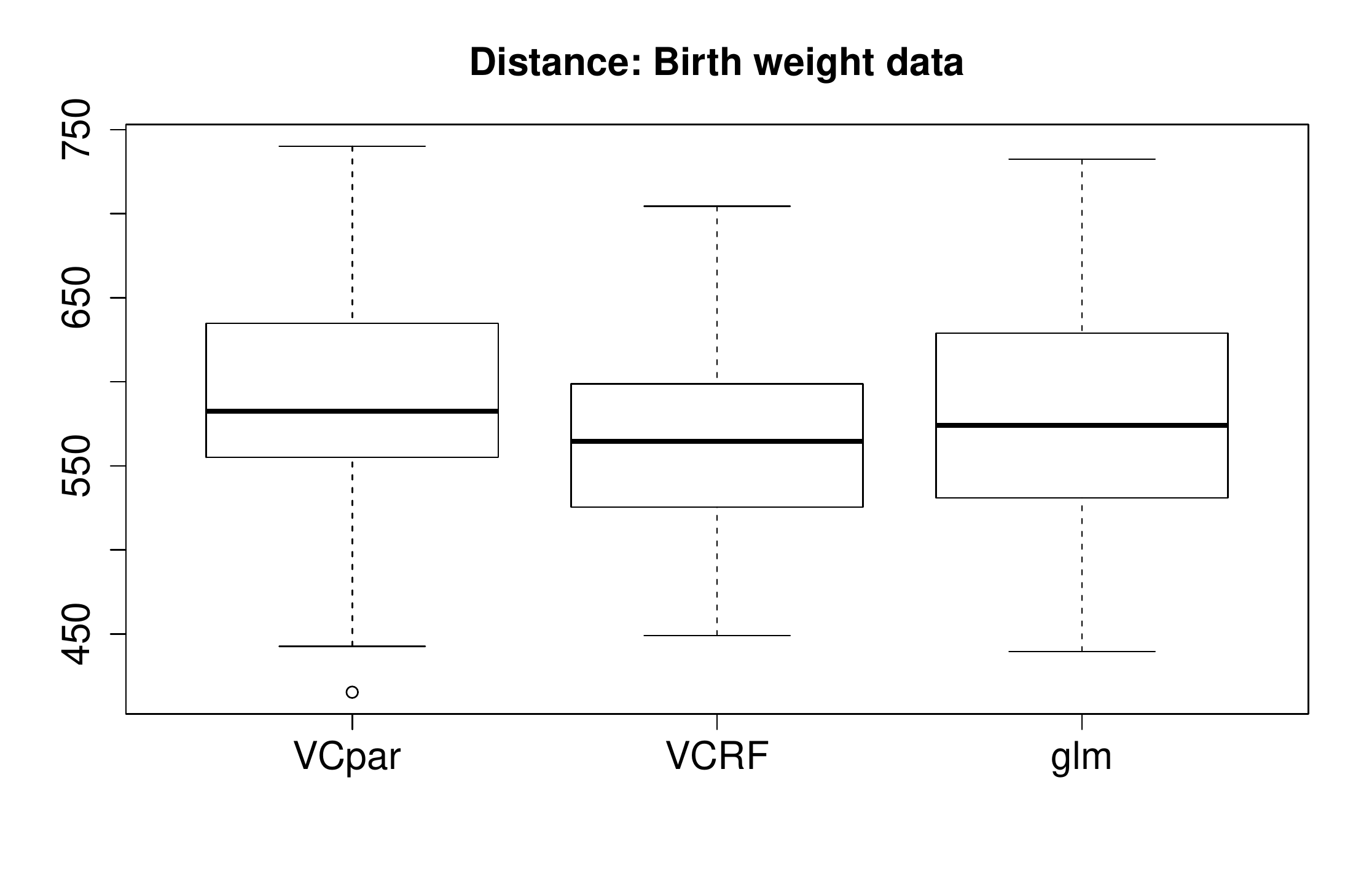}
\includegraphics[width=0.45\textwidth]{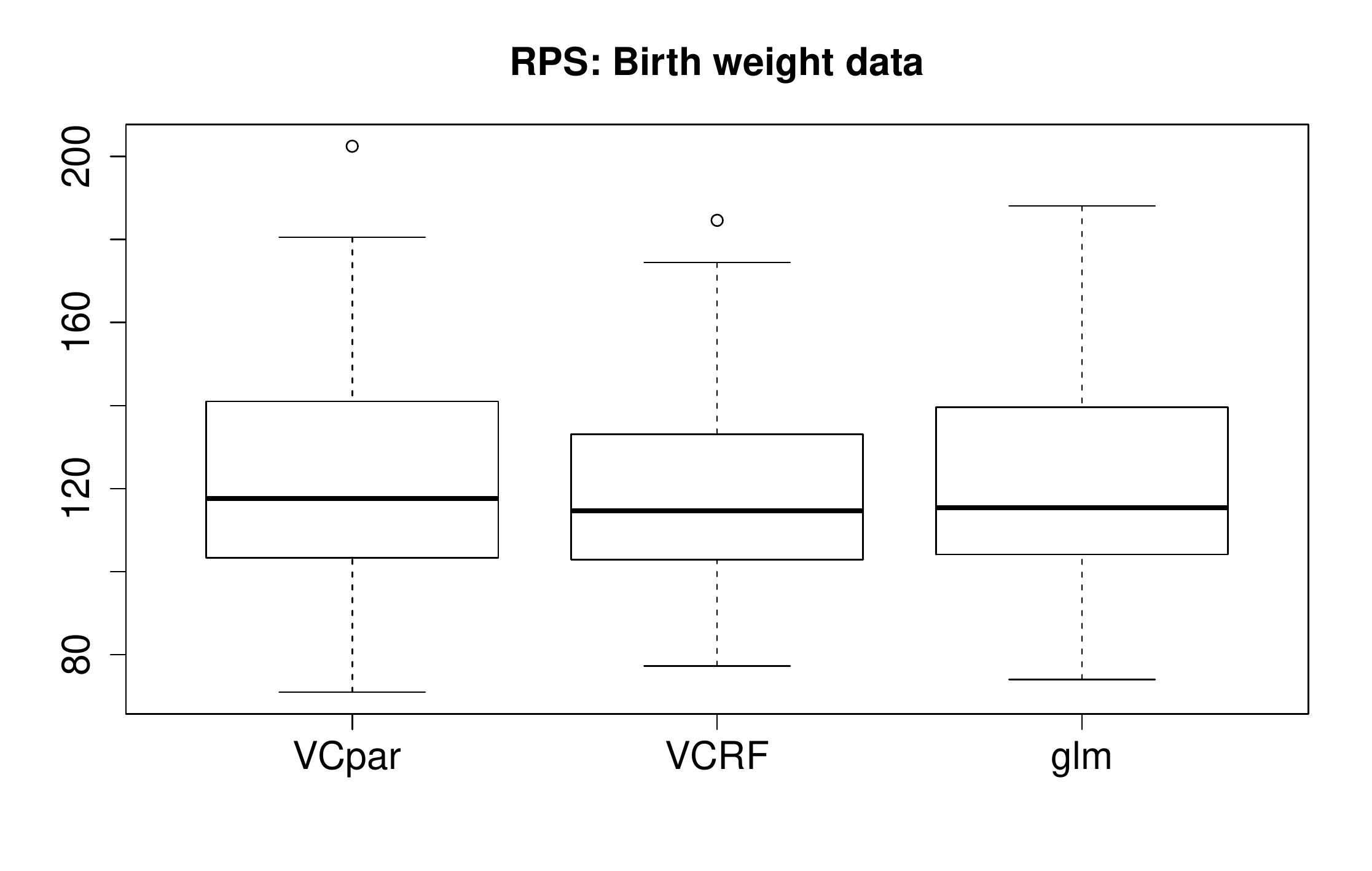}
\caption{Accuracy of prediction for several data sets, left column shows the accuracy in terms of distance, the right column shows the  ranked probability scores.}
\label{fig:pred}
\end{figure}

\section{Prediction}\label{sec:pred}

Beyond the identification of the explanatory variables that are relevant an important criterion of the usefulness of a  of regression model  is accuracy of prediction. In the following we consider briefly the benefits of using varying-threshold models to obtain better predictions.

With continuous dependent variables a  measure that suggests itself is the distance between observation $Y_i$ and prediction $\hat Y_i$,
\[
L_a(Y_i,\hat Y_i) = | Y_i-\hat Y_i|.
\]
A measure that uses not only the point prediction but  the whole predictive distribution is the ranked probability score (\citep{Gneitingetal:2007}). With the estimated distribution function denoted by $\hat F_i(.)$ it is given by
\[
L_{RPS}(Y_i,\hat F_i(.)) = \int (I(Y_i \le y) - \hat F_i(y))^2 dy,
\]
where $I(.)$ is the indicator function with $I(a)=1$ if $a$ is true and $I(a)=0$ otherwise.

Figure \ref{fig:pred} shows the prediction results for the three data sets, the ozone data, the housing data, and the birth weight data.
The data have been split 50 times with 4/5 of the data used as learning data and the rest as validation data. The boxplots show the performance in the validation data sets.
The left column shows the accuracy in terms of distance, the right column shows the obtained ranked probability scores.
The methods that are compared are
\begin{itemize}
\item VCpar: parametric varying coefficients model with linear predictor,
\item VCRF:  varying-thresholds model with random forest based on the program package \textit{randomForest},
\item glm: generalized linear model with normal distribution for response.
\end{itemize}
It is seen that for the air quality data and the housing data the varying coefficients models distinctly outperforms the classical regression model. In particular the random forests varying-coefficients model yields much better predictions than the other approaches. Only for the birth weight data there is not much to win in terms of better prediction when using a model that is more flexible than the classical regression model.

\section{Concluding Remarks}
Varying-coefficients models are a general class of models for continuous and ordinal dependent variables. In their linearly structured form they are generalizations of classical regression models, and can be used as diagnostic tools that can detect discrepancies between  the true data generating mechanism and the simplifying assumptions made by classical models with regard to the variation of parameters across thresholds. In more general nonparametric forms they gain additional flexibility by allowing for nonparametrically structured predictors. 

A strength of the approach is that all available program packages for binary responses can be used to fit the models, including generalized additive models and fitting routines that select variables. While variable selection methods like the lasso have been used in applications additive models have been omitted although their use is straightforward. In applications we also focused on continuous dependent variables, but an example with an ordinal response is given in the appendix.

For linearly structured models  software to compute the constrained maximum likelihood estimates and their standard errors has been written (to be published on GitHub). It fits linear models by using B-spines and ensures that the resulting parameter functions are admissible.




\bibliography{literatur}

\newpage
\section*{Appendix}
\subsubsection*{Equivalence of Linear Model and Varying Coefficients Model}
It has already been shown that  a varying coefficients model with  a linear intercept function follows if a linear regression model holds. 
Therefore we consider how to derive a regression model from a varying coefficients model.

It is assumed that a varying coefficients model with a linear intercept function,
\[
P(Y > y|\xb)=F(\alpha_0 + \alpha y+\xb^T\betab), 
\]
with $\alpha <0$ holds. Then, one obtains for the distribution function of $Y|\xb$
\[
F_{Y|\xb}(y|\xb) =  P(Y \le y|\xb)= 1- F(\alpha_0 + \alpha y+\xb^T\betab)= F(-\alpha_0 - \alpha y-\xb^T\betab), 
\] 
and for the density 
$f_{Y|\xb}(y|\xb) =  -\alpha f(-\alpha_0 - \alpha y-\xb^T\betab)$, 
where $f(y) = \partial F(y)\partial y$. The expectation of $Y|\xb$ is given by
\begin{align*}
\E(Y|\xb) &= \int -y \alpha f(-\alpha_0 - \alpha y-\xb^T\betab) dy =  \int [(-\eta-\alpha_0 -\xb^T\betab)/ \alpha] f(\eta)d\eta =\\
&=-\alpha_0/\alpha -\xb^T\betab/\alpha, 
\end{align*}
where $\eta=-\alpha_0 - \alpha y-\xb^T\betab$. This yields $\E(Y|\xb)=\gamma_0+\xb^T\gammab$ with $\gamma_0=-\alpha_0/\alpha$ $\gammab=-\betab/\alpha$. For the variance one obtains
\begin{align*}
\var(Y|\xb) &= \int -(y+ (\alpha_0 +\xb^T\betab)/\alpha) \alpha f(-\alpha_0 - \alpha y-\xb^T\betab) dy =  \int (\eta/\alpha)^2 f(\eta)d\eta =\\
&=\var_F /\alpha^2, 
\end{align*}
where $\var_F=\int \eta^2 f(\eta)d\eta$ is the variance corresponding to $F(.)$.

Thus one has a linear model $Y=\gamma_0+\xb^T\gammab +\sigma \varepsilon$ with $\epsilon$ having distribution function $F(.)$ and $\sigma^2=\var_F /\alpha^2$.

\subsubsection*{Ordinal Responses: Safety Data }
The package CUB \citep{iannario2018cub} contains the data set relgoods, which provides results of a survey aimed at measuring the subjective extent of feeling safe in the streets. The data were collected  in the metropolitan area of Naples, Italy. Every participant was asked to assess on a 10 point ordinal scale his/her personal score for  feeling  safe with large categories referring to feeling  safe.
There are $n=2225$ observations and five variables,   \textit{Gender} (0: male, 1: female),   the educational degree (\textit{EduDegree}; 1: compulsory school, 2: high school diploma, 3: Graduated-Bachelor degree, 4: Graduated-Master degree, 5: Post graduated),
\textit{WalkAlone} (1 = usually walking alone, 0 = usually walking in company), 
\textit{Neighbours} (Quality of the relationships with neighbors),  
\textit{Age}.

\begin{figure}[H]
\centering
\includegraphics[width=0.45\textwidth]{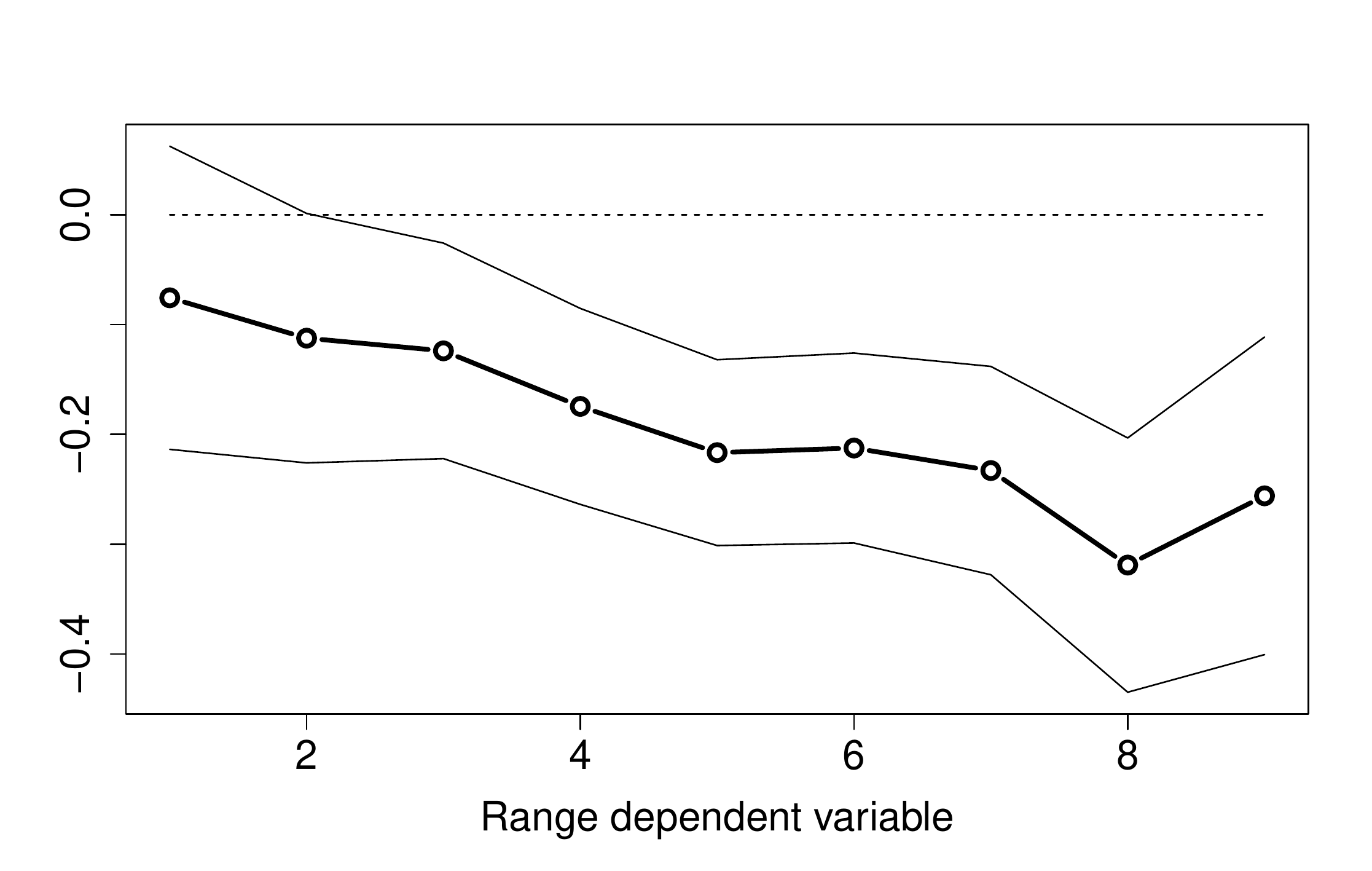} 
\includegraphics[width=0.45\textwidth]{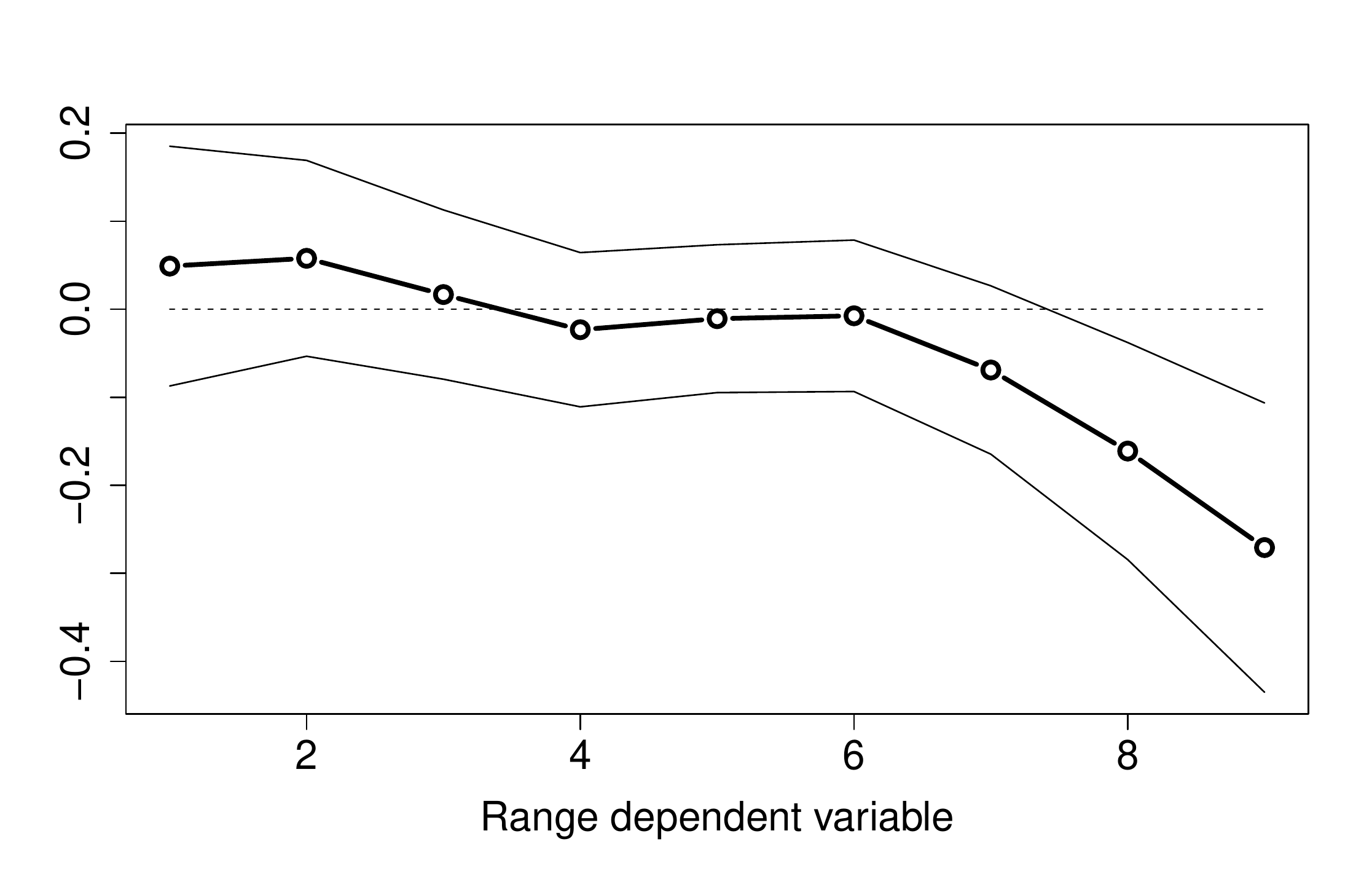}
\includegraphics[width=0.45\textwidth]{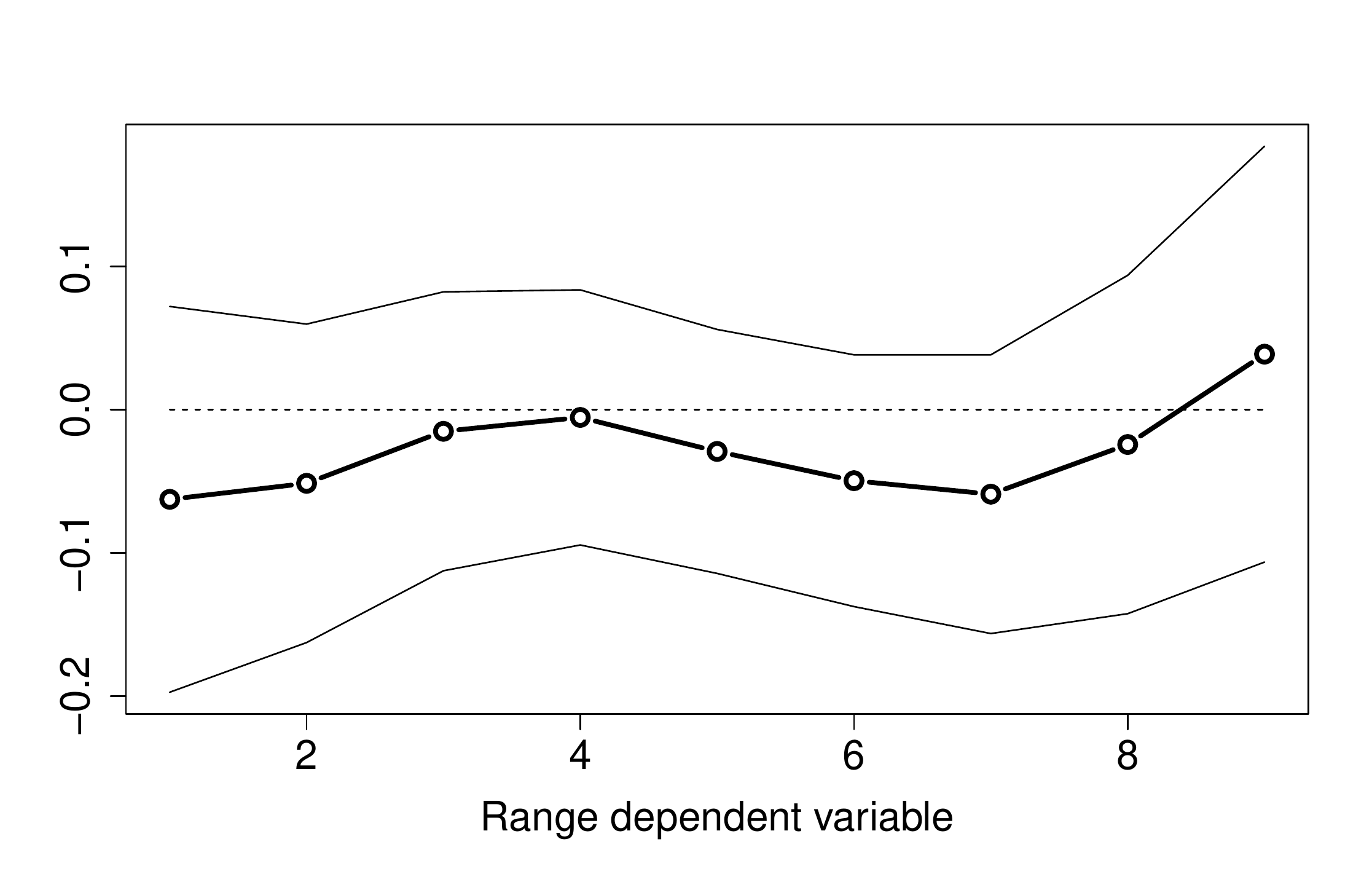}
\includegraphics[width=0.45\textwidth]{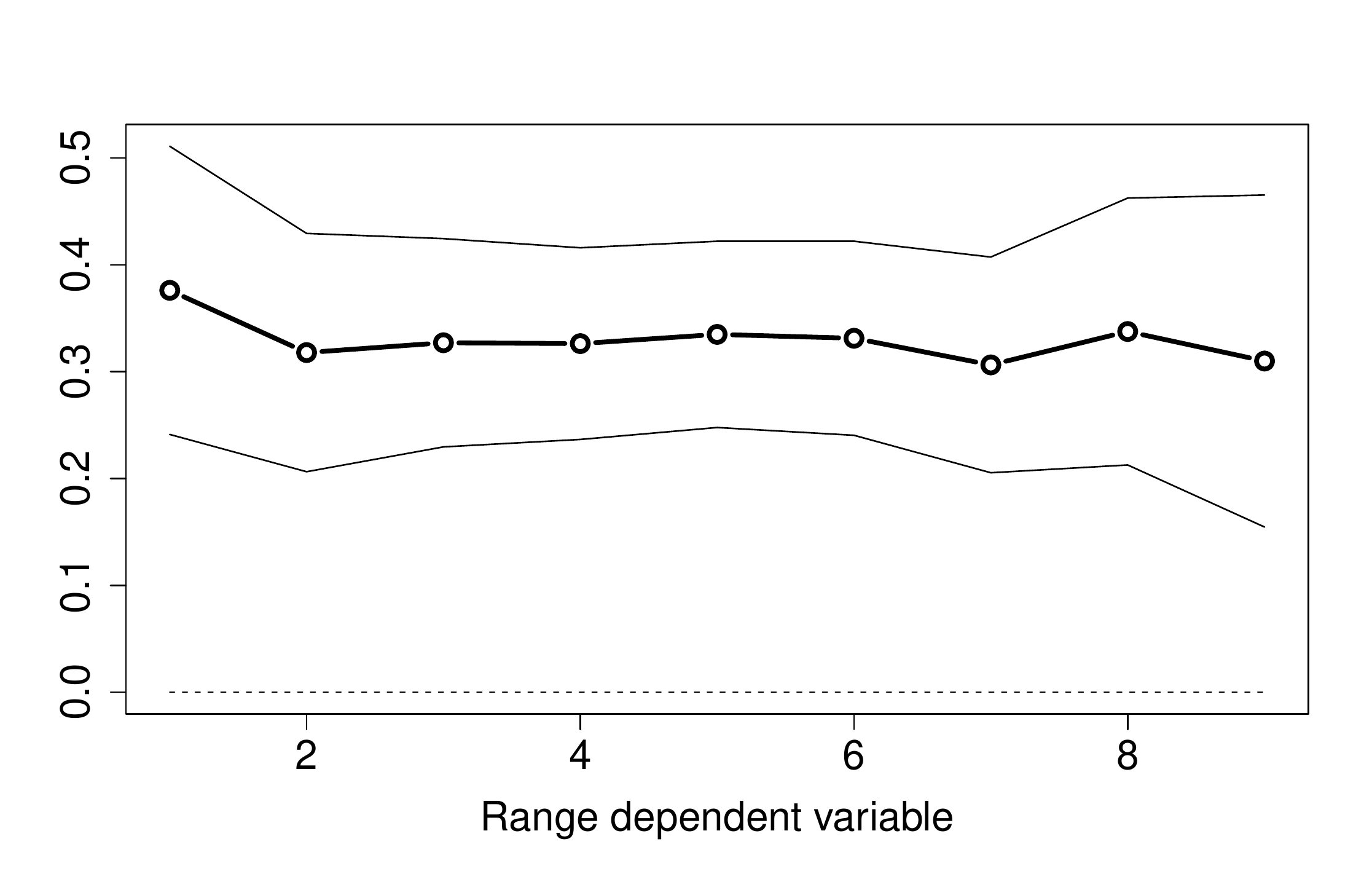}
\includegraphics[width=0.45\textwidth]{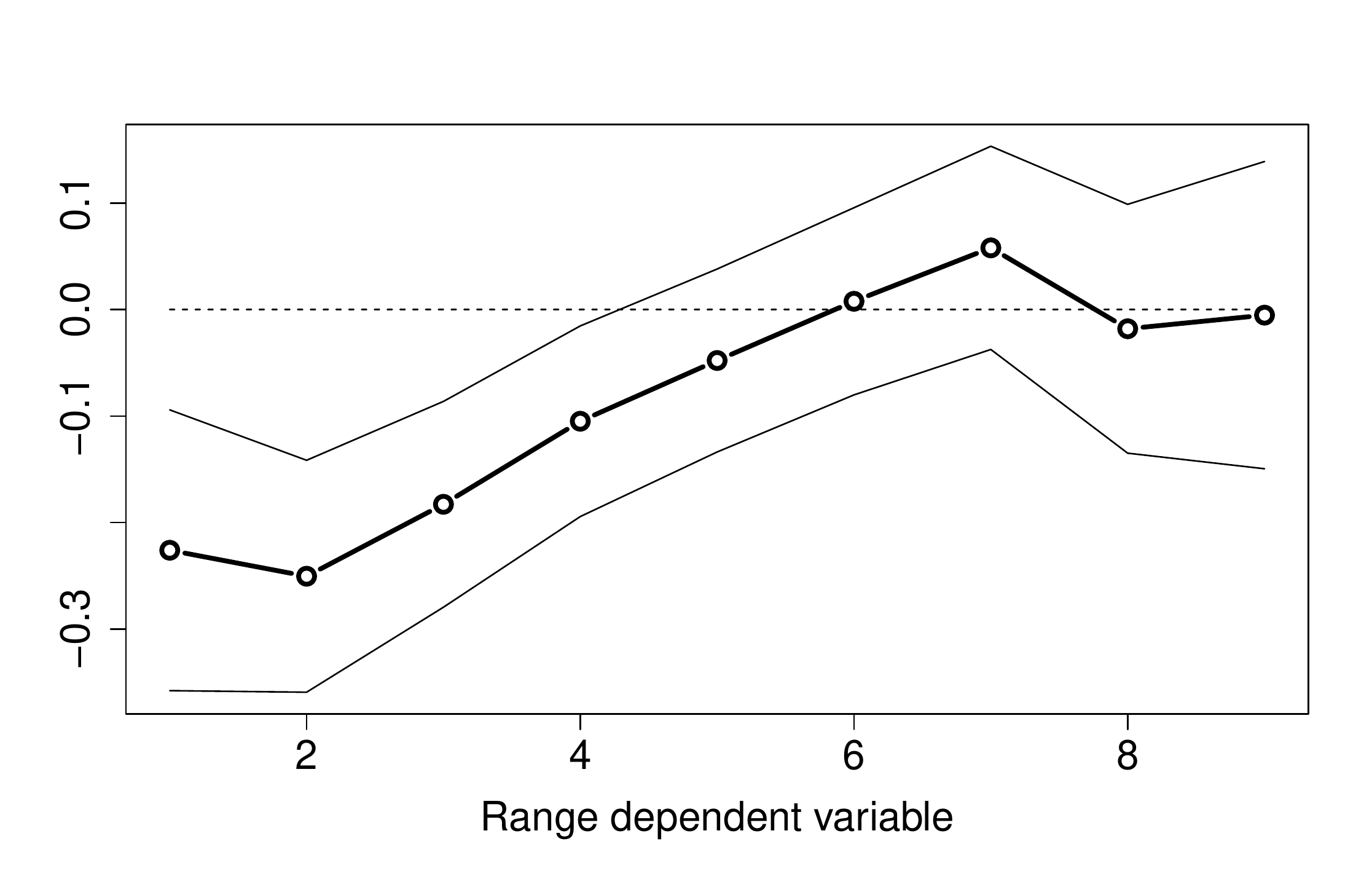}
\caption{Parameter functions for safety data for variables gender, educational degree, WalkAlone, Neighbours, Age (in that order)}.
\label{fig:safety}
\end{figure}

A varying-coefficient model with the logistic distribution function is fitted for the nine thresholds. It corresponds to a logistic cumulative model 
with category-specific parameters. Figure \ref{fig:safety} shows the resulting parameter functions. It is seen that effects are decreasing for gender and educational degree, increasing for age, constant but significant for Neighbours, and negligible for WalkAlone.

\end{document}

%% file: def.tex
\newcommand{\E}{\operatorname{E}}      
\newcommand{\var}{\operatorname{var}}

\newcommand{\alphab}{\boldsymbol{\alpha}}

\newcommand{\betab}{{\boldsymbol{\beta}}}

\newcommand{\deltab}{\boldsymbol{\delta}}

\newcommand{\gammab}{\boldsymbol{\gamma}}

\newcommand{\Phib}{{\boldsymbol{\Phi}}}

\newcommand{\ab}{\boldsymbol{a}}

\newcommand{\xb}{\boldsymbol{x}}

\newcommand{\zb}{\boldsymbol{z}}

\newcommand{\blanco}[1]{}

\def\d{\displaystyle}

%% file: ParametricQuantile3.bbl
\begin{thebibliography}{}

\bibitem[\protect\citeauthoryear{Agresti}{Agresti}{2010}]{Agresti:2009}
Agresti, A. (2010).
\newblock {\em Analysis of Ordinal Categorical Data, 2nd Edition}.
\newblock New York: Wiley.

\bibitem[\protect\citeauthoryear{Brant}{Brant}{1990}]{Brant:90}
Brant, R. (1990).
\newblock Assessing proportionality in the proportional odds model for ordinal
  logistic regression.
\newblock {\em Biometrics\/}~{\em 46}, 1171--1178.

\bibitem[\protect\citeauthoryear{Breiman}{Breiman}{1998}]{Breiman:98}
Breiman, L. (1998).
\newblock Arcing classifiers.
\newblock {\em Annals of Statistics\/}~{\em 26}, 801--849.

\bibitem[\protect\citeauthoryear{Breiman}{Breiman}{2001}]{Breiman:2001a}
Breiman, L. (2001).
\newblock Random forests.
\newblock {\em Machine Learning\/}~{\em 45}, 5--32.

\bibitem[\protect\citeauthoryear{Breiman, Friedman, Olshen, and Stone}{Breiman
  et~al.}{1984}]{BreiFrieOls:84}
Breiman, L., J.~H. Friedman, R.~A. Olshen, and J.~C. Stone (1984).
\newblock {\em Classification and Regression Trees}.
\newblock Monterey, CA: Wadsworth.

\bibitem[\protect\citeauthoryear{B\"{u}hlmann and Van De~Geer}{B\"{u}hlmann and
  Van De~Geer}{2011}]{buehlmann2011statistics}
B\"{u}hlmann, P. and S.~Van De~Geer (2011).
\newblock {\em Statistics for High-Dimensional Data: Methods, Theory and
  Applications}.
\newblock Springer-Verlag New York.

\bibitem[\protect\citeauthoryear{Cai, Fan, and Li}{Cai
  et~al.}{2000}]{cai2000efficient}
Cai, Z., J.~Fan, and R.~Li (2000).
\newblock Efficient estimation and inferences for varying-coefficient models.
\newblock {\em Journal of the American Statistical Association\/}~{\em
  95\/}(451), 888--902.

\bibitem[\protect\citeauthoryear{Cleveland and Devlin}{Cleveland and
  Devlin}{1988}]{CleDev:88a}
Cleveland, W.~S. and S.~J. Devlin (1988).
\newblock Locally weighted regression: An approach to regression analysis by
  local fitting.
\newblock {\em {J}ournal of the {A}merican {S}tatistical {A}ssociation\/}~{\em
  83}, 596--610.

\bibitem[\protect\citeauthoryear{Cleveland and Loader}{Cleveland and
  Loader}{1996}]{CleLoa:96}
Cleveland, W.~S. and C.~Loader (1996).
\newblock Smoothing by local regression: Principles and methods.
\newblock In W.~H\"ardle and M.~Schimek (Eds.), {\em Statistical Theory and
  Computational Aspects of Smoothing}, pp.\  10--49. Heidelberg:
  Physica-Verlag.

\bibitem[\protect\citeauthoryear{Cox}{Cox}{1995}]{Cox:95}
Cox, C. (1995).
\newblock Location-scale cumulative odds models for ordinal data: A generalized
  non-linear model approach.
\newblock {\em Statistics in Medicine\/}~{\em 14}, 1191--1203.

\bibitem[\protect\citeauthoryear{Diaz-Uriarte and de~Andres}{Diaz-Uriarte and
  de~Andres}{2006}]{DiaAnd:2006}
Diaz-Uriarte, R. and S.~A. de~Andres (2006).
\newblock Gene selection and classification of microarray data using random
  forest.
\newblock {\em Bioinformatics\/}~{\em 7}, 3.

\bibitem[\protect\citeauthoryear{DiCiccio, Efron, et~al.}{DiCiccio
  et~al.}{1996}]{diciccio1996bootstrap}
DiCiccio, T.~J., B.~Efron, et~al. (1996).
\newblock Bootstrap confidence intervals.
\newblock {\em Statistical science\/}~{\em 11\/}(3), 189--228.

\bibitem[\protect\citeauthoryear{Eilers and Marx}{Eilers and
  Marx}{2021}]{eilers2021practical}
Eilers, P.~H. and B.~D. Marx (2021).
\newblock {\em Practical Smoothing: The Joys of P-splines}.
\newblock Cambridge University Press.

\bibitem[\protect\citeauthoryear{Eilers and Marx}{Eilers and
  Marx}{1996}]{EilMar:96}
Eilers, P. H.~C. and B.~D. Marx (1996).
\newblock Flexible smoothing with {B}-splines and {P}enalties.
\newblock {\em Statistical Science\/}~{\em 11}, 89--121.

\bibitem[\protect\citeauthoryear{Fan and Zhang}{Fan and
  Zhang}{1999}]{fan1999statistical}
Fan, J. and W.~Zhang (1999).
\newblock Statistical estimation in varying coefficient models.
\newblock {\em Annals of Statistics\/}, 1491--1518.

\bibitem[\protect\citeauthoryear{Fitzenberger, Koenker, and
  Machado}{Fitzenberger et~al.}{2013}]{fitzenberger2013economic}
Fitzenberger, B., R.~Koenker, and J.~A. Machado (2013).
\newblock {\em Economic applications of quantile regression}.
\newblock Springer Science \& Business Media.

\bibitem[\protect\citeauthoryear{Galimberti, Soffritti, and Di~Maso}{Galimberti
  et~al.}{2012}]{galimberti2012classification}
Galimberti, G., G.~Soffritti, and M.~Di~Maso (2012).
\newblock Classification trees for ordinal responses in {R}: the rpartscore
  package.
\newblock {\em Journal of Statistical Software\/}~{\em 47}.

\bibitem[\protect\citeauthoryear{Gneiting and Raftery}{Gneiting and
  Raftery}{2007}]{Gneitingetal:2007}
Gneiting, T. and A.~Raftery ({2007}).
\newblock {Strictly proper scoring rules, prediction, and estimation}.
\newblock {\em Journal of the American Statistical Association\/}~{\em
  {102}\/}({477}), 359--376.

\bibitem[\protect\citeauthoryear{Gregorutti, Michel, and
  Saint-Pierre}{Gregorutti et~al.}{2017}]{gregorutti2017correlation}
Gregorutti, B., B.~Michel, and P.~Saint-Pierre (2017).
\newblock Correlation and variable importance in random forests.
\newblock {\em Statistics and Computing\/}~{\em 27\/}(3), 659--678.

\bibitem[\protect\citeauthoryear{Hapfelmeier, Hothorn, Ulm, and
  Strobl}{Hapfelmeier et~al.}{2014}]{hapfelmeier2014new}
Hapfelmeier, A., T.~Hothorn, K.~Ulm, and C.~Strobl (2014).
\newblock A new variable importance measure for random forests with missing
  data.
\newblock {\em Statistics and Computing\/}~{\em 24\/}(1), 21--34.

\bibitem[\protect\citeauthoryear{Hastie and Tibshirani}{Hastie and
  Tibshirani}{1993}]{HasTib:93}
Hastie, T. and R.~Tibshirani (1993).
\newblock Varying-coefficient models.
\newblock {\em Journal of the Royal Statistical Society\/}~{\em B 55},
  757--796.

\bibitem[\protect\citeauthoryear{Hastie, Tibshirani, and Friedman}{Hastie
  et~al.}{2001}]{HasTibFri:2001}
Hastie, T., R.~Tibshirani, and J.~H. Friedman (2001).
\newblock {\em The Elements of Statistical Learning}.
\newblock New York: Springer-Verlag.

\bibitem[\protect\citeauthoryear{Hosmer and Lemeshow}{Hosmer and
  Lemeshow}{1989}]{HosLem:89}
Hosmer, D.~H. and S.~Lemeshow (1989).
\newblock {\em Applied Logistic Regression}.
\newblock New York: Wiley.

\bibitem[\protect\citeauthoryear{Hosmer, Hosmer, Le~Cessie, and
  Lemeshow}{Hosmer et~al.}{1997}]{hosmer1997comparison}
Hosmer, D.~W., T.~Hosmer, S.~Le~Cessie, and S.~Lemeshow (1997).
\newblock A comparison of goodness-of-fit tests for the logistic regression
  model.
\newblock {\em Statistics in medicine\/}~{\em 16\/}(9), 965--980.

\bibitem[\protect\citeauthoryear{Hothorn, Kneib, and B{\"u}hlmann}{Hothorn
  et~al.}{2014}]{hothorn2014conditional}
Hothorn, T., T.~Kneib, and P.~B{\"u}hlmann (2014).
\newblock Conditional transformation models.
\newblock {\em Journal of the Royal Statistical Society: Series B: Statistical
  Methodology\/}, 3--27.

\bibitem[\protect\citeauthoryear{Hothorn and Lausen}{Hothorn and
  Lausen}{2003}]{HotLau:03}
Hothorn, T. and B.~Lausen (2003).
\newblock On the exact distribution of maximally selected rank statistics.
\newblock {\em Computational Statistics and Data Analysis\/}~{\em 43},
  121--137.

\bibitem[\protect\citeauthoryear{Hothorn, Moest, and Buehlmann}{Hothorn
  et~al.}{2018}]{hothorn2018most}
Hothorn, T., L.~Moest, and P.~Buehlmann (2018).
\newblock Most likely transformations.
\newblock {\em Scandinavian Journal of Statistics\/}~{\em 45\/}(1), 110--134.

\bibitem[\protect\citeauthoryear{Hothorn and Zeileis}{Hothorn and
  Zeileis}{2015}]{hothorn2015partykit}
Hothorn, T. and A.~Zeileis (2015).
\newblock partykit: A modular toolkit for recursive partytioning in r.
\newblock {\em The Journal of Machine Learning Research\/}~{\em 16\/}(1),
  3905--3909.

\bibitem[\protect\citeauthoryear{Hothorn and Zeileis}{Hothorn and
  Zeileis}{2021}]{hothorn2021predictive}
Hothorn, T. and A.~Zeileis (2021).
\newblock Predictive distribution modelling using transformation forests.
\newblock {\em Journal of Computational and Graphical Statistics\/}, 1--32.

\bibitem[\protect\citeauthoryear{Iannario, Piccolo, and Simone}{Iannario
  et~al.}{2020}]{iannario2018cub}
Iannario, M., D.~Piccolo, and R.~Simone (2020).
\newblock {CUB}: a class of mixture models for ordinal data. {R} package
  version 1.1.4, http://cran.r-project.org/package=cub.

\bibitem[\protect\citeauthoryear{Koenker, Chernozhukov, He, and Peng}{Koenker
  et~al.}{2017}]{koenker2017handbook}
Koenker, R., V.~Chernozhukov, X.~He, and L.~Peng (2017).
\newblock {\em Handbook of quantile regression}.
\newblock CRC press.

\bibitem[\protect\citeauthoryear{Koenker and Hallock}{Koenker and
  Hallock}{2001}]{koenker2001quantile}
Koenker, R. and K.~F. Hallock (2001).
\newblock Quantile regression.
\newblock {\em Journal of economic perspectives\/}~{\em 15\/}(4), 143--156.

\bibitem[\protect\citeauthoryear{Koenker, Ng, and Portnoy}{Koenker
  et~al.}{1994}]{koenker1994quantile}
Koenker, R., P.~Ng, and S.~Portnoy (1994).
\newblock Quantile smoothing splines.
\newblock {\em Biometrika\/}~{\em 81\/}(4), 673--680.

\bibitem[\protect\citeauthoryear{Liu, Mukherjee, Suesse, Sparrow, and Park}{Liu
  et~al.}{2009}]{liu2009graphical}
Liu, I., B.~Mukherjee, T.~Suesse, D.~Sparrow, and S.~K. Park (2009).
\newblock Graphical diagnostics to check model misspecification for the
  proportional odds regression model.
\newblock {\em Statistics in medicine\/}~{\em 28\/}(3), 412--429.

\bibitem[\protect\citeauthoryear{McCullagh}{McCullagh}{1980}]{McCullagh:80}
McCullagh, P. (1980).
\newblock Regression model for ordinal data (with discussion).
\newblock {\em Journal of the Royal Statistical Society\/}~{\em B 42},
  109--127.

\bibitem[\protect\citeauthoryear{Meinshausen}{Meinshausen}{2006}]{meinshausen2006quantile}
Meinshausen, N. (2006).
\newblock Quantile regression forests.
\newblock {\em Journal of Machine Learning Research\/}~{\em 7\/}(Jun),
  983--999.

\bibitem[\protect\citeauthoryear{Park, Mammen, Lee, and Lee}{Park
  et~al.}{2015}]{park2015varying}
Park, B.~U., E.~Mammen, Y.~K. Lee, and E.~R. Lee (2015).
\newblock Varying coefficient regression models: a review and new developments.
\newblock {\em International Statistical Review\/}~{\em 83\/}(1), 36--64.

\bibitem[\protect\citeauthoryear{Peterson and Harrell}{Peterson and
  Harrell}{1990}]{PetHar:90}
Peterson, B. and F.~E. Harrell (1990).
\newblock Partial proportional odds models for ordinal response variables.
\newblock {\em Applied Statistics\/}~{\em 39}, 205--217.

\bibitem[\protect\citeauthoryear{Strobl, Boulesteix, and Augustin}{Strobl
  et~al.}{2007}]{StrBouAug:2007}
Strobl, C., A.-L. Boulesteix, and T.~Augustin (2007).
\newblock Unbiased split selection for classification trees based on the gini
  index.
\newblock {\em Computational Statistics \& Data Analysis\/}~{\em 52}, 483--501.

\bibitem[\protect\citeauthoryear{Strobl, Boulesteix, Kneib, Augustin, and
  Zeileis}{Strobl et~al.}{2008}]{strobl2008conditional}
Strobl, C., A.-L. Boulesteix, T.~Kneib, T.~Augustin, and A.~Zeileis (2008).
\newblock Conditional variable importance for random forests.
\newblock {\em BMC bioinformatics\/}~{\em 9\/}(1), 307.

\bibitem[\protect\citeauthoryear{Strobl, Boulesteix, Zeileis, and
  Hothorn}{Strobl et~al.}{2007}]{strobl2007bias}
Strobl, C., A.-L. Boulesteix, A.~Zeileis, and T.~Hothorn (2007).
\newblock Bias in random forest variable importance measures: Illustrations,
  sources and a solution.
\newblock {\em BMC bioinformatics\/}~{\em 8\/}(1), 25.

\bibitem[\protect\citeauthoryear{Strobl, Malley, and Tutz}{Strobl
  et~al.}{2009}]{Strobetal:2009}
Strobl, C., J.~Malley, and G.~Tutz ({2009}).
\newblock {An Introduction to Recursive Partitioning: Rationale, Application
  and Characteristics of Classification and Regression Trees, Bagging and
  Random Forests.}
\newblock {\em {Psychological Methods}\/}~{\em {14}}, {323--348}.

\bibitem[\protect\citeauthoryear{Tutz}{Tutz}{2012}]{TutzBook2011}
Tutz, G. (2012).
\newblock {\em {Regression for Categorical Data}}.
\newblock Cambridge University Press.

\bibitem[\protect\citeauthoryear{Venables and Ripley}{Venables and
  Ripley}{2002}]{VenRip:2002}
Venables, W.~N. and B.~D. Ripley (2002).
\newblock {\em Modern Applied Statistics with S. Fourth edition}.
\newblock New York: Springer--Verlag.

\bibitem[\protect\citeauthoryear{Waldmann, Kneib, Yue, Lang, and
  Flexeder}{Waldmann et~al.}{2013}]{waldmann2013bayesian}
Waldmann, E., T.~Kneib, Y.~R. Yue, S.~Lang, and C.~Flexeder (2013).
\newblock Bayesian semiparametric additive quantile regression.
\newblock {\em Statistical Modelling\/}~{\em 13\/}(3), 223--252.

\bibitem[\protect\citeauthoryear{Williams}{Williams}{2016}]{williams2016understanding}
Williams, R. (2016).
\newblock Understanding and interpreting generalized ordered logit models.
\newblock {\em The Journal of Mathematical Sociology\/}~{\em 40\/}(1), 7--20.

\bibitem[\protect\citeauthoryear{Wu and Liu}{Wu and Liu}{2009}]{wu2009variable}
Wu, Y. and Y.~Liu (2009).
\newblock Variable selection in quantile regression.
\newblock {\em Statistica Sinica\/}, 801--817.

\bibitem[\protect\citeauthoryear{Yu and Jones}{Yu and
  Jones}{1998}]{yu1998local}
Yu, K. and M.~Jones (1998).
\newblock Local linear quantile regression.
\newblock {\em Journal of the American statistical Association\/}~{\em
  93\/}(441), 228--237.

\end{thebibliography}
